\documentclass[10pt, journal, letterpaper]{IEEEtran}

\usepackage[utf8]{inputenc}
\usepackage{graphicx}
\usepackage{listings}
\usepackage{amsmath}
\usepackage{amssymb}
\usepackage{stackrel}
\usepackage{framed}
\usepackage{float}
\usepackage{caption}
\usepackage{subcaption}
\usepackage{array}

\usepackage{algorithm}
\usepackage{cite}
\usepackage{amsthm}
\usepackage{hyperref}
\newtheorem{lemma}{Lemma}

\newtheorem{theorem}{Theorem}
\newtheorem{definition}{Definition}

\usepackage{algorithmic}
\usepackage[shortlabels]{enumitem}
\usepackage{bigints}
\usepackage{xcolor}

\newcommand{\eps}{\epsilon}

\newcommand{\R}{\mathcal{R}}

\renewcommand{\t}{\theta}

\newif\ifhideappendix
\hideappendixfalse 

\ifhideappendix
   \usepackage[1-14]{pagesel}
\fi
\usepackage{etoolbox}
\makeatletter
\def\@IEEEBIOskipN{0pt}
\makeatother
\begin{document}
\title{Optimal Oblivious Load-Balancing for Sparse Traffic in Large-Scale Satellite Networks}
\author{Rudrapatna Vallabh Ramakanth and Eytan Modiano
\thanks{R Vallabh Ramakanth and Eytan Modiano are affiliated with the Laboratory of Information and Decision Systems (LIDS), Massachusetts Institute of Technology, Cambridge, MA, 02139. A preliminary version of this paper appeared in the conference proceedings of IEEE INFOCOM 2026 -- Digital Object Identifier (DOI): \url{10.1109/INFOCOM59046.2026.11571543}.\\
The authors would like to thank SES S.A. for their gracious support toward this research.\\
Email: \{vallabhr, modiano\}@mit.edu
}}
\date{}
\maketitle

\begin{abstract}
 Oblivious load-balancing involves routing traffic from sources to destinations using predetermined paths independent of the traffic to minimize the maximum load on any link in the network.
This paper addresses oblivious load-balancing in large-scale Low-Earth Orbit satellite networks which are modeled as torus networks with sparse traffic localized to certain hotspot regions.
For an $N\times N$ torus network with at most $k$ active source-destination pairs, we establish that any oblivious routing scheme incurs a worst-case load of at least $\sqrt{2k}/4$ for $1<k \leq N^2/2$,
and $N/4$ for $N^2/2\leq k\leq N^2$, by using a linear programming formulation. 
Moreover, we demonstrate that the celebrated Valiant Load-Balancing scheme is suboptimal under sparse traffic and construct an optimal oblivious load-balancing scheme that achieves the proposed lower bound. Further, we identify a $\sqrt{2}$ multiplicative gap between the worst-case load of an optimal non-oblivious routing and the worst-case load of any optimal oblivious routing. 
We extend the load-balancing results to $N\times M$ tori with unequal link capacities along the vertical and horizontal directions, which model satellite network topologies more realistically. Additionally, we extend the result to higher-dimensional torus networks, also known as $N-$ary $m-$cubes, which are relevant to datacenter and interconnection networks.
\end{abstract}
\begin{IEEEkeywords}
satellite networks, load-balancing, torus networks, sparse traffic, oblivious routing
\end{IEEEkeywords}
\section{Introduction}
The performance of a data communication network critically depends on the underlying routing policy. Designing a good routing policy is a challenging task, and it is strongly linked to how accurately the network operator knows the underlying traffic pattern. Indeed, if network operators could predict the dynamics of traffic accurately, they can appropriately route the traffic in order to balance the load on all the links. Unfortunately, this is far from reality, first, because it is extremely difficult to predict future traffic and second, because modifying routing tables rapidly poses a large computational overhead to the network. Moreover, satellite network operators prefer setting routing tables beforehand for easier network management. Furthermore, a static routing policy can set a benchmark for any dynamic routing scheme that adapts to the underlying traffic. Naturally, this leads us to asking what is the best static routing policy that can effectively balance the load for any underlying traffic pattern. 

\begin{figure}[!hbt]
    \centering
\includegraphics[width=0.55\linewidth]{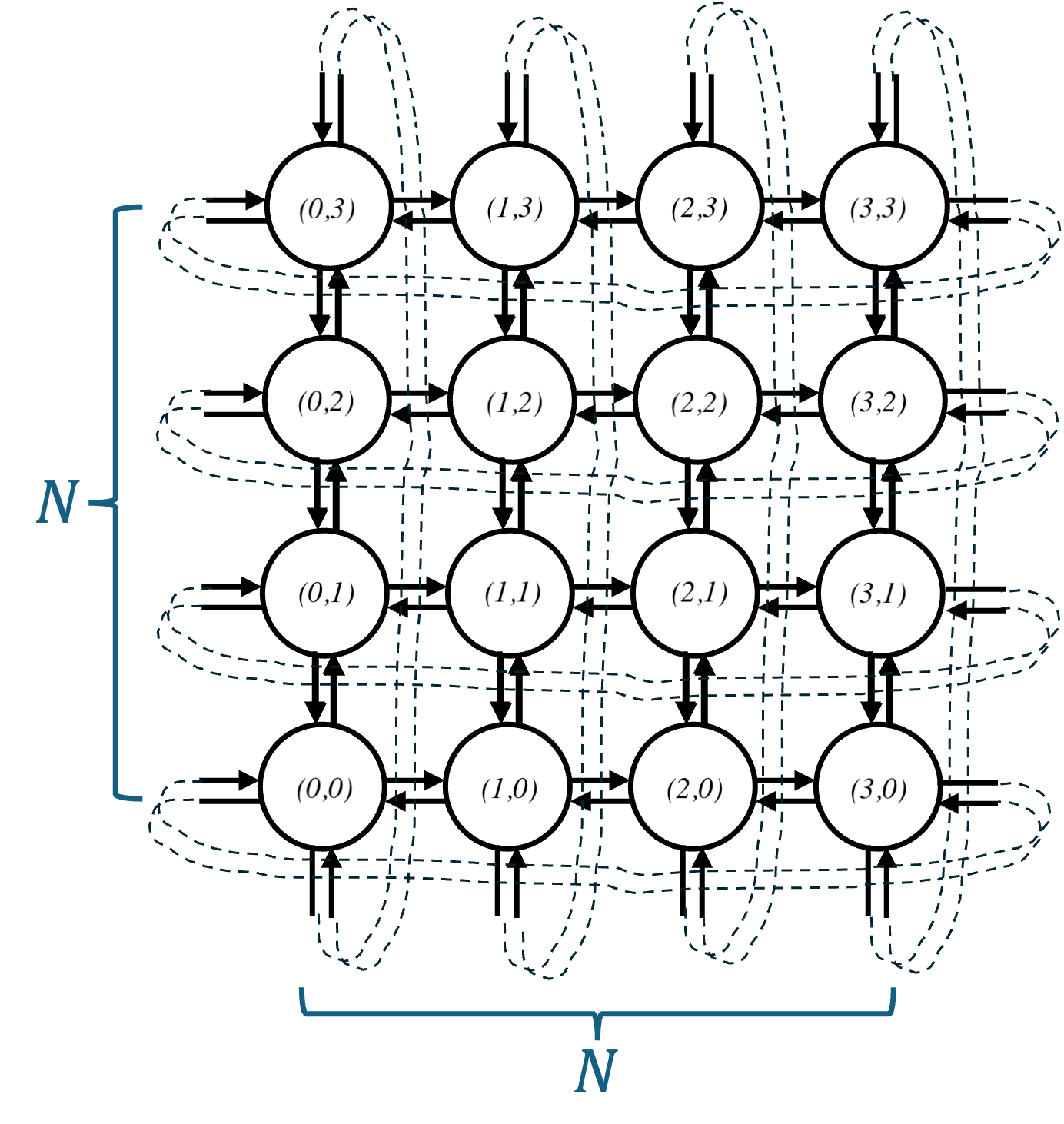}
    \caption{$N\times N$ torus with $N=4$.}
    \label{fig:torus}
\end{figure}

Oblivious routing policies are predetermined routes from sources to destinations that are independent of the underlying traffic, which aim to minimize the maximum load on any link. In this work, we are primarily motivated by the problem of oblivious load-balancing of traffic in large-scale Low-Earth Orbit (LEO) satellite constellations designed for data communication. 
We consider a satellite network in which each satellite maintains four intersatellite links (ISLs): two to neighboring satellites in the same orbit and two to satellites in adjacent orbits. These ISLs are relatively stable and easy to maintain \cite{handley_delay_2018}. A common dynamic model for such networks is the snapshot model \cite{werner_dynamic_1997, ekici_distributed_2001}.
For simplicity, we assume that at each snapshot the network can be modeled as a toroidal mesh \cite{sun_capacity_2003, sun_routing_2004, Ramakanth2025} (see Fig. \ref{fig:torus}). This assumption is reasonable because the ISL topology remains approximately constant over routing timescales \cite{handley_delay_2018}. While the satellite layer is relatively regular and static, RF links to ground nodes are highly dynamic \cite{handley_delay_2018}. We therefore abstract the RF layer by assuming that traffic enters and exits directly at satellite nodes, and focus exclusively on load-balancing within the satellite layer. Routing in the RF layer is handled independently by each satellite and is beyond the scope of this work.

While there have been multiple works studying load-balancing in satellite networks \cite{kim_traffic_1998, ekici_distributed_2001, sun_capacity_2003, neely_dynamic_2003, liu_low-complexity_2015, deng_distance-based_2023, Ramakanth2025}, they mostly do so with restrictive assumptions on traffic or using queue-based estimates of traffic. Our focus is on developing predetermined routing strategies that guarantee optimal worst-case load for sparse traffic, with only minimal assumptions on the underlying traffic pattern over the intersatellite network. In general, we assume that the satellite network can be modeled as an $N\times M$ torus network with different capacities along the vertical directions and horizontal directions. We first present our results for the $N\times N$ torus network for ease of presentation, and then move on to the general $N\times M$ case in later sections.
Oblivious routing policies for structured networks have been widely researched in the past \cite{applegate_making_2006, chitavisutthivong_optimal_2023}. For $N\times N$ torus networks, the optimal oblivious routing problem has been studied in \cite{towles_throughput-centric_2003, dally_principles_2004, ramanujam_randomized_2013}. However, prior works mostly optimize for load over a class of traffic called the hose model\cite{duffield_flexible_1999} wherein every node can produce a maximum of one unit of traffic destined to any other node, and every node can also receive a maximum of one unit of traffic from any other node. Under this class of traffic, the celebrated Valiant Load-Balancing\cite{valiant_scheme_1982, zhang-shen_designing_2005} scheme applied to an $N\times N$ torus is known to be one of the simplest optimal oblivious routing strategies\cite{dally_principles_2004, ramanujam_randomized_2013}. However, this is true only when \emph{all} nodes in the network generate traffic.

\begin{figure}[h!]
    \centering
\includegraphics[width=0.8\linewidth]{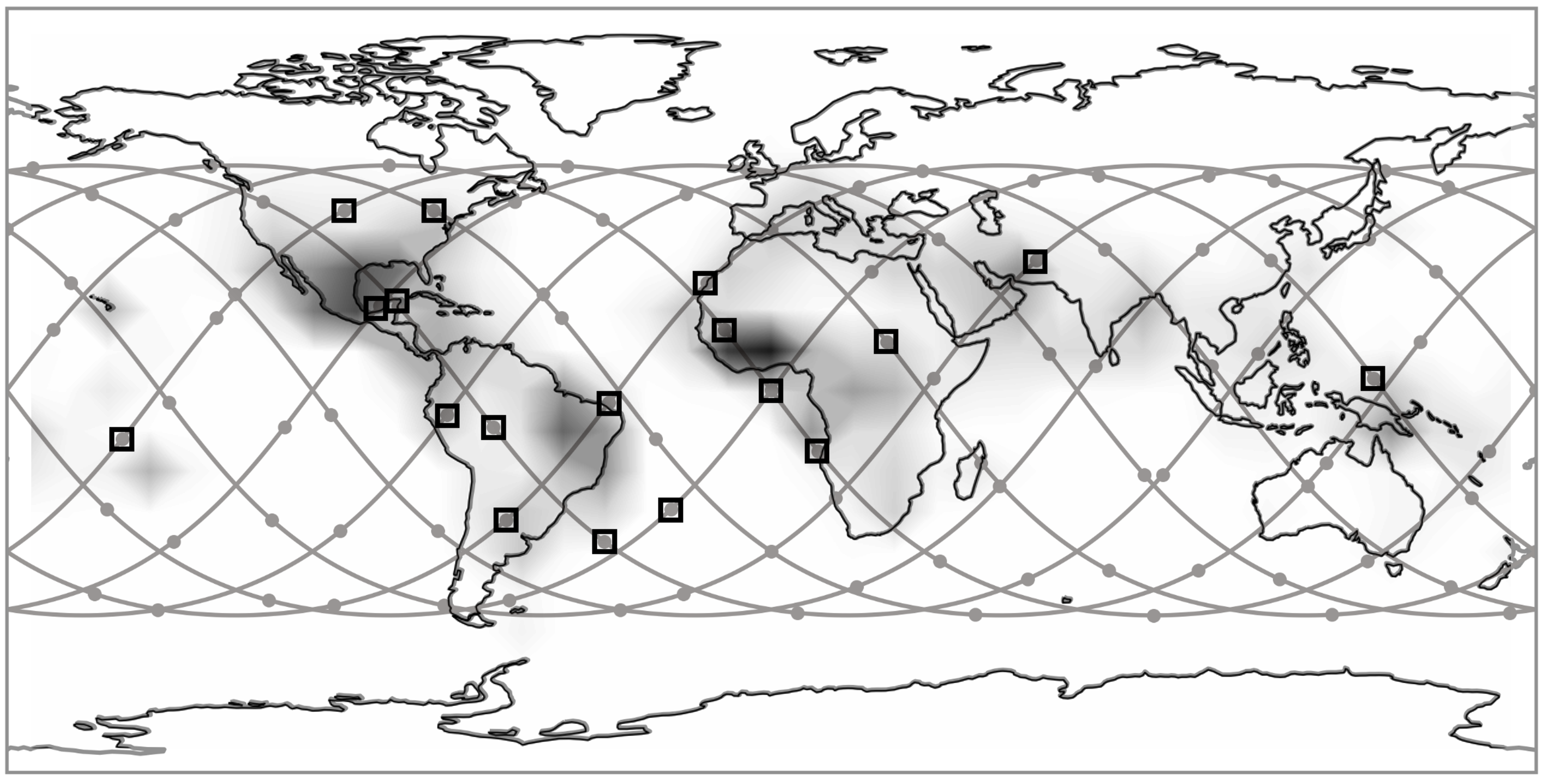}
    \caption{World traffic heatmap at particular time of day, overlaid with a 10 orbit constellation with 100 satellites. Dark regions indicate high traffic. Satellites that would experience high traffic demand are boxed. Data provided by Satellite Telecom Operator SES S.A.}
    \label{fig:constellation_traffic}
\end{figure}

The novelty in our work is in understanding how to effectively distribute \emph{sparse} traffic over a torus network. In other words, we assume that a maximum of $k$ nodes in the network are responsible for the bulk of the traffic in the network. We call this the $k-$sparse traffic model.
The constraint of sparse traffic is of particular interest because in many real-world networks, a small fraction of the nodes produce the bulk of the traffic. It is especially true in the context of satellite networks given the widely heterogeneous distribution of land and population over the world, where most of the data traffic comes from localized population hotspots. 
We demonstrate this in Figure \ref{fig:constellation_traffic}, using an example of aggregate regional traffic demands provided to us by a satellite telecom operator. An example Walker-Delta constellation of 100 satellites with an inclination of $53^\text{o}$ and 10 orbital planes is superimposed over the Earth. It is easy to see that only a small number of satellites localized to certain locations (hotspots) experience significantly high traffic in this constellation. 
These hotspots can shift over time, making it critical for this hotspot traffic to be effectively distributed over the network.

In this work, we are primarily interested in the case where $1 < k \leq N^2/2$. When $k \geq N^2/2$, the traffic is no longer ``sparse" and the Valiant Load-Balancing Scheme can be shown to be optimal.
The contributions of our work are summarized below.
\begin{enumerate}[1), leftmargin=*, nosep]
    \item We show a rather surprising result that the Valiant Load-Balancing scheme is suboptimal under $k-$sparse traffic (see Section \ref{subsec:VLB_suboptimality}).
    \item We characterize the universal lower bound on the maximum link load of any oblivious routing scheme under $k-$sparse traffic. We show that no oblivious policy can achieve a maximum link load of lower than approximately $\sqrt{2k}/4$ for all $k-$sparse traffic matrices. We do this by proposing a particular traffic matrix, which we call the Split-Diamond traffic, that incurs a load of $\sqrt{2k}/4$ for all routing policies that satisfy certain symmetry conditions (see Theorem \ref{thm:optimal_oblivious_lower_bound}).
    \item We construct an optimal routing policy, which we call the Local Load-Balancing (LLB) scheme, that achieves a maximum link load of approximately $\sqrt{2k}/4$ for all $k-$sparse traffic matrices (see Theorem \ref{thm:optimal_oblivious_upper_bound}). 
    \item  Lastly, we extend our results to the general case of an $N\times M$ torus with unequal link capacities along the vertical and horizontal directions in Section \ref{sec:asymmetric_tori}, and to $N-$ary $m-$cubes in Section \ref{sec:higher_dims}.
\end{enumerate}
The rest of the paper is organized as follows. In Section \ref{sec:notation_prob_formulation}, we describe the model of the satellite network, routing policies and classes of traffic in detail, and formulate the oblivious maximum link load problem. Section \ref{sec:prelim} develops a weak cut-based lower bound for the maximum link load of any routing policy under $k-$sparse traffic, demonstrates the suboptimality of the Valiant Load-Balancing Scheme, and describes a sufficient condition on the optimal routing policy. The main results of the paper, the lower and upper bounds, are presented in Sections \ref{sec:worst_case_lb} and \ref{sec:routing_ub} respectively.
We extend our work to $N\times M$ tori with unequal link capacities along the vertical and horizontal directions in Section \ref{sec:asymmetric_tori}, and to higher dimensions in Section \ref{sec:higher_dims}.
Finally, we conclude the paper in Section \ref{sec:results}.
Proofs in this paper are omitted to highlight the key ideas of this work while keeping space limitations in mind.
Finally, although this work is focused primarily on satellite networks, the results are also applicable to torus networks commonly used in other domains such as interconnection networks, optical networks and datacenter networks.
\newcommand{\D}{\mathcal{D}}
\newcommand{\MLU}{\textsc{MaxLoad}}
\section{Model and Problem Formulation}\label{sec:notation_prob_formulation}
\subsection{Network Model and Notation}\label{subsec:network_model}
We consider an $N\times N$ square 2D-torus (see Fig. \ref{fig:torus}), which we denote in graph notation as $G(V,E)$, where $V$ is the set of nodes and, $E$ is the set of directional links in the network. 
We label each node $i \in V$ by a vector $(i_x, i_y) \in \{0,...,N-1\}\times \{0,..., N-1\}$. In other words, $V = \{0,...,N-1\}\times \{0,..., N-1\}$. For any $u,v \in \{0,...,N-1\}\times \{0,..., N-1\}$, we define the ``$+$" operation as 
$$
    u + v \triangleq \big((u_x + v_x) \bmod N, (u_y + v_y) \bmod N\big).
$$
For any $u,v \in \{0,...,N-1\}\times \{0,..., N-1\}$, we define the ``$-$" operation similarly. We also denote $0$ to be the origin node wherever appropriate, and $-v \triangleq 0 - v$ for any $v \in V$. It is important to note that operators $+$ and $-$ form a closure over $V$, i.e., they form a map from $V\times V \to V$.  Next, we define $e_1 = (1, 0)$ and $e_2 = (0,1)$. Then, the set $A = \{e_1, e_2, -e_1, -e_2\}$ denotes the set of directions along which we have links from any node $i \in V$. In other words, every node $i$ in the network has outgoing links to $\{i+e_1, i+e_2, i-e_1, i-e_2\}$. Similarly, every node $i$ also has incoming links from $\{i+e_1, i+e_2, i-e_1, i-e_2\}$. Therefore, we can denote every link in the network using the starting node $i$ and the direction $e \in A$ along which it is directed. More specifically, the set of links can be represented as $E = \{(i,i+e) \;|\; i \in V, e \in A\}$. Every directed edge in $E$ has capacity $c_{i,i+e}$ which we assume to be  $1$ for all $(i,i+e) \in E$. Finally, we use the terms ``link" and ``edge" interchangeably to refer to $(i,i+e) \in E$.

\subsection{Routing Policy}\label{subsec:routing1}
We adopt a flow based analysis for routing policies, which also permits fractional flows. This implies that traffic from a source node to a destination node can potentially be routed along different paths.
We denote node $s \in V$ to be the source of traffic and $t \in V\backslash\{0\}$ to be the displacement vector. In other words, $s$ is the source node and $s+t$ is the destination of the traffic.
Based on this notation, any routing policy can be described as a vector of flow variables $f^{s,s+t}_{i,i+e}$, such that
\begin{equation}
\sum_{e \in A} f^{s,s+t}_{i,i+e} - \sum_{e \in A} f^{s,s+t}_{i+e,i} = 
\begin{cases}
    0 & i \neq s,s+t\\
    1 & i = s\\
    -1 & i = s+t
\end{cases} \qquad \forall s, \forall t, \forall i \label{eqn:flow_conservation}
\end{equation}
and,
\begin{equation}
0 \leq f^{s,s+t}_{i, i+e} \leq 1, \forall s \in V, \forall t \in V\backslash\{0\}, \forall (i,i+e) \in E. \label{eqn:flow_positivity}
\end{equation}
Here, $f^{s,s+t}_{i,i+e}$ denotes the fraction of traffic from source node $s$ to destination node $s+t$ routed along the link $(i,i+e)$. The constraint in equation \eqref{eqn:flow_conservation} is the flow conservation constraint. Note that we need $O(|V \times V \times E|) = O(N^6)$ flow variables to describe any routing policy in general. Later, we will show that we do not need so many flow variables to describe the optimal oblivious routing policy because of the inherent symmetry of the network. Consequently, this reduces the number of required flow variables to represent the optimal oblivious policy to $O(|V\times E|) = O(N^4)$. We denote the set of all valid routing policies as $\R = \{f \; | \; f \text{ satisfies constraint \eqref{eqn:flow_conservation} and \eqref{eqn:flow_positivity} }\}$. Note that the set $\R$ is a convex polyhedron, defined by only linear constraints.

\subsection{Maximum Load on any Link}\label{subsec:max_load_def}
Our objective is to develop routing policies agnostic to the underlying traffic that balance the maximum load on every link in the best possible manner.  In other words, we want to minimize the maximum amount of traffic routed on any link. We formally define this notion. Let $\D$ be a class of traffic demands, and let $d\in\D$ be a particular traffic demand, where $d_{s,s+t}$ denotes the traffic demand from node $s$ to node $s+t$. Let $f \in \R$ be a particular routing policy. We define the maximum link load of routing policy $f$ under traffic matrix $d$, i.e., $\MLU(f,d)$ as follows.
\begin{align*}
\MLU(f,d) \triangleq \max_{(i,i+e) \in E} \; \sum_{\substack{s \in V\\t \in V\backslash\{0\}}} d_{s,s+t} f^{s,s+t}_{i,i+e}.
\end{align*}
The optimal oblivious routing for minimizing the maximum load on any link, over a class of traffic matrices $\D$, can be posed as a minmax optimization problem.  The optimal oblivious policy is then given by $f^*$ where
\begin{equation}
    \max_{d \in \D} \MLU(f^*, d) = \min_{f \in \R} \max_{d \in \D} \MLU(f, d). \label{eqn:minmax}
\end{equation}
\textbf{Remark.} \textit{The solution $f^*$ guarantees that the load never exceeds $\max_{d \in \D} \MLU(f^*, d)$ for any traffic demand in $\D$. Thus, traffic may vary over time, yet the load guarantee remains valid as long as demands stay within the class $\D$. This property is particularly valuable for satellite networks.}

If $\D$ is a convex polytope of traffic matrices, then one can pose the above optimization problem as a linear program\cite{dally_principles_2004, towles_throughput-centric_2003, applegate_making_2006}.  Therefore, for any convex polytope of traffic matrices $\D$, the above program can be efficiently solved for moderately sized networks. However, this can get computationally challenging as the network size increases. 
In the next subsection, we define a special class of traffic matrices that are sparse, which are of practical importance.

\subsection{The Class of Sparse Traffic Matrices}\label{subsec:sparse_traffic_def}
In this work, we start with the hose model which is quite popular in literature\cite{duffield_flexible_1999,kodialam_maximum_2006,towles_throughput-centric_2003}. It is a practical model for traffic, which assumes that any node in the network can produce or sink traffic only up to a finite maximum capacity. In our hose model, without loss of generality, we assume that each node can source or sink one unit of traffic, that is,
\begin{equation}
    \sum_{t \in V\backslash\{0\}} d_{s,s+t} \leq 1 \quad\forall s \in V\\ \label{eqn:source_rate_constraint}
\end{equation}
\begin{equation}
    \sum_{t \in V\backslash\{0\}} d_{s-t,s} \leq 1 \quad\forall s \in V. \label{eqn:sink_rate_constraint}
\end{equation}
However, in reality, it is often found that only a few nodes in the network produce large amounts of traffic. In other words, the traffic is sparse or limited. To that end, we impose an additional constraint that the traffic matrices under consideration should have at most $k$ sources and $k$ sinks.
In other words,
\begin{align}
&\sum_{s \in V} \mathbf{1}({\text{node $s$ sources non-zero traffic})} \leq k\nonumber\\
&\sum_{s \in V} \mathbf{1}({\text{node $s$ sinks non-zero traffic})} \leq k, \label{eqn:sparse_constraint}
\end{align}
where $\mathbf{1}({A})$ is the indicator function for condition $A$. A traffic matrix $d$ is said to be $k-$sparse if it satisfies constraints \eqref{eqn:source_rate_constraint}, \eqref{eqn:sink_rate_constraint} and \eqref{eqn:sparse_constraint}. We denote the set of all $k-$sparse traffic matrices as $\D'_k$ defined as
$$
\D'_k = \{d \;| \; d \text{ is $k-$sparse}\}.
$$
Note that $\D'_k$ is highly non-convex.
We can also define a notion of $k-$limited traffic matrices. When a traffic matrix $d$ satisfies  
\begin{equation}
\sum_{s, t} d_{s,s+t} \leq k, \label{eqn:total_traffic_constraint}
\end{equation}
along with constraints \eqref{eqn:source_rate_constraint} and \eqref{eqn:sink_rate_constraint}, it is said to be $k-$limited. This is because the total traffic demand is lesser than $k$ units.
 We denote the set of all $k-$limited traffic matrices as $\D_k$, where
$$
\D_k = \{d \;| \; d \text{ is $k-$limited}\}.
$$
Note that $\D_k$ is a convex polyhedron, as it is defined by only linear constraints. One can intuitively think of class $\D_k$ as a convex relaxation of class $\D'_k$. It is also easy to see that $\D'_k \subset \D_k$. However, more interestingly, both classes of traffic are equivalent for $\MLU$. 
\begin{lemma}\label{lemma:k_sparse_k_limited_equivalence}
    For any valid routing policy $f$, $\max_{d \in \D_k} \MLU(f, d) = \max_{d \in \D'_k} \MLU(f, d)$.
\end{lemma}
The proof of Lemma \ref{lemma:k_sparse_k_limited_equivalence} comes from showing that the vertices of the polytope of $\D_k$ are in $\D'_k$, i.e., $\text{vertices}(\D_k) \subset \D'_k$.
Using this fact and the principles of linear programming, it is easy to see that for any routing policy, there is a worst-case traffic in the set $\D_k$ that would also be the worst-case traffic in the set $\D'_k$. Consequently, an oblivious routing policy is optimal over traffic class $\D_k$ if and only if it is an optimal oblivious routing policy over traffic class $\D'_k$. The details are deferred to Appendix \ref{proof:vertices}.
Hence, our mathematical development depends on, and applies to, the convex class of $k-$limited traffic matrices $\D_k$, which also captures our $k-$sparse traffic of interest.


\subsection{Problem Formulation}\label{subsec:problem_formulation}
We want to understand the performance and characterize optimal oblivious routing policies when the traffic is known to be $k-$limited. Formally, we want to find the optimal value of the worst-case maximum load over all $k-$limited traffic given by 
\begin{align}
\theta^* = \min_{f \in \R} \max_{d \in \D_k} \MLU(f, d), \label{eqn:main_problem1}
\end{align}
and also find the optimal routing policy $f^*$ that achieves $\MLU(f^*, d) \leq \theta^*$ for all $d \in \D_k$. 

The above problem is in fact a linear program, as $\R$ and $\D_k$ are convex polyhedra. This is because \eqref{eqn:main_problem1} can be rewritten as
\begin{subequations}
\begin{align}
    \theta^* =  \min_{f} \quad &\theta \nonumber \\
                \text{s.t.} \quad &\max_{d \in \D_k}\sum_{s,t}d_{s,s+t}f^{s,s+t}_{i,i+e} \leq \theta \quad \forall i,e \label{eqn:max_constraint1}\\
                & f \text{ satisfies linear constraints \eqref{eqn:flow_conservation} and \eqref{eqn:flow_positivity}.}
\end{align}
\end{subequations}

Observe that constraint \eqref{eqn:max_constraint1} requires finding a maximum over the set $\D_k$ for every feasible routing policy. 
This means that we have to solve a linear program every time we need to ensure that constraint \eqref{eqn:max_constraint1} is satisfied. 
However, this can be easily circumvented using strong duality-based techniques  \cite{applegate_making_2006, towles_throughput-centric_2003, chitavisutthivong_optimal_2023, kodialam_maximum_2006}. These works employ strong duality to repose the oblivious max-load problem as a linear program.
Although the problem of finding the optimal routing is a linear program, the number of constraints and variables becomes prohibitively large to computationally solve even for a moderately sized network.
To that end, we aim to bound the performance of oblivious routing for sparse traffic by exploiting the structure in the problem.
In the next section, we develop a preliminary cut-based lower bound for the maximum load. Furthermore, we reduce the search space for optimal routing policies by invoking a sufficiency condition for optimality.

\section{Preliminary Analysis}
\label{sec:prelim}
We develop some preliminary bounds on the max load in \eqref{eqn:main_problem1} and properties of the optimal routing $f^*$. We first start with a lower bound of the minmax optimization problem.

\newcommand{\mincut}{\textsc{MinCut}}
\subsection{Cut-Based Lower Bound}\label{subsec:cut_based_lb}
\begin{figure}
     \centering
     \includegraphics[width=0.5\linewidth]{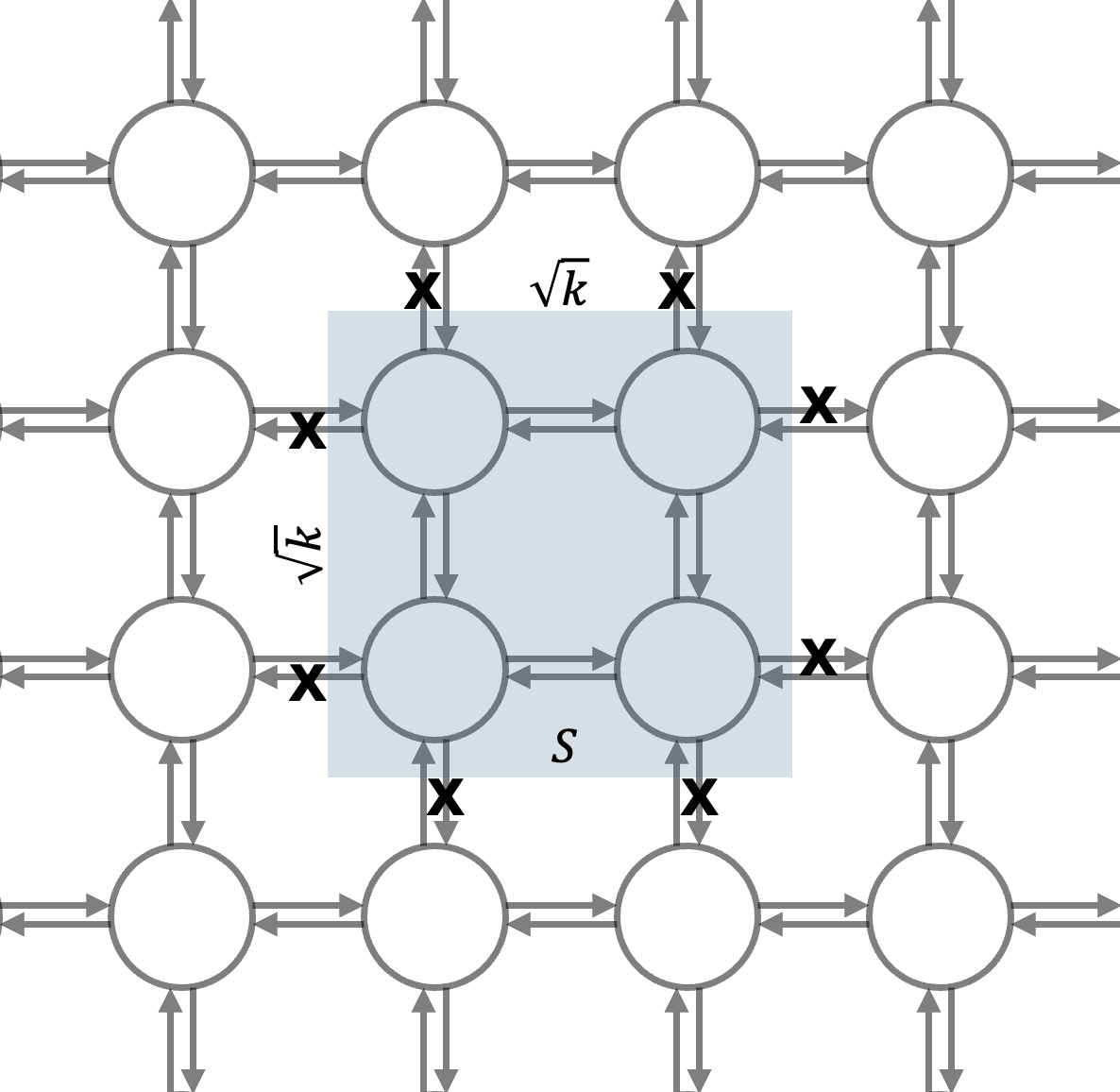}
     \caption{The minimum cut set for $k = 4$. The set of vertices $S$ is highlighted. The edges to be cut are marked with a cross.}
     \label{fig:cut_set}
 \end{figure}

Let us split the nodes in the network $V$ into two sets $S$ and $T = V\backslash S$, such that all the nodes in $S$ source one unit of traffic, and the destinations are in $T$. Clearly, all the traffic from $S$ destined to $T$ must flow through a cut set of $S$ and $T$, denoted as $C(S,T)$. Formally, a cut set $C(S,T)$ is a set of edges that are directed from $S$ to $T$ such that, once removed, there exists no directed path from any node in $S$ to any node in $T$. 
For the constructed traffic matrix to be in $\D_k$, we require the total number of sources $|S| \leq k$. 
For any routing policy $f$, the load must be at least
$$
\max_{d \in \D_k}\MLU(f,d) \geq \frac{k}{\min_{|S| = k} |C(S, T)|}.
$$
 From the result in \cite{sun_capacity_2003} on minimum sized cuts in a torus, for $k \leq N^2/4$, the minimum cut of a vertex set of size $k$ is roughly $4\sqrt{k}$. This is because, the minimum cut set of size $k$ is roughly a square of side length $\sqrt{k}$\cite{sun_capacity_2003}. This is visually depicted in Figure \ref{fig:cut_set}. 
 
 Consequently, the maximum load over $\D_k$ will be at least $\sqrt{k}/4$ for \emph{any} routing policy, even non-oblivious ones.\\
 \textbf{Remark.} \textit{This lower bound is not tight for oblivious routing policies. In Section \ref{sec:worst_case_lb}, we develop a tighter lower bound of $\sqrt{2k}/4$ on the maximum load of any oblivious routing.}
\subsection{Suboptimality of the Valiant Load-Balancing Scheme}\label{subsec:VLB_suboptimality}
We demonstrate the suboptimality of Valiant Load-Balancing (VLB) scheme for $k-$sparse traffic when $k$ is small. To do this, we construct a traffic matrix such that the maximum link load achieved by the VLB scheme is bounded away from the maximum link load developed in Section \ref{sec:routing_ub}. 

The VLB scheme is easily understood as a two phase routing scheme. First, the traffic from every source is  uniformly distributed to all the nodes in the network, which we call the intermediate nodes, along their respective shortest paths. Second, the intermediate nodes forward the traffic to their respective destinations along respective shortest paths. 

\begin{figure}[!t]
    \centering
    \includegraphics[width=0.5\linewidth]{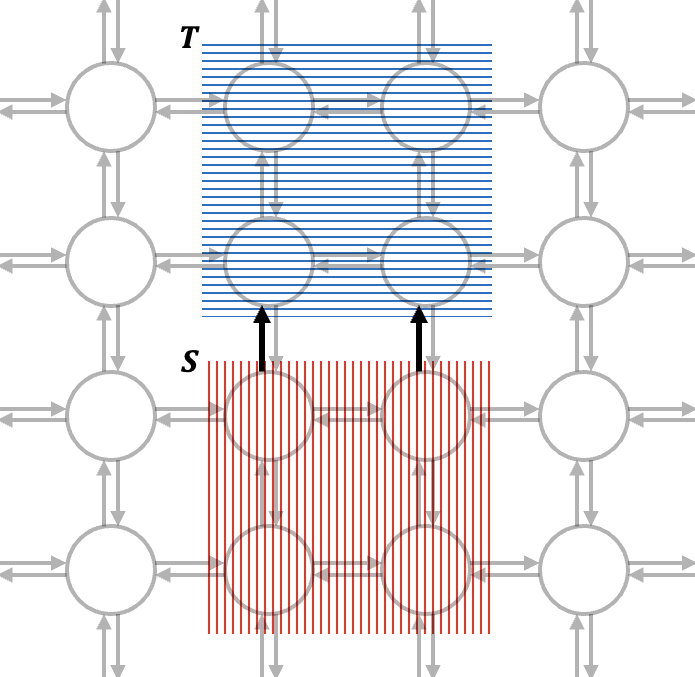}
    \caption{Set of source nodes $S$ is shaded with red vertical lines. Set of sink nodes $T$ is shaded with blue horizontal lines. The dark edges in the figure lie in the cut-set of both $S$ and $T$. These edges would see a load of at least $\frac{2\sqrt{k}}{4}(1-k/N^2)$ under the VLB scheme.}
    \label{fig:VLB_suboptimality}
\end{figure}

Now, consider a sparse traffic matrix as shown in Figure \ref{fig:VLB_suboptimality}, wherein a set $S$ of $k$ sources, arranged as a square, must send traffic across to an adjacent set $T$ of $k$ sinks, also arranged as a square. For this traffic matrix, the VLB scheme would first distribute all the traffic from $S$ uniformly to all nodes in the network, then uniformly aggregate traffic from all nodes in the network to $T$.
This implies that a fraction of $1-k/N^2$ of the total traffic must exit set $S$, which leads to the each link in the minimum cut-set of $S$ having to support at least $\frac{k(1-k/N^2)}{4 \sqrt{k}}$ units of traffic. Similarly, each link in the minimum cut-set of $T$ will have to support at least $\frac{k(1-k/N^2)}{4 \sqrt{k}}$ units of traffic (because $1-k/N^2$ of the total traffic must end up inside $T$ through the minimum cut-set). Now, the two dark links shown in Figure \ref{fig:VLB_suboptimality} would see this load twice since they lie in the minimum cut-set of both $S$ and $T$. Therefore, these links have to support a total traffic of at least $2 \times \frac{k(1-k/N^2)}{4 \sqrt{k}}$. This tells us that the worst-case maximum link load for sufficiently sparse traffic is at least $\frac{2\sqrt{k}}{4}(1-k/N^2)$ under the VLB scheme. This is clearly bounded away from the minimum achievable worst-case link load of $\sqrt{2k}/4$, shown in Sections \ref{sec:worst_case_lb} and \ref{sec:routing_ub}.    

However, we show that the VLB scheme is an optimal oblivious routing for $k-$sparse traffic when $k \geq N^2/2$ and achieves a maximum link load of $N/4$ for all traffic in $\D_k$, $k \geq N^2/2$. 
\begin{lemma}VLB is an optimal oblivious routing for $k-$sparse traffic when $k \geq N^2/2$.
\end{lemma}
\begin{proof}\label{proof:VLB_optimality_N_4}
Observe that when $k = N^2/2$,  the worst-case load would be at least $\frac{N^2/2}{\textsc{BisectionCut}} = \frac{N^2/2}{2N} = N/4$, where the $\textsc{BisectionCut}$ is the minimum number of links you need to remove such that \textbf{no} traffic from a particular half of the network can reach the other half. It is easy to see that the worst-case load is non-decreasing as $k$ increases.
However, it is well-known that Valiant Load-Balancing achieves a maximum link load of approximately $N/4$ \cite{dally_principles_2004,ramanujam_randomized_2013} for any traffic $d$ such that $\sum_{t} d_{s,s+t} \leq 1$, $\forall s\in V$ and $\sum_{t} d_{s-t,s} \leq 1$, $\forall s\in V$. Therefore, Valiant Load-Balancing is an optimal oblivious load balancing scheme when $k \geq N^2/2$.
\end{proof}
\subsection{Sufficiency Condition for Optimality}\label{subsec:automorphism_sufficiency}
 \begin{figure*}[!hbt]
     \centering
     \begin{subfigure}{0.45\textwidth}
         \centering
         \includegraphics[width=0.7\textwidth]{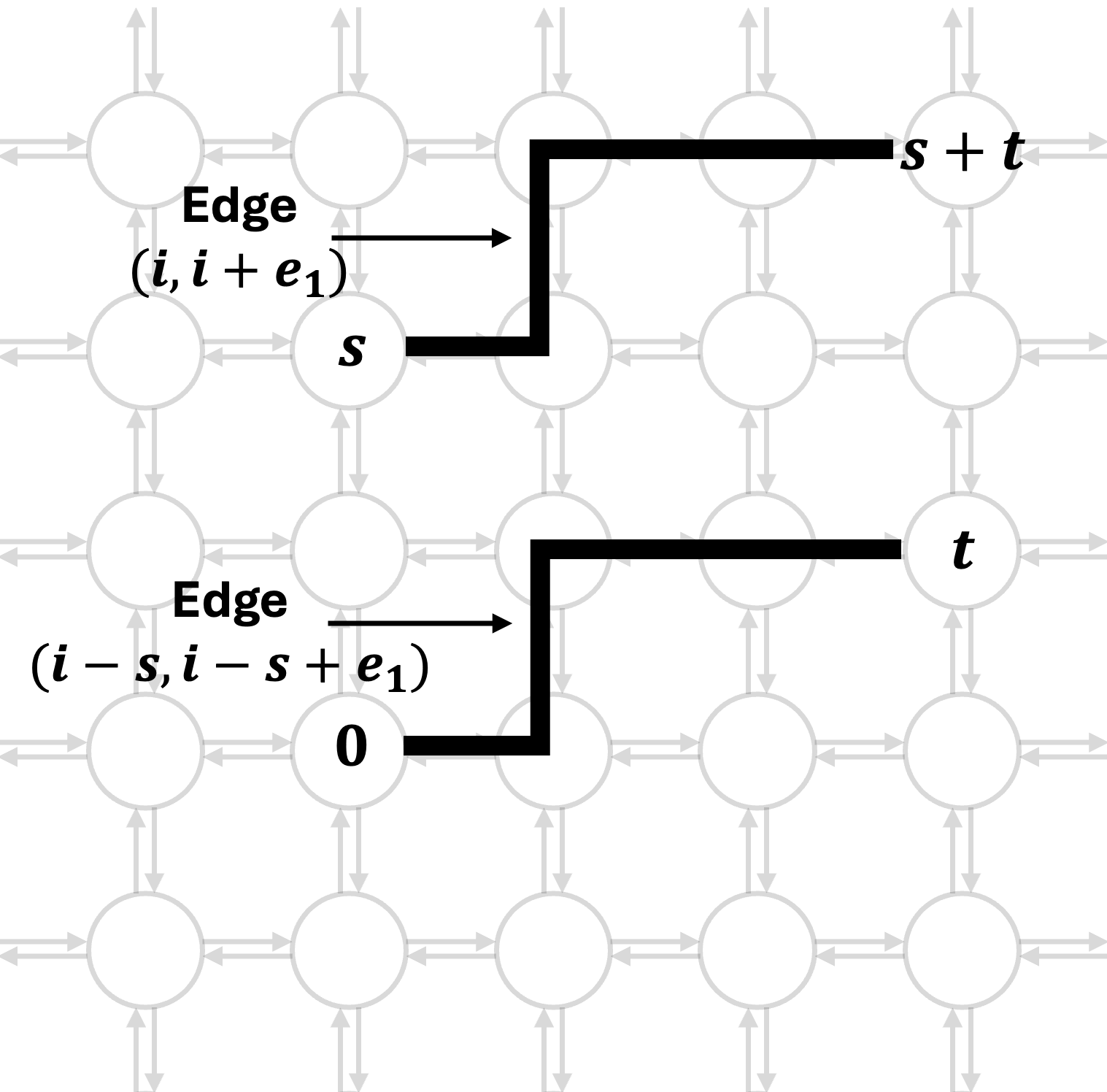}
         \caption{Due to translation invariance, the route from origin $0$ to destination $t$, is the same as the translated route from source $s$ to destination $s+t$. The flow along edge $(i,i+e)$ from $s$ to $s+t$ is $f^{s,s+t}_{i,i+e} = 1$. The flow along edge $(i-s,i-s+e)$ from $0$ to $t$ is $g^{t}_{i-s,i-s+e} = 1$.}
         \label{fig:translation_sym}
     \end{subfigure}
     \hfill
     \begin{subfigure}{0.45\textwidth}
         \centering
         \includegraphics[width=0.75\textwidth]{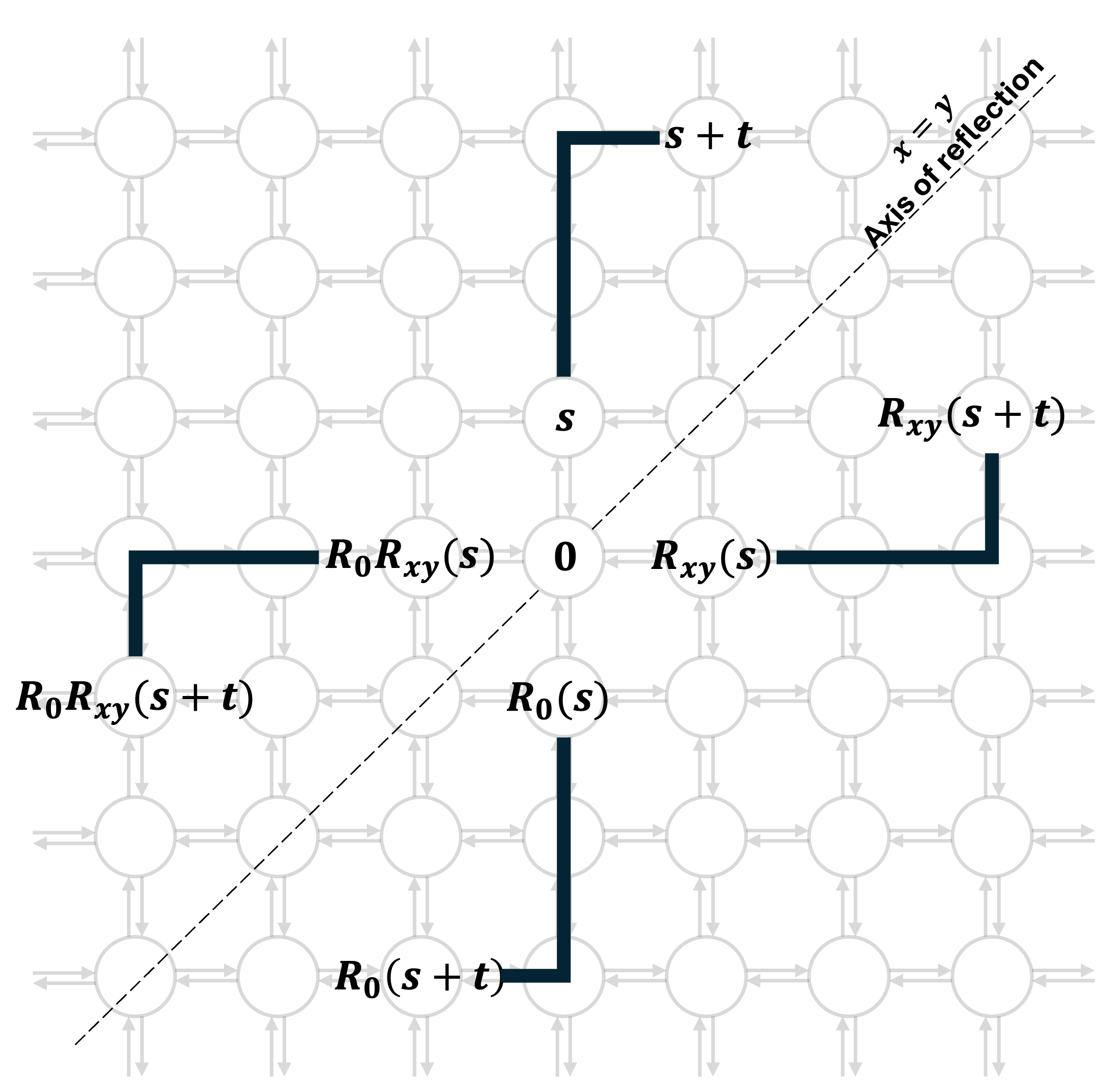}
         \caption{Due to the $R_{xy}$ reflection automorphism, the route from origin $s$ to destination $s+t$ is the same as the reflected route from source $R_{xy}(s)$ to destination $R_{xy}(s+t)$. Similarly, the case for the $R_{0}$ and $R_0R_{xy}$ automorphism is also shown.}
         \label{fig:reflection_sym}
     \end{subfigure}
     \caption{Symmetries of Routing Policies}
\end{figure*}
In this section, we describe sufficient conditions on the optimal routing in problem \eqref{eqn:main_problem1} (Lemma \ref{lemma:symmetry_sufficiency_condition}). The sufficient condition will drastically reduce the number of constraints and variables and further simplify the problem. Then, the sparse oblivious routing problem is reposed with fewer constraints and variables in Lemma \ref{lemma:reduced_problem}. 

The $N\times N$ torus network exhibits high degree of symmetry. Intuitively, symmetry means that the network looks the same even after translation or reflection transformations. 
We formalize this notion of symmetry using automorphisms\cite{chitavisutthivong_optimal_2023}.
\begin{definition}[Automorphism \cite{chitavisutthivong_optimal_2023}] A mapping $\phi: V \to V$ is an automorphism of the network $G(V,E)$ when
\begin{enumerate}[(a)]
\item $\phi$ is invertible
 \item $(i, i+e) \in E \implies (\phi(i), \phi(i+e)) \in E$.
\end{enumerate}
 \end{definition}
Observe that $V$ and $E$ are closed under any automorphism. Moreover, the traffic class $\D_k$ is closed under any automorphism $\phi$. This is because if $d \in \D_k$ then, $d' \in \D_k$, where $d'_{s,s+t} \triangleq d_{\phi(s),\phi(s+t)}\; \forall s, t$.
Similarly, the class of routing policies is also closed under automorphism $\phi$ since if $f \in \R$ then $f' \in \R$, where $f'^{s,s+t}_{i,i+e} \triangleq f^{\phi(s),\phi(s+t)}_{\phi(i),\phi(i+e)}\; \forall s, t, i, e$. Automorphisms of the network allow us to formulate a sufficient condition for optimality, stated in Lemma \ref{lemma:symmetry_sufficiency_condition}.
\begin{lemma}[{Adopted from \cite[Theorem~1]{chitavisutthivong_optimal_2023}}]\label{lemma:symmetry_sufficiency_condition}
There exists an optimal oblivious routing policy such that it is invariant to all automorphisms of the network. In other words, there exists an optimal routing policy $f^*$ such that ${f^*}^{s,s+t}_{i,i+e} = {f^*}^{\phi(s),\phi(s+t)}_{\phi(i), \phi(i+e)}$, for every automorphism $\phi$ of the network.
\end{lemma}

The proof of Lemma \ref{lemma:symmetry_sufficiency_condition} comes from modifying \cite[Theorem~1]{chitavisutthivong_optimal_2023} for max load.
\label{proof:suff_condition}
\begin{proof}[Proof of Lemma \ref{lemma:symmetry_sufficiency_condition}]
The proof is immediate by invoking convexity of $\max_{d \in \D_k}\MLU(\cdot,d)$. Let $f$ be any valid routing. Let $\Phi$ be a set of automorphisms of the network, and define $\bar f$ as $\bar f ^{s,s+t}_{i,i+e} = \frac{1}{|\Phi|}\sum_{\phi \in \Phi} f^{\phi(s),\phi(s+t)}_{\phi(i), \phi(i+e)}$. $\bar f$ is a valid routing policy due to the convexity of $\R$. Then, observe that 
\begin{align*}
    \max_{d \in \D_k}&\MLU(\bar f, d) = \max_{d \in \D_k}\max_{(i,i+e) \in E}\sum_{\substack{s\in V\\t \in V\backslash\{0\}}} d_{s,s+t} \bar f^{s,s+t}_{i,i+e}\\
    &= \max_{d \in \D_k}\max_{(i,i+e) \in E}\frac{1}{|\Phi|}\sum_{\phi \in \Phi}\sum_{\substack{s\in V\\t \in V\backslash\{0\}}} d_{s,s+t} f^{\phi(s),\phi(s+t)}_{\phi(i),\phi(i+e)}\\
    &\leq \frac{1}{|\Phi|}\sum_{\phi \in \Phi}\max_{d \in \D_k}\max_{(i,i+e) \in E}\sum_{\substack{s\in V\\t \in V\backslash\{0\}}} d_{s,s+t} f^{\phi(s),\phi(s+t)}_{\phi(i),\phi(i+e)}\\
    &\stackrel{(*)}{=} \frac{1}{|\Phi|}\sum_{\phi \in \Phi}\max_{d \in \D_k}\max_{(i,i+e) \in E}\sum_{\substack{s\in V\\t \in V\backslash\{0\}}} d_{s,s+t} f^{s,s+t}_{i,i+e}\\
    &= \max_{d \in \D_k}\MLU(f, d).
\end{align*}
Equality $(*)$ follows from the closure of $E$, $\D_k$ and $\R$ under $\phi$.
Hence, the performance of any routing policy $f$ is no better than a routing policy $\bar f$ obtained by $f$ averaged over a set of automorphisms. It is easy to see that routing $\bar f$ is invariant under any automorphism in $\Phi$. Therefore, even for any optimal routing policy $f^*$, we can always construct an automorphism invariant routing policy $\bar f^*$ that can be obtained by averaging over a set of automorphisms which is no worse in maximum load.
\end{proof}

The sufficient condition tells us that it is enough to search for the optimal routing within routing policies that satisfies certain symmetry constraints. This drastically reduces the search space as it provides more structure to the linear program that can be exploited.


Now, we describe certain automorphisms of the torus network, namely, the translation and reflection automorphisms. 
\begin{enumerate}
    \item The translation automorphism by vector $v \in V$ is the map $T_{v}(i) \triangleq i-v, \; \forall i \in V$.
    \item The reflection automorphism about $x=y$ axis, denoted by $R_{xy}(\cdot)$ is defined as $R_{xy}(i) \triangleq (i_y, i_x)$, where the $x$ and $y$ coordinates are interchanged $\forall i \in V$.
    \item The reflection automorphism about the origin $0$, denoted by $R_0(\cdot)$, is defined as $R_0(i) \triangleq -i$, $\forall i \in V$.
\end{enumerate}
It is not difficult to see that all these maps preserve the structure of the network and are valid automorphisms for the $N \times N$ torus. 
Observe that the composition of reflections $R_0(R_{xy}(\cdot))$, denoted as $R_0R_{xy}$, is also a valid automorphism for the $N\times N$ torus. These automorphisms let us reduce the search space of the optimal routing.

On applying Lemma \ref{lemma:symmetry_sufficiency_condition} with the above automorphisms of the $N\times N$ torus, we can conclude that there is an optimal routing scheme that is invariant to translations, i.e.,
$$
f^{s, s+t}_{i,i+e} = f^{T_s(s),T_s(s+t)}_{T_s(i), T_s(i+e)} = f^{0, t}_{i-s, i-s+e}.
$$
In other words, the result implies that it is sufficient to only specify the routes from the origin $0$ to every other node $t$. The route from node $s$ to node $s+t$ would only be a translation of the route from the origin $0$ to node $t$. This idea is visualized in Figure \ref{fig:translation_sym}. Similarly, there is an optimal routing scheme that is invariant to reflections. That is to say, for a reflection automorphism $\phi$, the route from node $\phi(s)$ to node $\phi(s) + \phi(t)$ is the reflection of the route from node $s$ to node $s+t$. In equations,
$$
f^{s, s+t}_{i,i+e} = f^{\phi(s),\phi(s)+\phi(t)}_{\phi(i), \phi(i)+\phi(e)}.
$$
This idea is visualized in Figure \ref{fig:reflection_sym}. 
In essence, Lemma \ref{lemma:symmetry_sufficiency_condition} tells us that we only need to find load minimizing paths from the origin to a small subset of nodes. Every other route would be a translation or reflection of these routes due to the symmetry in the network. The sufficient condition also lets us simplify notation for the routing policy because we can drop the explicit dependence of the routing policy on the source node $s$. 
 Now, we denote the routing policy as $g^{t}_{i,i+e}$, which represents the fraction of traffic from the origin to node $t$ routed along the edge $(i,i+e)$. 
 Using this characterization, one can obtain the fraction of any $(s,s+t)$ traffic routed along any edge $(i,i+e)$ using the relation $f^{s,s+t}_{i,i+e} = g^{t}_{i-s, i-s+e}$.
 Additionally, due to reflection invariance, $$
  g^t_{i,i+e} = g^{R_{xy}(t)}_{R_{xy}(i),R_{xy}(i+e)} =g^{R_{0}(t)}_{R_{0}(i),R_{0}(i+e)}.
 $$  
Furthermore, without loss of generality, it suffices to only check the maximum load on the $(0,e_1)$ edge because of the symmetry of the network. This in turn reduces the number of constraints in problem \eqref{eqn:main_problem1}.
We repose the optimization problem in \eqref{eqn:main_problem1} with reduced number of variables and constraints. 

\begin{lemma}\label{lemma:reduced_problem}
The optimal oblivious routing problem for $k-$limited traffic in \eqref{eqn:main_problem1} is equivalent to 
{
\thinmuskip=0mu
\thickmuskip=0mu
\medmuskip=0mu
\begin{subequations}
\begin{align}
    \min_{g, \theta} \quad &\theta \\
                \text{s.t.} \quad &\max_{d \in \D_k}\sum_{s,t}d_{s,s+t}g^{t}_{-s,-s+e_1} \leq \theta \label{eqn:max_constraint_sym}\\
                &\begin{aligned}\label{eqn:sym_flow_conservation}
                   \sum_{e\in A} g^{t}_{i,i+e} - &\sum_{e\in A} g^{t}_{i-e,i} = \begin{cases}
                    0 & i \neq 0, t\\
                    1 & i = 0\\
                    -1 & i = t
                \end{cases}
                \end{aligned}
                &&\forall i, t\\
                & 
                \begin{aligned}
                    g^t_{i,i+e} &= g^{R_{xy}(t)}_{R_{xy}(i),R_{xy}(i+e)} =g^{R_{0}(t)}_{R_{0}(i),R_{0}(i+e)}
                \end{aligned} &&\forall i,e, t\label{eqn:flow_symmetry}\\
                & 0 \leq g^t_{i,i+e_1} \leq 1 &&\forall i, t. \label{eqn:sym_flow_positivity}
\end{align}\label{eqn:problem_with_suff}
\end{subequations}
}
\end{lemma}
\vspace{-15pt}
The proof of Lemma \ref{lemma:reduced_problem} is provided in Appendix \ref{proof:reduced_problem}.
The above problem can be concisely stated as a bilinear minmax optimization problem
\begin{equation}
    \theta^* = \min_{g \in \R_{sym}} \max_{d \in \D_k} \sum_{s,t} d_{s,s+t}g^{t}_{-s,-s+e_1},\label{eqn:minmax_with_suff}
\end{equation}
where $\R_{sym}$ is the class of routing policies which satisfy translation and reflection invariance, i.e., $$\R_{sym} \triangleq \{g \; | \; g \text{ satisfies \eqref{eqn:sym_flow_conservation}, \eqref{eqn:flow_symmetry} and \eqref{eqn:sym_flow_positivity}}\}.$$
Lemma \ref{lemma:reduced_problem} exploits the structure of the torus to give us the max load problem as a bilinear optimization problem.
From von Neumann's minimax equivalence theorem \cite{v_neumann_zur_1928} for bilinear objectives (or equivalently, the strong duality of linear programs), we also have the following equivalence.
\begin{align}
    \theta^* 
    &=  \max_{d \in \D_k} \min_{g \in \R_{sym}} \sum_{s,t} d_{s,s+t}g^{t}_{-s,-s+e_1}.\label{eqn:maxmin_with_suff}
\end{align}
We use this equivalence principle to first derive lower bound on $\theta^*$, and then describe a routing policy that achieves this lower bound for all $k-$limited traffic.

\section{Worst-Case Load Lower Bound} 
\label{sec:worst_case_lb}
\begin{figure}[!b]
    \centering
\includegraphics[width=0.7\linewidth]{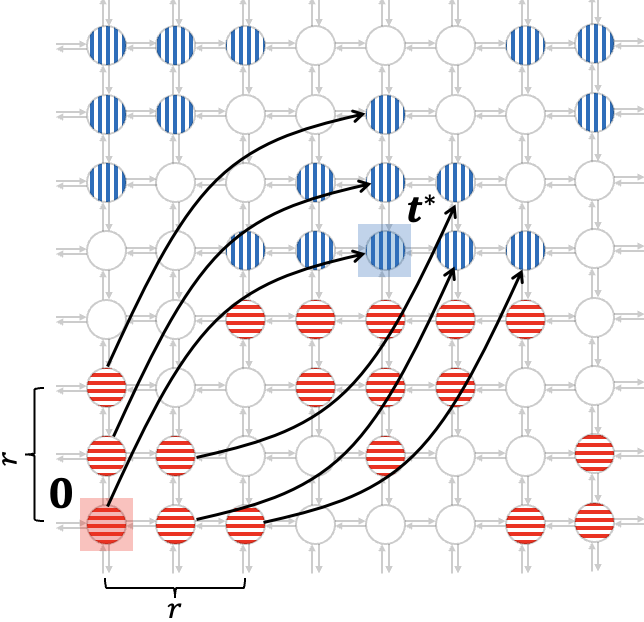}
    \caption{Split-Diamond traffic with $r=3$ for an $8\times8$ torus network. The origin $0$ and node $t^*$ are marked by boxes. Here, $t^* = (4,4)$. The source nodes of traffic are shaded with red horizontal lines and the destinations are shaded with blue vertical lines. Few of the source-destination pairs are marked with dark arrow lines. Every source node $s$ has traffic to destination $s+t^*$.  There is a total of $2r^2 = 18$ units of traffic in this traffic matrix.}
    \label{fig:split_diamond_traf}
\end{figure}
In this section, we provide a lower bound on the maximum load $\theta^*$ as described in the previous sections. To that end, we construct a specific traffic matrix which we call the Split-Diamond traffic matrix, denoted by $d^{SD(r)}$. Here, $r \in \mathbb{N}$ is a parameter and it determines the ``size" of the Split-Diamond traffic matrix. Using this construction, we show that any routing policy in $\R_{sym}$ would suffer a load of roughly $\sqrt{2k}/4$ under the Split-Diamond traffic matrix of appropriate ``size". The Split-Diamond traffic construction allows us to tighten the lower bound we obtained in Section \ref{subsec:cut_based_lb} by a factor of $\sqrt{2}$ for symmetric routing policies. Recall that in Section \ref{subsec:cut_based_lb}, we showed a universal lower bound of $\sqrt{k}/4$. 

To describe the Split-Diamond traffic matrix, we first need a distance metric on the network. 
We use the following distance metric $\Delta(.,.)$ on $V$.
\begin{align}\label{eqn:Delta_defn}
\Delta(u,v) = &\min\{(u_x - v_x) \bmod N, (v_x - u_x) \bmod N \}\nonumber\\
+ &\min\{(u_y - v_y) \bmod N, (v_y - u_y) \bmod N \}.
\end{align}
The distance metric $\Delta(\cdot,\cdot)$ counts the minimum number of hops required to go from node $u$ to node $v$. 
We construct the Split-Diamond traffic using the distance metric above.
When $N$ is even, the Split-Diamond traffic matrix with parameter $r$ is defined as
\begin{align}
    d^{SD(r)}_{s,s+t} = 
    \begin{cases}
        1 &  s \in \mathcal{E}_1(r)\cup \mathcal{E}_2(r) \text{ and }  t = t^*\\
        0 & \text{otherwise},
    \end{cases}
\end{align}
where {
\thinmuskip=1mu
\medmuskip=1mu
\begin{align*}
    \mathcal{E}_1(r) &= \left\{j \in V \;\; | \; \Delta(j,0) < r \text{ and }  j_y \leq \frac{N}{2}-1 \right\}\\
    \mathcal{E}_2(r) &= \left\{j \in V \; | \;\Delta\left(j,\left(\frac{N}{2}, \frac{N}{2}\right)\right) \leq r \text{ and } j_y \leq \frac{N}{2}-1\right\},
\end{align*}
}
\noindent
and, $t^* = \left(\frac{N}{2}, \frac{N}{2}\right)$ is the farthest node away from the origin.
For simplicity, we only describe the case for $N$ even. See Figure \ref{fig:split_diamond_traf} for an example. When $N$ is odd, the construction is similar with some minor modifications.
Observe that in the above traffic matrix, every source has traffic destined only to the node farthest away from it. Intuitively, the Split-Diamond traffic matrix is a collection of locally clustered sources and destinations, where all the source-destination pairs are $N$ hops away. The clusters have a diamond shape, and hence the name. The parameter $r$ determines the size of this cluster.  
It is easy to see that the total traffic in the Split-Diamond traffic matrix is $\sum_{s,t} d^{SD(r)}_{s, s+t} = 2r^2$. Therefore, the Split-Diamond traffic with $r \leq \sqrt{k/2}$ belongs to traffic class $\D_k$.
Next, we formally state the lower bound on $\theta^*$ due to Split-Diamond traffic of appropriate size in Theorem \ref{thm:optimal_oblivious_lower_bound}. 

\begin{theorem}\label{thm:optimal_oblivious_lower_bound}
For an $N\times N$ torus network, $k\leq N^2/2$\\
(a) When $2k$ is a perfect square, all routing policies $g \in \R_{sym}$ incur a maximum link load of at least $\sqrt{2k}/4$ under the Split-Diamond traffic $d^{SD(r)} \in \D_k$ with $r=\sqrt{k/2}$.\\
(b) When $2k$ is not a perfect square, all routing policies $g \in \R_{sym}$ incur a maximum link load of at least $\sqrt{2k'}/4$ for a particular $d \in \D_k$, where $k'$ is the largest integer such that $2k'$ is a perfect square and $k' \leq k$.
\end{theorem}
\noindent
Theorem \ref{thm:optimal_oblivious_lower_bound} immediately gives us a lower bound on $\theta^*$.
\begin{align*}
    \frac{\sqrt{2k}}{4} &\lesssim 
    \max_{d \in \D_k}\min_{g \in \R_{sym}} \sum_{s,t} d_{s,s+t}g^{t}_{-s,-s+e_1} = \theta^*.
\end{align*}
Remarkably, Theorem \ref{thm:optimal_oblivious_lower_bound} gives us a lower bound that is $\sqrt{2}$ times stronger compared to the lower bound developed in Section \ref{subsec:cut_based_lb}. 

Proving part (a) of Theorem \ref{thm:optimal_oblivious_lower_bound} involves inspecting the dual of the inner minimization problem in the maxmin optimization in \eqref{eqn:maxmin_with_suff}. Namely, we look at the dual of $\min_{g \in \R_{sym}} \sum_{s,t} d_{s,s+t} g^{t}_{-s,-s+e_1}$ for any traffic matrix $d \in \D_{k}$, when $2k$ is a perfect square. 

From strong duality, the problem \eqref{eqn:maxmin_with_suff} is equivalent to
\begin{equation}
    \begin{aligned}
        \max_{d, \nu} &\quad \frac{1}{4}\sum_{t} \nu^t_t\\
        \text{s.t.} & \quad \nu^t_{i+e} - \nu^t_{i} \leq \lambda(d,t,i,e) &&\forall i \in V \; \; \forall t \in V\backslash\{0\}\\
        &\quad \nu^t_0 = 0 && \forall t \in V\backslash\{0\}\\
        &\quad d \in \D_k,
    \end{aligned}\label{eqn:min_dual_main}
\end{equation}
where $
\lambda(d,i,t,e)$ is interpreted as the cost of using edge $(i,i+e)$ for a route from the origin $0$ to destination $t$, that depends on traffic matrix $d$.
On further inspection, we observe that the above problem corresponds to finding the maximum minimum cost path from the origin to every destination $t \in V\backslash\{0\}$, where the  cost of each edge for a particular destination $t$ is $\lambda(d,t,i,e)$. In other words,
$$
\theta^* = \frac{1}{4}\max_{d \in \D_k}\min_{\bar{\mathcal{P}}} \sum_{t} \sum_{(i,i+e) \in \mathcal{P}_t}\lambda(d,t,i,e),
$$
where $\bar{\mathcal{P}}$ is the collection of paths from the node $0$ to every $t \in V \backslash\{0\}$. 

To tackle the above maxmin problem, it suffices to construct a traffic matrix in $\D_k$ such that any reasonable path from the origin to a destination $t$ incurs the same cost. As a consequence, the sum cost of any path under this traffic matrix would give us a lower bound on the maxmin problem.

Indeed, the Split-Diamond traffic matrix is carefully constructed to achieve this. We omit the technical details in the interest of space. First, we note that $d^{SD(r)}$ only assigns non-zero traffic (and hence, non-zero cost) to $t = t^* \triangleq (N/2, N/2)$, i.e., the farthest node from the origin for simplicity. Then, the source nodes of traffic, $s$ (and consequently, destinations $s+t^*$) are carefully selected such that any reasonable path from $0$ to $t^*$ incurs a cost of $2r$, where $r \in \mathbb{N}$ is the parameter of the Split-Diamond traffic matrix. As mentioned earlier, for the Split-Diamond traffic with parameter $r$, $\sum_{s,t}d^{SD(r)}_{s,s+t} = 2r^2$ must be lesser than or equal to $k$ for $d^{SD(r)}$ to belong to $\D_k$. The sum cost of the objective under Split-Diamond traffic of size $r$ is therefore $2r/4$. We maximize this sum cost under the constraint $2r^2 \leq k, r \in \mathbb{N}$. When $2k$ is a perfect square, the best choice of $r$ would be $\sqrt{k/2}$ and the value of the objective is $\sqrt{2k}/4$. Since $d^{SD(r)}$ is a feasible solution, we arrive at $\theta^* \geq \sqrt{2k}/4$ when $2k$ is a perfect square. The proof for part (b) of Theorem \ref{thm:optimal_oblivious_lower_bound} follows immediately from the result of part (a) and convexity of the problem. The detailed proof can be found in Appendix \ref{proof:optimal_oblivious_lower_bound}.

Recall from Section \ref{subsec:automorphism_sufficiency} that there always exists an optimal oblivious routing policy that is automorphism invariant. Consequently, Theorem \ref{thm:optimal_oblivious_lower_bound} lets us conclude that any oblivious routing policy would suffer a maximum load of $\sqrt{2k}/4$ under the Split-Diamond traffic. Although the performance of any translation and reflection invariant routing policy is $\sqrt{2k}/4$ under the Split-Diamond traffic with $r=\sqrt{k/2}$, it does not imply that the Split-Diamond traffic is the worst-case traffic for \emph{any} routing policy. If we do not restrict ourselves to the class of automorphism invariant routing policies, we can achieve a load lower than $\sqrt{k}/4$ for the Split-Diamond traffic.


\section{Optimal Oblivious Routing for Sparse Traffic}\label{sec:routing_ub}
In this section, we describe a routing policy that achieves the lower bound on $\theta^*$. For that reason, we develop a certain structured routing policy, which we call the Local Load Balancing (LLB) scheme. The LLB scheme is parametrized by $r \in \mathbb{N}$, $r<N/2$. We denote the routing policy as $g^{LLB(r)}$. 
The routing policy is specifically designed to load balance any $k-$sparse traffic effectively in the network for an appropriate choice of $r$. Later, we show that the LLB scheme achieves a maximum link load of at most about $\sqrt{2k}/4$ for any traffic matrix in $\D_k$.

\begin{figure}[!t]
\centering
     \begin{subfigure}{0.45\textwidth}
        \centering
     \includegraphics[width=0.6\textwidth]{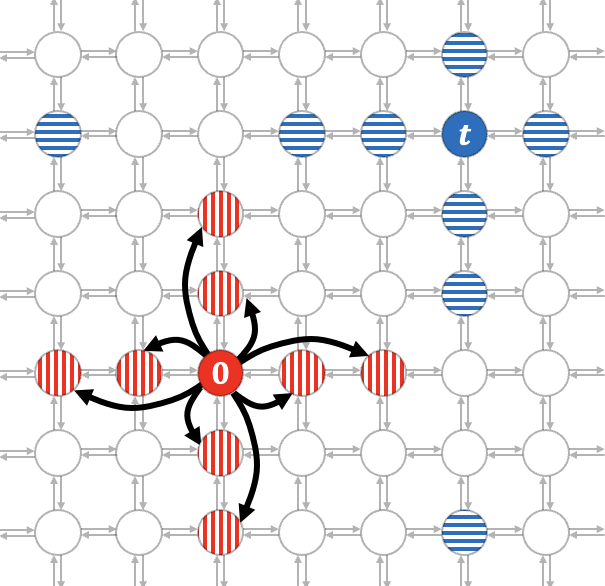}
         \caption{First phase - Distribute source traffic among stem nodes along the shortest path. The source (origin) is highlighted in dark red and the stem nodes $S_r(0)$ are shaded with red vertical lines.}
         \label{fig:LLB_phase1}
     \end{subfigure}
     \hfill
     \begin{subfigure}{0.45\textwidth}
         \centering
        \includegraphics[width=0.65\textwidth]{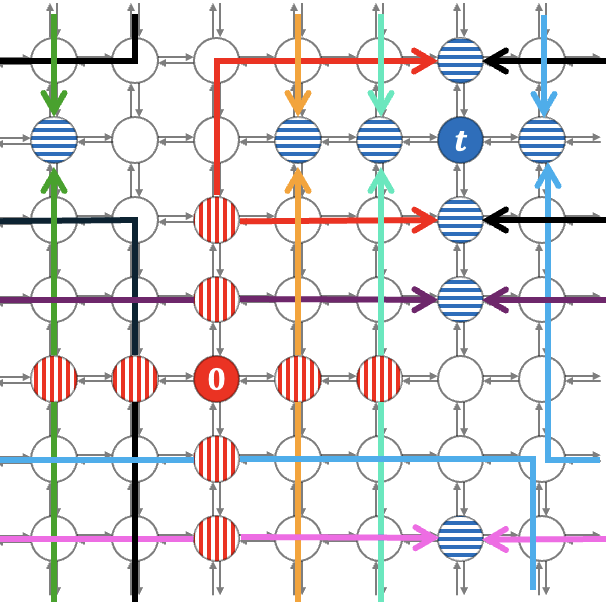}
         \caption{Second phase - Route traffic from $S_r(0)$ to $S_r(t)$ along edge-disjoint paths with two edge-disjoint paths emanating from each node in $S_r(0)$ and terminating at each node in $S_r(t)$.}
         \label{fig:LLB_phase2}
     \end{subfigure}
     \begin{subfigure}{0.45\textwidth}
         \centering
        \includegraphics[width=0.65\textwidth]{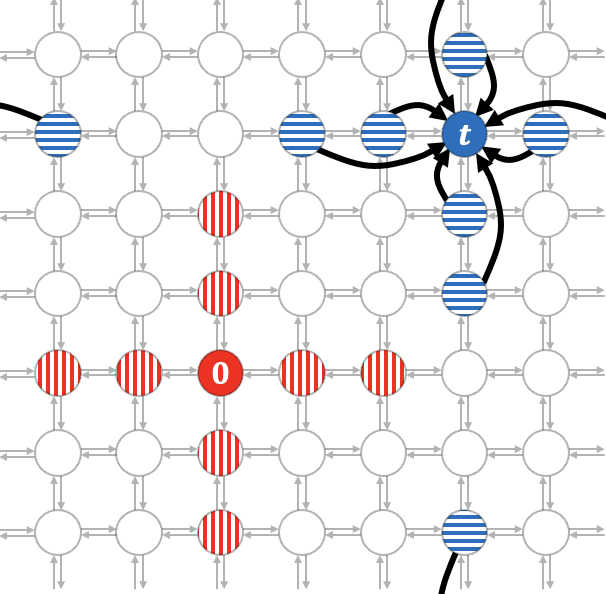}
         \caption{Third phase - Aggregate traffic from the stem nodes $S_r(t)$ to the destination node $t$ along the shortest path. The destination node is highlighted in dark blue and the stem nodes $S_r(t)$ are shaded with blue horizontal lines.}
         \label{fig:LLB_phase3}
     \end{subfigure}
     \caption{Example of Local Load Balancing Routing Scheme with parameter $r=2$.}
     \label{fig:LLB_example}
\end{figure}

Before we describe the LLB scheme, we define some terminology. A \textit{stem} $S_r(i)$ of size $r$ at node $i$ is the set of nodes that differ from node $i$ only at a coordinate and are at most $r$ hops away from node $i$. Mathematically, the stem is the set of nodes
$$S_r(i) \triangleq \{j \in V \;|\; 0 < \Delta(i,j) \leq r \text{ and } j_x = i_x \text{ or } j_y = i_y\}.$$
The stem of a node is displayed in the example in Figure \ref{fig:LLB_phase1} and \ref{fig:LLB_phase3}. For the $N \times N$ torus, the stem of size $r$ has $4r$ nodes. \\
\textbf{Remark.} \textit{One can also view the \textit{stem} $S_r(i)$ as the collection of $r$ node neighbors of node $i$ along the positive and negative vertical and horizontal directions respectively. We will henceforth call the nodes in the stem along each of these directions as ``legs". There are 4 legs in the stem.}

Now, we describe the Local Load Balancing (LLB) scheme with parameter $r$. The routing policy can be viewed as being split into three phases. For traffic from the origin $0$ to any destination $t \in V \backslash\{0\}$ such that $S_r(0) \cap S_r(t)$ is empty,\\
$\bullet$ \underline{\textit{First phase}}: Distribute traffic equally among the nodes in $S_{r}(0)$ along the shortest path. In other words, $1/4r$ fraction of traffic is first sent from the origin to every node in $S_r(0)$.\\
$\bullet$ \underline{\textit{Second phase}}: Route traffic from each node in the stem $S_r(0)$ to each node in the stem $S_r(t)$ along edge-disjoint paths, such that each node in $S_r(0)$ sends equal traffic along 2 unique edge-disjoint paths and each node in $S_r(t)$ receives traffic along 2 edge-disjoint paths. The paths in the second phase \textbf{must not} use the edges used in the first and third phases. Each path carries $1/8r$ fracton of traffic.\\
$\bullet$ \underline{\textit{Third phase}}: Aggregate traffic from the nodes in stem $S_r(t)$ to the final destination of the traffic, $t$, along the shortest path. As a consequence of the first two stages, $1/4r$ fraction of traffic needs to be aggregated from each node in $S_r(t)$.

Recall that it is sufficient for us to describe a set of routes from the origin $0$ to every node $t = (t_x,t_y)$ such that $t_x \leq t_y \leq N/2$. Then, from the consequences of automorphism invariance, the route from any source $s$ to any destination $s+t$ can be obtained from the transformation of the aforementioned set of routes. When there is overlap between the stems $S_r(0)$ and $S_r(t)$, minor modifications need to be made in the first and third phase, which we omit in the interest of space. Refer to Appendix \ref{appendix:llb_modification} for more detail.
The LLB scheme with $r=2$ is visualized in Figure \ref{fig:LLB_example} for an $N\times N$ torus with $N=7$.

We call it a ``local" load-balancing scheme as it distributes traffic among local nodes, namely the stem nodes around the origin and destination. This is unlike the popular Valiant Load-Balancing algorithm which distributes traffic to every other node in the network along shortest paths before aggregating it back at the destination from every node in the network along shortest paths.

It is important to highlight that the second phase of LLB requires existence of $8r$ edge-disjoint paths from $S_r(0)$ to $S_r(t)$ with two edge-disjoint paths emanating from every node in $S_r(0)$ and two edge-disjoint paths terminating at every node in $S_r(t)$. It is always possible to find such edge-disjoint paths for $r \leq N/2$, as in Lemma \ref{lemma:mengers_thm_for_routing}.

\begin{lemma}\label{lemma:mengers_thm_for_routing} In an $N\times N$ torus network, there exist $8r$ edge-disjoint paths from every node in stem $S_r(0)$ to every node in stem $S_r(t)$ such that two edge-disjoint paths emanate from each node in $S_r(0)$ and two edge-disjoint paths terminate at each node in $S_r(t)$, when $r < N/2$ and the destination $t \in V\backslash\{0\}$ is such that $S_r(0) \cap S_r(t)$ is empty.
\end{lemma}
Lemma \ref{lemma:mengers_thm_for_routing} comes from a special case of the celebrated Max-Flow Min-Cut theorem (also known as Menger's Theorem\cite{lawler_combinatorial_1976}) to count the maximum number of edge-disjoint paths from a set of nodes to another set of nodes. The proof requires showing that the cut size of $S_r(i)$ is always $8r + 4$ for $r < N/2$. This is intuitive from the geometry of $S_r(0)$, $S_r(t)$ and the torus. Consequently, there must exist at least $8r$ paths from $S_r(0)$ to $S_r(t)$. These paths can be easily found using Max-Flow algorithms \cite{dinics_max_flow, edmonds_karp_max_flow, ford_fulkerson_max_flow}. (More details in Appendix \ref{proof:mengers_thm}). 

\begin{theorem}\label{thm:optimal_oblivious_upper_bound} For an $N\times N$ torus network, $k\leq N^2/2$\\
(a) When $2k$ is a perfect square, the LLB scheme with parameter $r=\sqrt{k/2}$ achieves a maximum link load of $\sqrt{2k}/4$ for any traffic matrix in $\D_k$.\\
(b) When $2k$ is not a perfect square, the LLB scheme with parameter $r = \sqrt{k'/2}$ achieves a maximum link load of $\sqrt{2k'}/4$ where $k'$ is the smallest integer such that $2k'$ is a perfect square and $k \leq k'$.
\end{theorem}
\noindent
From Theorem \ref{thm:optimal_oblivious_upper_bound}, we can conclude that
\begin{align*}
    \theta^* &= \min_{g \in \R_{sym}} \max_{d \in \D_k} \sum_{s,t} d_{s,s+t} g^t_{-s,-s+e_1} \lesssim \frac{\sqrt{2k}}{4},
\end{align*}
because the LLB scheme is a feasible routing, i.e., $g^{LLB} \in \mathcal{R}_{sym}$.
Therefore, Theorems \ref{thm:optimal_oblivious_lower_bound} and \ref{thm:optimal_oblivious_upper_bound}  tell us that the LLB scheme is an optimal oblivious routing scheme for sparse traffic when $2k$ is a perfect square. 
To prove Theorem \ref{thm:optimal_oblivious_upper_bound}, we first greedily maximize the load any link can experience under the LLB policy and show that it is bounded above by $\sqrt{2k}/4$ when $2k$ is a perfect square. Later, we use the result of part (a) to show part (b).  
The details are omitted here in the interest of space and can be found in Appendix \ref{proof:optimal_oblivious_upper_bound}. 

\section{Extension to General $N \times M$ Tori}
\label{sec:asymmetric_tori}
\begin{figure}[!hbt]
    \centering
    \includegraphics[width=0.6\linewidth]{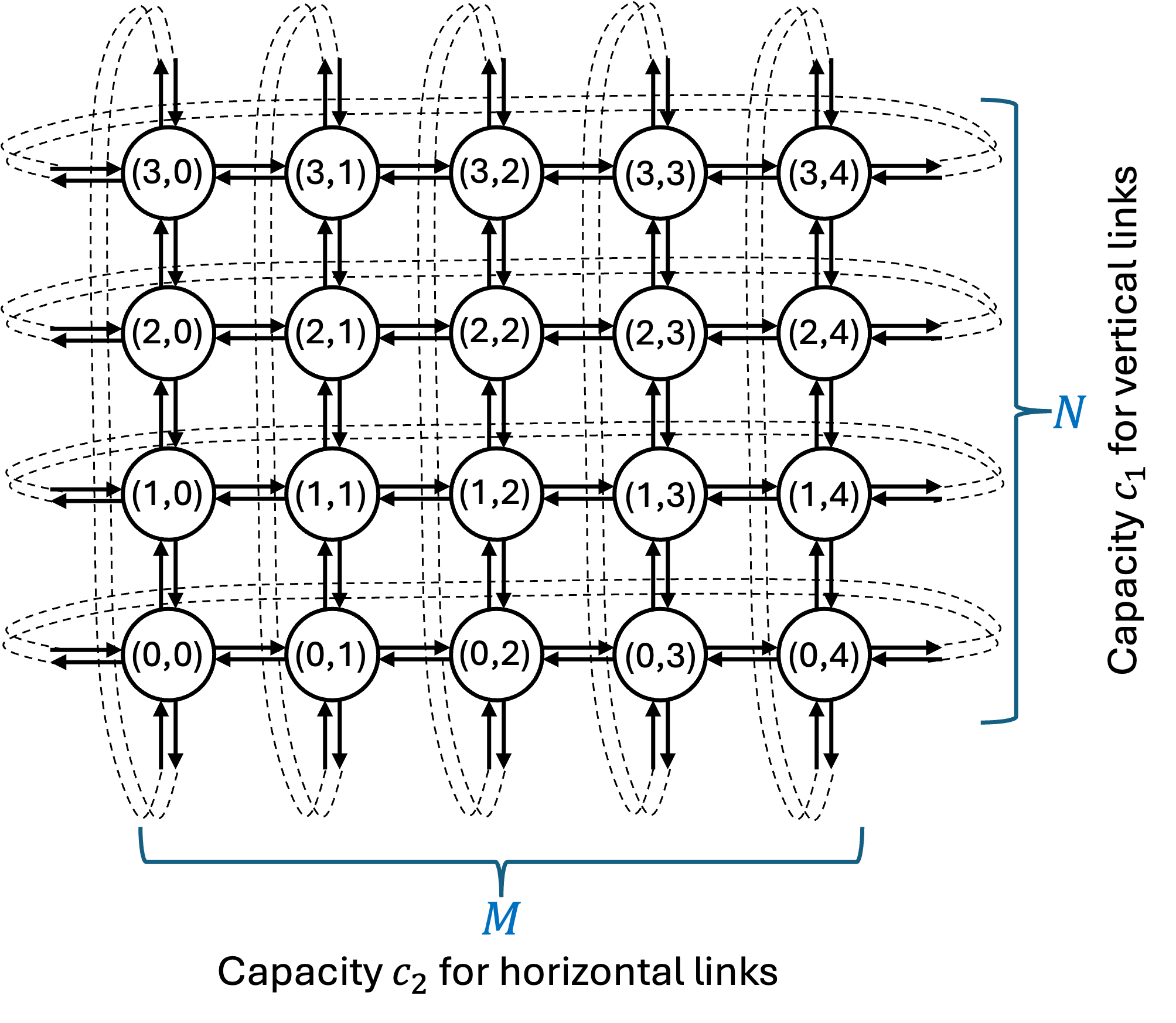}
    \caption{$N\times M$ torus with $N=4$ and $M=5$.}
    \label{fig:NxMtorus}
\end{figure}

The network topology considered in the previous sections of the paper is too restrictive. 
In this section, we relax certain restrictions on the topology and establish optimal oblivious load-balancing bounds for $N\times M$ tori with capacities $c_1$ along the vertical links and $c_2$ along the horizontal links, as shown in Figure \ref{fig:NxMtorus}. 
This topology more realistically models satellite networks with $M$ orbital planes and $N$ satellites per plane, where intra-orbit and inter-orbit links have capacities $c_1$ and $c_2$, respectively. However, the load-balancing insights from the $N\times N$ case still smoothly carry over.

The $N\times M$ torus has links at each node along the directions in set $A = \{e_1, -e_1, e_2, -e_2\}$, where $e_1 = (0,1)$ is the positive vertical direction, $-e_1 = (0,-1)$ is the negative vertical direction, $e_2 = (1,0)$ is the positive horizontal direction, and $-e_2 = (-1,0)$ is the negative horizontal direction.
The links in the vertical directions, that is, along $e_1$ and $-e_1$ have a capacity of $c_1$, and the links in the horizontal directions, that is, along $e_2$ and $-e_2$ have a capacity of $c_2$. The links along $\pm e_1$ and $\pm e_2$ are not identical, unlike the case in the previous sections. 
We want to characterize an optimal routing strategy $f^*$ such that 
\begin{equation}
    \max_{d \in \D_k}\MLU(f^*, d) = \min_{f\in \R}\max_{d \in \D_k}\MLU(f, d)\label{eqn:min_max_NxM}
\end{equation}
for the $N\times M$ torus.

Observe that the $N\times M$ torus exhibits translation symmetry and symmetry under reflection about the origin. However, unlike the $N\times N$ torus, \textbf{the $N\times M$ torus does not exhibit symmetry under reflection about the $x=y$ axis}. This is the only major difference between the analysis in the previous sections and in this section. Using the automorphisms of the general $N\times M$ torus, we can formulate a reduced version of problem \eqref{eqn:min_max_NxM}, similar to that of Lemma \ref{lemma:reduced_problem} for an $N\times N$ torus. Like in Section \ref{subsec:automorphism_sufficiency}, we will again denote the reduced routing policy as $g^{t}_{i,i+e}$. Essentially, $g^{t}_{i,i+e}$ represents the fraction of traffic from the origin to node $t$ routed along the edge $(i,i+e)$. Using this characterization, one can obtain the fraction of any $(s,s+t)$ traffic routed along any edge $(i,i+e)$ simply by using the translation transformation $f^{s,s+t}_{i,i+e} = g^{t}_{i-s, i-s+e}$. It is sufficient to characterize an optimal route from the origin to every destination $t$ to describe an optimal route for all source-destination pairs. The reduced problem for the general $N\times M$ torus is stated in Lemma \ref{lemma:reduced_problem_for_NxM}.
\begin{lemma}\label{lemma:reduced_problem_for_NxM}
The optimal oblivious routing problem for $k-$limited traffic in \eqref{eqn:min_max_NxM} is equivalent to 
{
\thinmuskip=0mu
\thickmuskip=0mu
\medmuskip=0mu
\begin{subequations}
\begin{align}
    \min_{g, \theta} \quad &\theta \\
                \text{s.t.} \quad &\max_{d \in \D_k}\sum_{s,t}d_{s,s+t}g^{t}_{-s,-s+e_1} \leq c_1 \theta \label{eqn:max_constraint_NxM_1}\\
                &\max_{d \in \D_k}\sum_{s,t}d_{s,s+t}g^{t}_{-s,-s+e_2} \leq c_2 \theta \label{eqn:max_constraint_NxM_2}\\
                &\begin{aligned}\label{eqn:sym_flow_conservation_NxM}
                   \sum_{e\in A} g^{t}_{i,i+e} - &\sum_{e\in A} g^{t}_{i-e,i} = \begin{cases}
                    0 & i \neq 0, t\\
                    1 & i = 0\\
                    -1 & i = t
                \end{cases}
                \end{aligned}
                &&\forall i, t\\
                & 
                \begin{aligned}
                    g^t_{i,i+e} &=g^{R_{0}(t)}_{R_{0}(i),R_{0}(i+e)}
                \end{aligned} &&\forall i,e, t\label{eqn:flow_symmetry_NxM}\\
                & 0 \leq g^t_{i,i+e_1} \leq 1, \quad 0\leq g^t_{i,i+e_2} \leq 1 &&\forall i, t. \label{eqn:sym_flow_positivity_NxM}
\end{align}\label{eqn:problem_with_suff_NxM}
\end{subequations}
}   
\end{lemma}
\noindent
Problem \eqref{eqn:problem_with_suff_NxM} can be concisely stated as 
\begin{equation}
\theta^* =  \min_{g \in \R'_{sym}}\max_{d \in \D_k} \max_{j\in\{1,2\}}\frac{1}{c_j}\sum_{s,t}d_{s,s+t}g^{t}_{-s,-s+e_j},\label{eqn:min_max_concise_NxM}
\end{equation}
where $\R'_{sym}$ is the class of routing policies which satisfy translation and reflection invariance, i.e., $$\R'_{sym} \triangleq \{g \; | \; g \text{ satisfies \eqref{eqn:sym_flow_conservation_NxM}, \eqref{eqn:flow_symmetry_NxM} and \eqref{eqn:sym_flow_positivity_NxM}}\}.$$
\subsection{Worst-Case Load Lower Bound for the $N\times M$ torus}\label{sec:worst_case_lb_NxM}
We use insights developed from the proof of Theorem \ref{thm:optimal_oblivious_lower_bound} to construct a lower bound for \eqref{eqn:problem_with_suff_NxM}.
For the general $N\times M$ torus, we consider a generalized version of the Split-Diamond traffic in order to develop lower bounds on the load that any routing policy in $\R'_{sym}$ would have to incur. This is formally stated in Theorem \ref{thm:asymmetric_tori_lb}.

\begin{theorem}\label{thm:asymmetric_tori_lb}
For an $N\times M$ torus with capacities $c_1$ along the vertical links and $c_2$ along the horizontal links\\
a) When $k \leq L^2/2$, all routing policies $g \in \R'_{sym}$ incur a maximum load of at least $\frac{\sqrt{2k}}{4\sqrt{c_1c_2}} - \eps'$,\\
b)  When $L^2/2 \leq k \leq NM/2$, all routing policies $g \in \R'_{sym}$ incur a maximum load of at least $\frac{k}{2L\sqrt{c_1c_2}}$,\\
c) When $k \geq NM/2$, all routing policies $g \in \R'_{sym}$ incur a maximum load of at least $\frac{NM}{4L\sqrt{c_1c_2}}$,\\
where $L = \min(\sqrt{c_2/c_1}N, \sqrt{c_1/c_2}M)$.
\end{theorem}
\noindent
Theorem \ref{thm:asymmetric_tori_lb} immediately gives us a lower bound on
\begin{align*}
    \min_{f \in\R}\max_{d \in \D_k}\MLU(f, d)
    \gtrsim \begin{cases}
        \frac{\sqrt{2k}}{4\sqrt{c_1c_2}} & k \leq \frac{L^2}{2} \\
        \frac{k}{2 L \sqrt{c_1c_2}} & \frac{L^2}{2} \leq k \leq \frac{NM}{2}\\
        \frac{NM}{4L\sqrt{c_1c_2}} &  k \geq \frac{NM}{2}
    \end{cases}.
\end{align*}
$L$ is to be interpreted as the normalized size of smallest dimension of the torus. It is in fact the minimum bisection bandwidth of the general $N\times M$ torus normalized by 2 times the geometric mean of capacities along each dimension, i.e., $2\sqrt{c_1c_2}$.
We prove Theorem \ref{thm:asymmetric_tori_lb} in Appendix \ref{proof:asymmetric_tori_lb}.

\subsection{Optimal Oblivious Routing in the $N\times M$ torus}\label{subsec:routing_NxM_ub}
In this subsection, we propose a routing policy that achieves the lower bound on max load described in Theorem \ref{thm:asymmetric_tori_lb}. We define a generalization of the LLB scheme in Section \ref{sec:routing_ub} which we call the Generalized Local Load Balancing (GLLB) scheme. The GLLB scheme is parametrized by $r_1, r_2 \in \mathbb{N}$ and $r_1 \leq N/2, r_2 \leq M/2$. We denote the routing policy as $g^{GLLB(r_1,r_2)}$. 
The routing policy is specifically designed to near-optimally load balance any $k-$sparse traffic effectively in the $N\times M$ network for an appropriate choice of $r_1$ and $r_2$ when $k \leq L^2/2$. 
The structure of the GLLB policy is inherently dependent on the geometry of the network. As a result, it must be formulated on a case-by-case basis, with its precise form determined by the parameters $r_1, r_2$, the network size $N, M$, and link capacities $c_1$ and $c_2$. When $k \geq L^2/2$, we have a simple routing strategy that achieves the lower bound. For the detailed description of GLLB, refer to Appendix \ref{appendix:GLLB}.

\newcommand{\ceil}[1]{\lceil #1 \rceil}
\begin{theorem}\label{thm:asymmetric_tori_ub}
For an $N\times M$ torus with capacities $c_1$ along the vertical links and $c_2$ along the horizontal links\\
a) When $k \leq L^2/2$, Generalized Local Load-Balancing $g^{GLLB(r_1,r_2)}$, with parameter $r_1 = \ceil{ \sqrt{c_1k/2c_2}}$ and $r_2 = \ceil{ \sqrt{c_2k/2c_1}}$, achieves a load of at most $\frac{\sqrt{2k}}{4\sqrt{c_1c_2}} + \eps'$ for any traffic in $\D_k$,\\
b)  When $k \geq L^2/2$, there exists a routing strategy that achieves a load of at most $\min\{\frac{k}{2L\sqrt{c_1c_2}}, \frac{NM}{4L\sqrt{c_1c_2}}\}$. for any traffic in $\D_k$,\\
where $L = \min(\sqrt{c_2/c_1}N, \sqrt{c_1/c_2}M)$.
\end{theorem}
Theorem \ref{thm:asymmetric_tori_ub} immediately gives us an upper bound on
\begin{align*}
    \min_{f \in\R}\max_{d \in \D_k}\MLU(f, d)
    \lesssim \begin{cases}
        \frac{\sqrt{2k}}{4\sqrt{c_1c_2}} & k \leq \frac{L^2}{2} \\
        \frac{k}{2 L \sqrt{c_1c_2}} & \frac{L^2}{2} \leq k \leq \frac{NM}{2}\\
        \frac{NM}{4L\sqrt{c_1c_2}} &  k \geq \frac{NM}{2}
    \end{cases}.
\end{align*}
The proof of Theorem \ref{thm:asymmetric_tori_ub} can be found in Appendix \ref{proof:asymmetric_tori_ub}.
\section{Extension to $N$-ary $m$-cube topologies}\label{sec:higher_dims}
Our results can also be straightforwardly extended to high-dimensional torus networks, or, $N$-ary $m-$cubes \cite{dally_principles_2004} which are relevant to interconnection networks and datacenter networks. An $N$-ary $m-$cube consists of $N^m$ nodes arranged in a $m-$dimensional cube with $N$ nodes along each direction connected in a ring-like fashion. Each link in this topology has the same link capacity. We state the minimum max-load for the optimal oblivious load-balancing policy (when $k \leq N^m/m!$) below. 
\begin{theorem}
    For an $N$-ary $m$-cube with equal unit link capacities along all directions,
\begin{equation*}
    \min_{f \in \R}\max_{d \in \D_k} \MLU(f, d) \approx \frac{(m!)^{1/m}k^{1/m}}{2m},
\end{equation*}
when $k \leq (N^m/m!)$.
\end{theorem}
The proof of this theorem needs establishing a lower bound and upper bound of $\min_{f \in \R}\max_{d \in \D_k} \MLU(f, d)$. The lower bound comes from extending arguments of Theorem \ref{thm:optimal_oblivious_lower_bound} to higher dimensions. The upper bound comes from constructing a load-balancing algorithm similar to LLB in two dimensions. It can be described in three phases. As in LLB, we will have $r$ source stem nodes and destination stem nodes along each dimension. Here, $r = (m!)^{1/m}k^{1/m}/2$. There are a total of $2mr$ source and destination stem nodes respectively.\\
\underline{\textit{First phase}}: Spread the traffic from the source to $r$ source stem nodes along the shortest paths in each dimension. Each source stem node receives $1/(2mr)$ of the traffic.\\
\underline{\textit{Second phase}}: Route the traffic from the source stem nodes to the destination stem nodes along edge disjoint paths. 
\underline{\textit{Third phase}}: Aggregate the traffic from the $r$ destination stem nodes to the destination along shortest paths in each dimension. \\
The detailed proof is omitted in the interest of space.

\section{Numerical Results}
\label{sec:results}
We evaluate the performance of the proposed LLB and GLLB routing schemes under different $k$-sparse traffic matrices. We compare LLB with Equal-Cost Multipath (ECMP) routing, which distributes traffic equally among shortest paths, and Valiant Load Balancing (VLB). We additionally benchmark LLB and GLLB against the optimal \textbf{non-oblivious} routing for each given traffic matrix. Specifically, for a traffic matrix $d$, we define
$
\mathrm{OPT}(d) = \arg\min_f \MLU(f,d),
$
where the routing $f$ is optimized with knowledge of the realized traffic matrix $d$. Thus, unlike LLB, GLLB, ECMP, and VLB, OPT is traffic-dependent and is not an oblivious routing scheme.

\subsection{$N\times N$ Torus with Equal Link Capacities}
\begin{figure}[!hb]
    \centering
    \begin{subfigure}{0.45\textwidth}
        \centering
        \includegraphics[width=\linewidth]{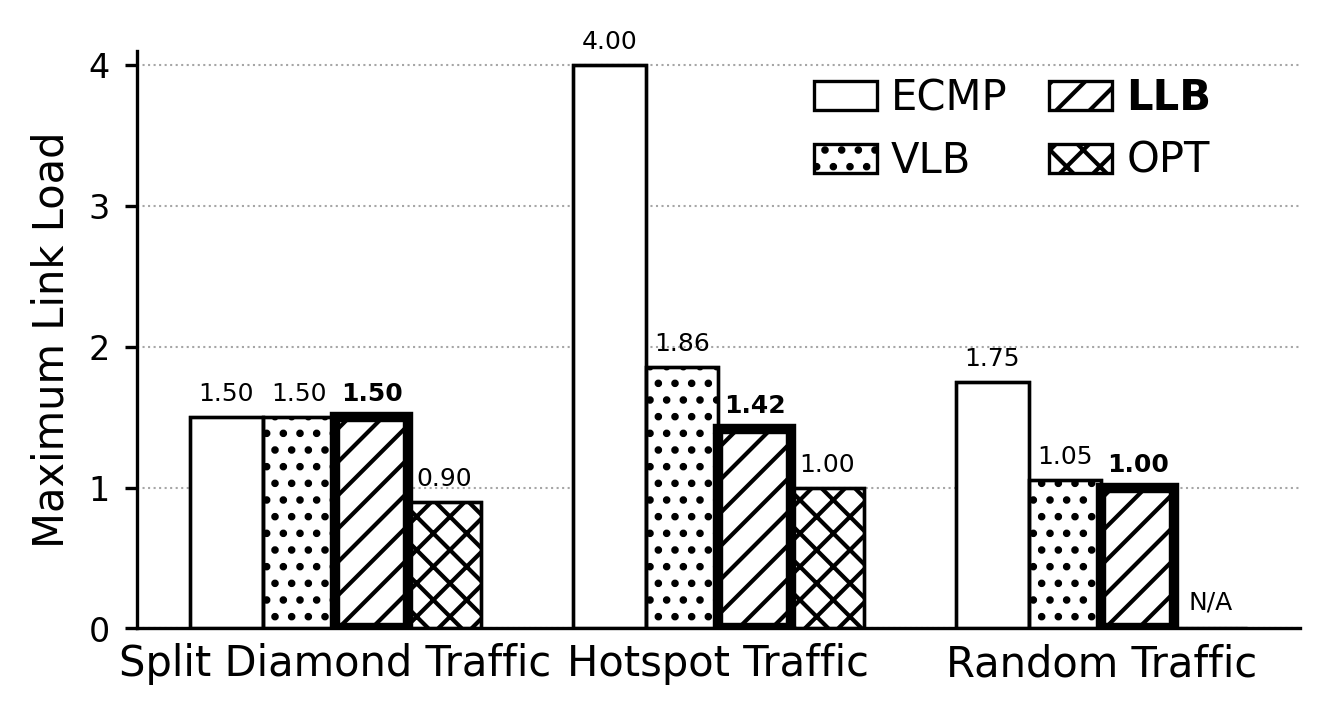}
        \caption{Maximum Link Loads of routing policies.}
        \label{fig:NxN_link_loads}
    \end{subfigure}
    \hfill
    \begin{subfigure}{0.45\textwidth}
        \centering
        \includegraphics[width=\linewidth]{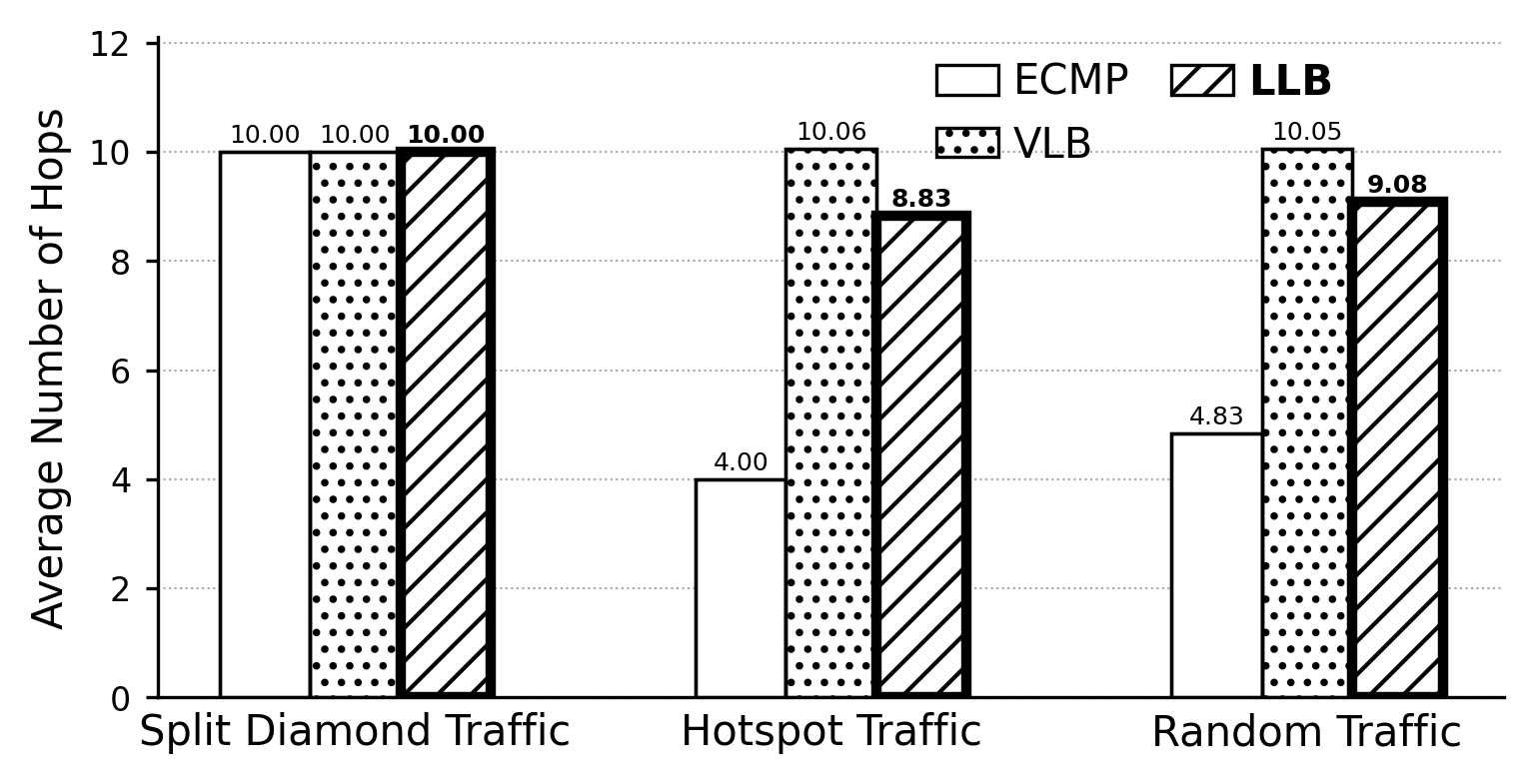}
        \caption{Average number of hops from source to destination.}
        \label{fig:NxN_num_hops}
    \end{subfigure}
    \caption{Numerical Results in $10\times 10$ torus with $k=18$.}
    \label{fig:NxN_results}
\end{figure}

We first consider a $10\times10$ torus with unit-capacity links in both directions and $k=18$. We evaluate the routing schemes under three $k$-sparse traffic patterns: Split-Diamond traffic with $r=\sqrt{k/2}=3$, Hotspot traffic, and Random traffic. Hotspot traffic consists of localized, square-like clusters of sources and destinations that are spatially adjacent, as illustrated in Fig.~\ref{fig:VLB_suboptimality}. For Random traffic, the $k$ sources and destinations are selected uniformly at random. We report the average maximum link load over 1000 independent random traffic instances. Since OPT is optimized separately for each traffic matrix, we do not report a corresponding average value for the random-traffic experiment. The resulting maximum link loads and average source-destination hop counts are shown in Fig.~\ref{fig:NxN_link_loads}.

For Split-Diamond traffic, ECMP, LLB, and VLB all achieve the same maximum link load of
$
\frac{\sqrt{2k}}{4}=1.5.
$
This follows from the automorphism invariance of these oblivious routing schemes and the symmetry of the Split-Diamond traffic matrix. In contrast, OPT can exploit knowledge of the realized traffic matrix and achieves a substantially lower maximum link load.
Under Hotspot traffic, LLB achieves a lower maximum link load than both ECMP and VLB. This illustrates the benefit of local load balancing for spatially concentrated traffic. Unlike VLB, which distributes traffic broadly across the network, LLB can exploit the locality of the demand while still balancing the resulting link loads. Under Random $k$-sparse traffic, LLB and VLB exhibit similar average maximum link loads over the 1000 traffic instances.
Figure~\ref{fig:NxN_link_loads} compares the average source-destination hop count of LLB, ECMP, and VLB. Since ECMP uses only shortest paths, its hop count provides a baseline for evaluating the additional path length incurred by the load-balancing schemes. LLB achieves comparable, and in some cases smaller, average hop counts than VLB while providing improved load balancing for the considered sparse traffic patterns. The hop count achieved by LLB depends on the particular collection of edge-disjoint paths returned by the max-flow algorithm. In these experiments, we use Dinic's max-flow algorithm \cite{dinics_max_flow} to construct the edge-disjoint paths.

\vspace{-1em}
\subsection{$N\times M$ Torus with Unequal Link Capacities}
We next consider a $10\times7$ torus with unequal link capacities. Specifically, the vertical and horizontal links have capacities $c_1=2$ and $c_2=1$ respectively. This emulates a satellite network with 10 orbital planes, with 7 satellites each, and the intra-orbit ISLs having twice the capacity of the inter-orbit ISLs. We compare GLLB with ECMP and VLB under $k$-sparse traffic with $k=16$. We also compare each oblivious routing scheme with the optimal non-oblivious routing
$
\mathrm{OPT}
$
for each specified traffic matrix $d$.

\begin{figure}[!hb]
    \centering
    \begin{subfigure}{0.45\textwidth}
        \centering
        \includegraphics[width=\linewidth]{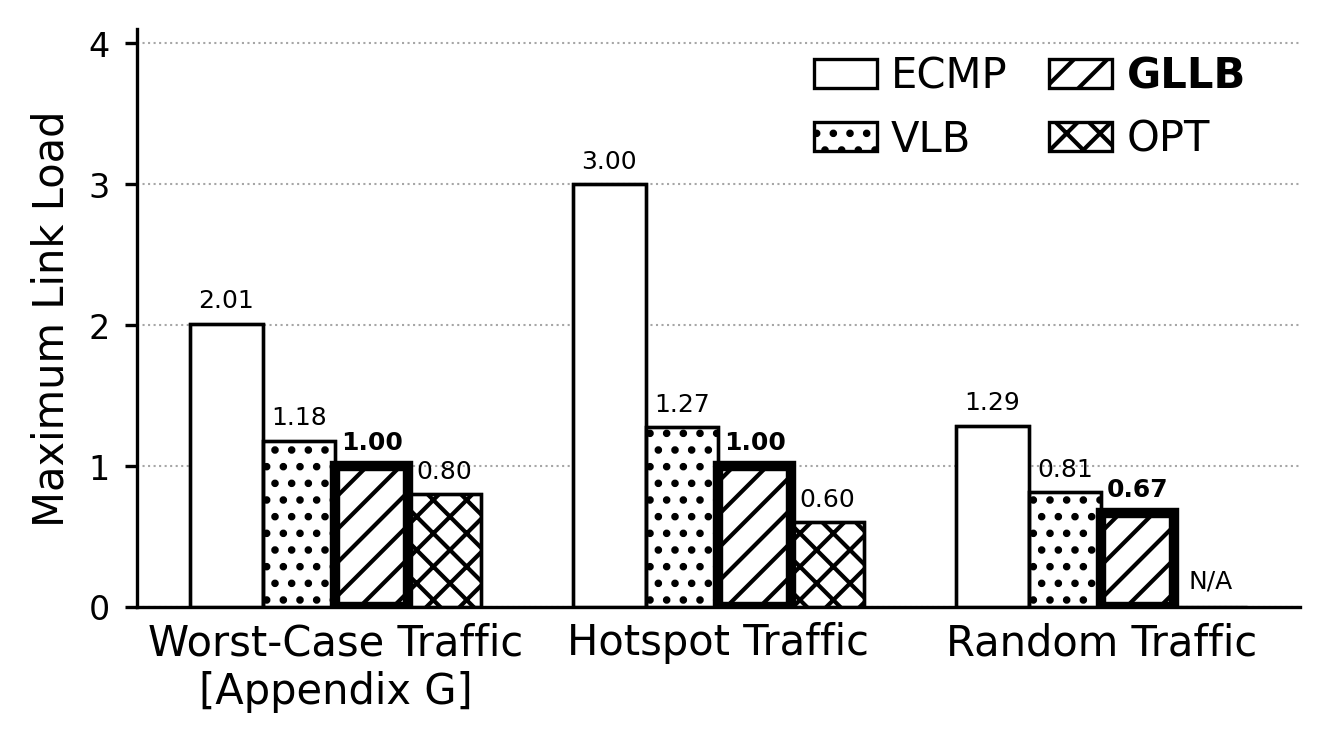}
        \caption{Maximum Link Loads of routing policies.}
        \label{fig:NxM_link_loads}
    \end{subfigure}
    \hfill
    \begin{subfigure}{0.45\textwidth}
        \centering
        \includegraphics[width=\linewidth]{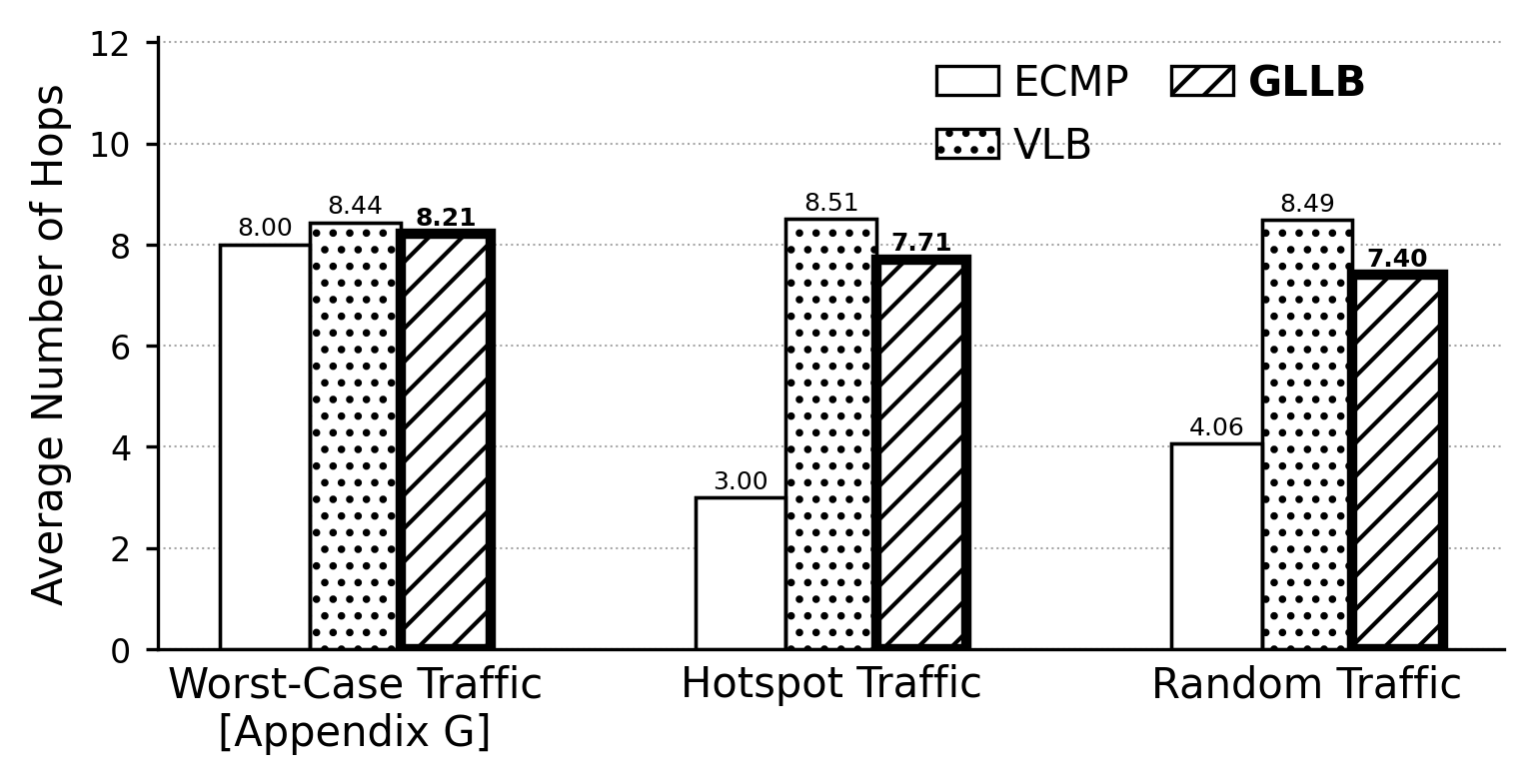}
        \caption{Average number of hops from source to destination.}
        \label{fig:NxM_num_hops}
    \end{subfigure}
    \caption{Numerical Results in $10\times 7$ torus with $k=16$.}
    \label{fig:NxM_results}
\end{figure}

We consider three traffic patterns: the worst-case traffic matrix for oblivious routing described in Appendix~\ref{proof:asymmetric_tori_lb}, Hotspot traffic, and Random traffic. As in the previous experiment, Hotspot traffic consists of spatially localized source-destination clusters, while for Random traffic the $k$ sources and destinations are selected uniformly at random. We report the average maximum link load over 1000 random traffic instances. The corresponding maximum link loads and average source-destination hop counts are shown in Fig.~\ref{fig:NxM_link_loads}.

Across the considered traffic patterns, GLLB achieves a lower maximum link load than ECMP and VLB. In addition, the observed GLLB load is consistent with the theoretical bound
$
\frac{\sqrt{2k}}{4\sqrt{c_1c_2}}=1.
$
As expected, OPT can further reduce the maximum link load by optimizing the routing for the particular realized traffic matrix.

Figure~\ref{fig:NxM_num_hops} compares the average hop counts of GLLB, ECMP, and VLB. GLLB maintains hop counts comparable to, and in some cases lower than, those of VLB while achieving lower maximum link loads in the considered experiments. As before, the particular hop count obtained by GLLB depends on the edge-disjoint paths generated by the max-flow procedure. We use Dinic's max-flow algorithm \cite{dinics_max_flow} for this path construction.

\vspace{-1.2em}
\section{Conclusion}
In this paper, we studied oblivious load balancing for sparse traffic in torus networks, motivated by large-scale LEO satellite networks. For an $N\times N$ torus with at most $k$ active source-destination pairs, we established tight lower bounds on the worst-case maximum link load and constructed LLB, an optimal oblivious routing scheme that achieves these bounds. We showed that VLB can be suboptimal under sparse, localized traffic, while LLB exploits traffic locality to achieve lower link loads without significantly increasing path lengths. Finally, we extended our results to asymmetric $N\times M$ tori with unequal link capacities and higher-dimensional tori, broadening the applicability of the framework to heterogeneous satellite and interconnection networks. We also provide an efficient and descriptive mechanism to construct these traffic oblivious load-balancing paths.
On a higher level, we demonstrate that preserving traffic locality when it is sparse is highly advantageous, and globally distributing traffic can be suboptimal in terms of load and number of hops.

\section*{Acknowledgment}
We would like to thank our sponsor SES S.A. for their gracious support toward this research. We want to especially thank Joel Grotz and Valvanera Moreno for their valuable insights and feedback over the course of this work.
\vspace{-1em}

\bibliographystyle{IEEEtran}
\bibliography{references}
 \vskip -0.45\baselineskip 
\begin{IEEEbiography}[{\includegraphics[width=1in,height=1.25in,clip,keepaspectratio]{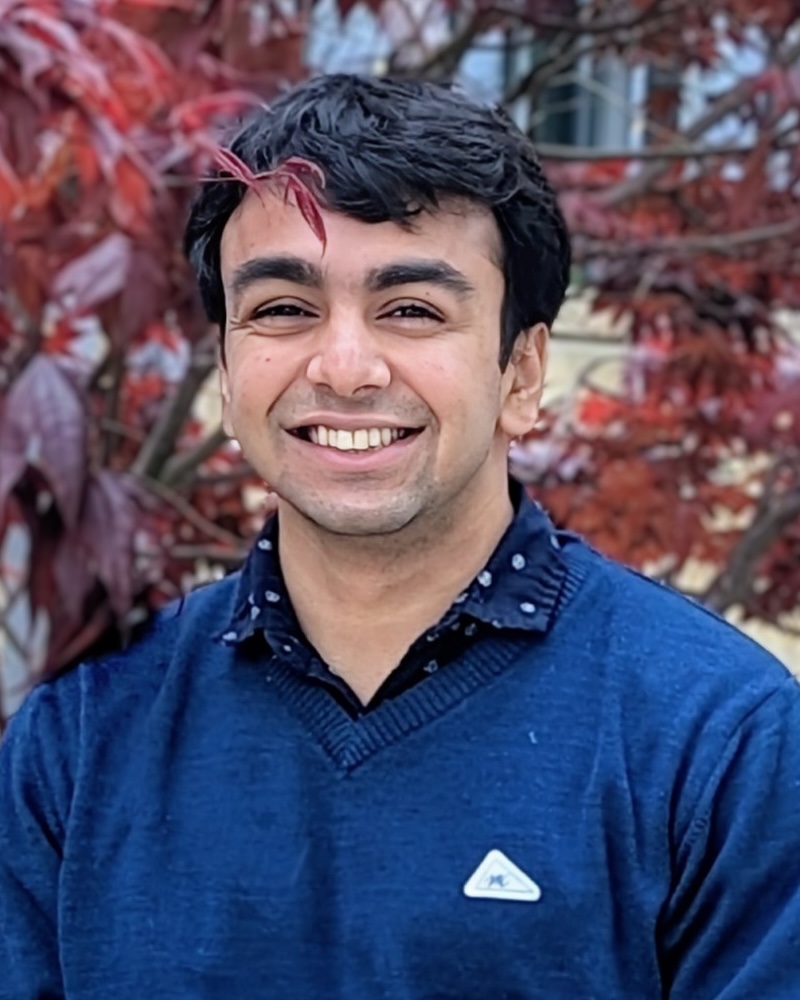}}]{Rudrapatna Vallabh Ramakanth}
is a PhD candidate in the Department of Aeronautics and Astronautics and the Laboratory of Decision and Information Systems at MIT, under Prof. Eytan Modiano. He received his SM from MIT in 2024. Prior to joining MIT, he received his MTech and BTech in Electrical Engineering from the Indian Institute of Technology Madras in 2022. 
\end{IEEEbiography}
\begin{IEEEbiography}[{\includegraphics[width=1in,height=1.25in,clip,keepaspectratio]{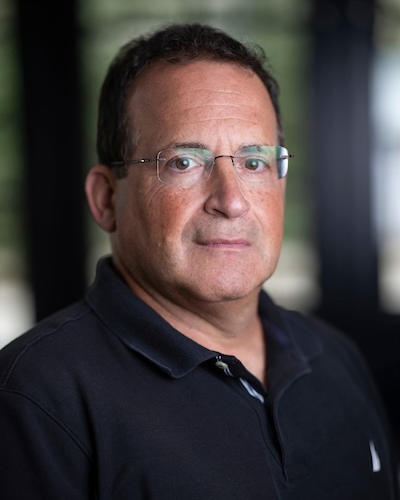}}]{Eytan Modiano}
is The Richard C. Maclaurin Professor in the Department of Aeronautics and Astronautics and the Laboratory for Information and Decision Systems (LIDS) at MIT.  Prior to Joining the faculty at MIT in 1999, he was a Naval Research Laboratory Fellow between 1987 and 1992, a National Research Council Post Doctoral Fellow during 1992-1993, and a member of the technical staff at  MIT Lincoln Laboratory between 1993 and 1999.  Eytan Modiano received his B.S. degree in Electrical Engineering and Computer Science from the University of Connecticut at Storrs in 1986 and his M.S. and PhD degrees, both in Electrical Engineering, from the University of Maryland, College Park, MD, in 1989 and 1992 respectively.  His research is on modeling, analysis and design of communication networks and protocols. He received the INFOCOM Achievement Award (2020) for contributions to the analysis and design of cross-layer resource allocation algorithms for wireless, optical, and satellite networks.   He is the co-recipient of the INFOCOM 2018 Best paper award, the MobiHoc 2018 best paper award, the MobiHoc 2016 best paper award, the Wiopt 2013 best paper award, and the Sigmetrics 2006 best paper award.  He was the Editor-in-Chief for IEEE/ACM Transactions on Networking (2017-2020), and served as Associate Editor for IEEE Transactions on Information Theory and IEEE/ACM Transactions on Networking.  He was the Technical Program co-chair for  IEEE Wiopt 2006, IEEE Infocom 2007, ACM MobiHoc 2007, and DRCN 2015; and general co-chair of Wiopt 2021.  He had served on the IEEE Fellows committee in 2014 and 2015, and is a Fellow of the IEEE and an Associate Fellow of the AIAA.
\end{IEEEbiography}

\appendices
\section{LLB When Source and Destination Stems Overlap}\label{appendix:llb_modification}
\begin{figure*}
\centering
     \begin{subfigure}{0.3\textwidth}
        \centering
     \includegraphics[width= 0.9\textwidth]{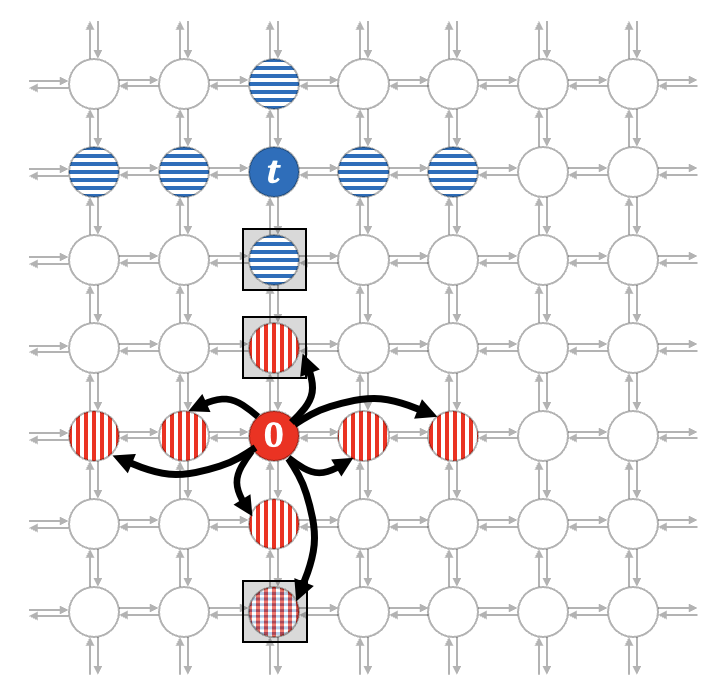}
         \caption{First phase - Distribute source traffic among stem nodes along the shortest path. The source (origin) is highlighted in dark red and the modified stem nodes $S^I_r(0)$ are shaded with red vertical lines. The boxed nodes are common to both the stems $S_r(0)$ and $S_r(t)$. Additionally, the node with the checkered pattern is part of both modified stems, $S^I_r(0)$ and $S^I_r(t)$.}
     \end{subfigure}
     \hfill
     \begin{subfigure}{0.3\textwidth}
         \centering
        \includegraphics[width= 0.9\textwidth]{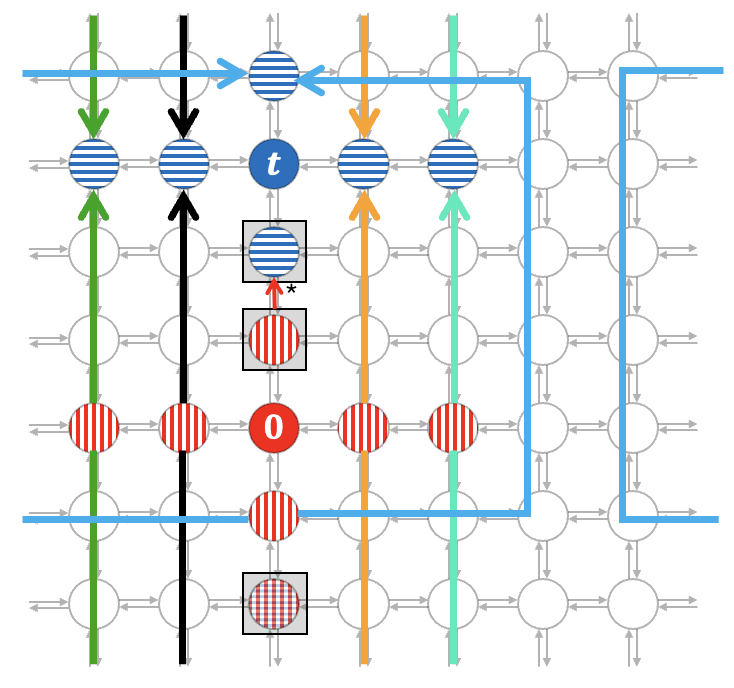}
         \caption{Second phase - Route traffic from $S^I_r(0)$ to $S^I_r(t)$ along edge-disjoint paths with two edge-disjoint paths emanating from each node in $S^I_r(0)$ and terminating at each node in $S^I_t(t)$. Here the path marked with $\ast$ denotes the residual traffic. Note that since the boxed node belongs to both modified stems, we do not need to explicitly forward traffic from it.}
     \end{subfigure}
     \hfill
     \begin{subfigure}{0.3\textwidth}
         \centering
        \includegraphics[width= 0.9\textwidth]{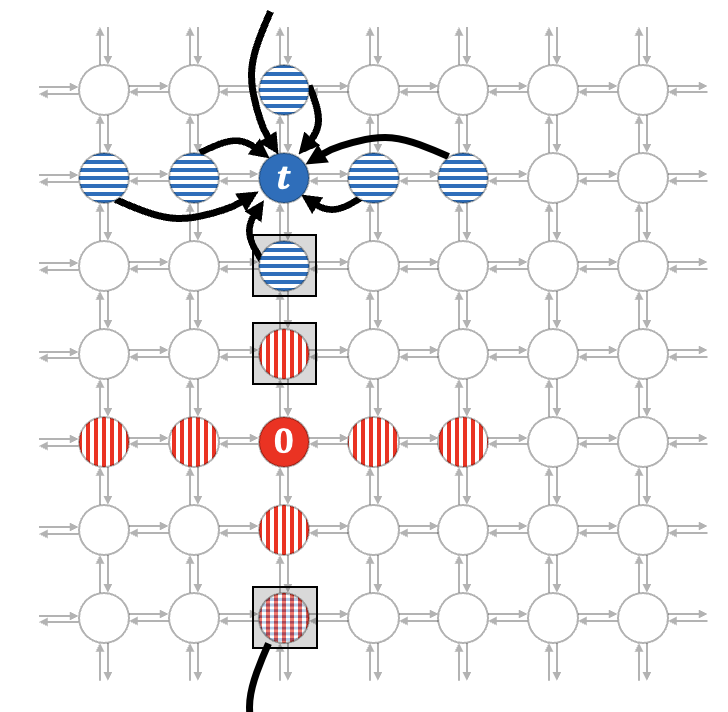}
         \caption{Third phase - Aggregate traffic from the stem nodes $S^I_r(t)$ to the destination node $t$ along the shortest path. The destination node is highlighted in dark blue and the stem nodes $S^I_r(t)$ are shaded with blue horizontal lines. The boxed nodes were common to both the initial stems $S_r(0)$ and $S_r(t)$.}
     \end{subfigure}
     \caption{LLB Routing Scheme with parameter $r=2$ with destination $t=(0,3)$ in a $7\times 7$ torus. This is a Type-I overlap.}
     \label{fig:cases_of_LLB_overlap_type_I}
\end{figure*}
\begin{figure*}
\centering
     \begin{subfigure}{0.3\textwidth}
        \centering
     \includegraphics[width=0.9\textwidth]{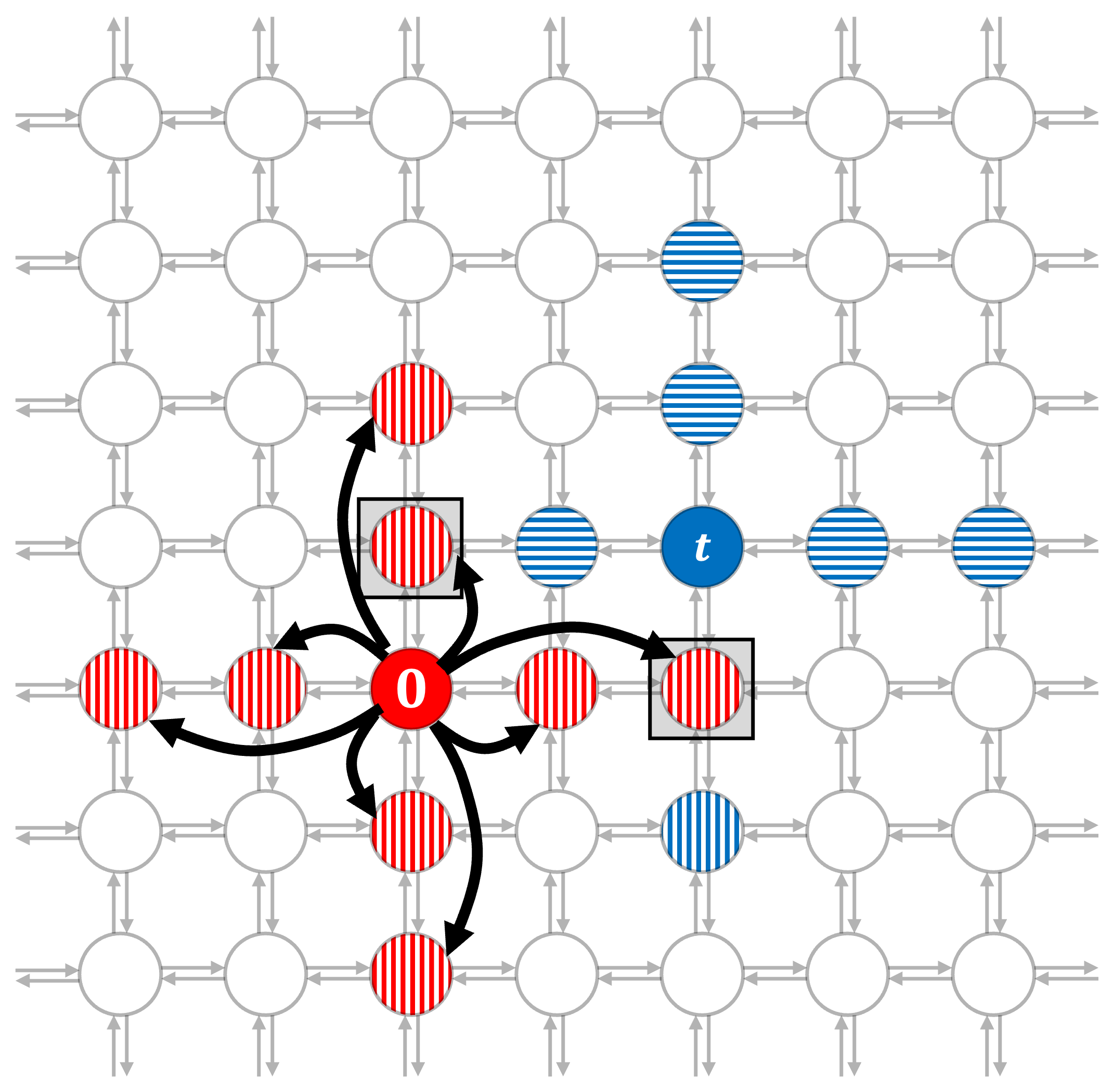}
         \caption{First phase - Distribute source traffic among stem nodes along the shortest path. The source (origin) is highlighted in dark red and the stem nodes $S_r(0)$ are shaded with red vertical lines. The boxed nodes are common to both the stems $S_r(0)$ and $S_r(t)$.}
     \end{subfigure}
     \hfill
     \begin{subfigure}{0.3\textwidth}
         \centering
        \includegraphics[width=0.9\textwidth]{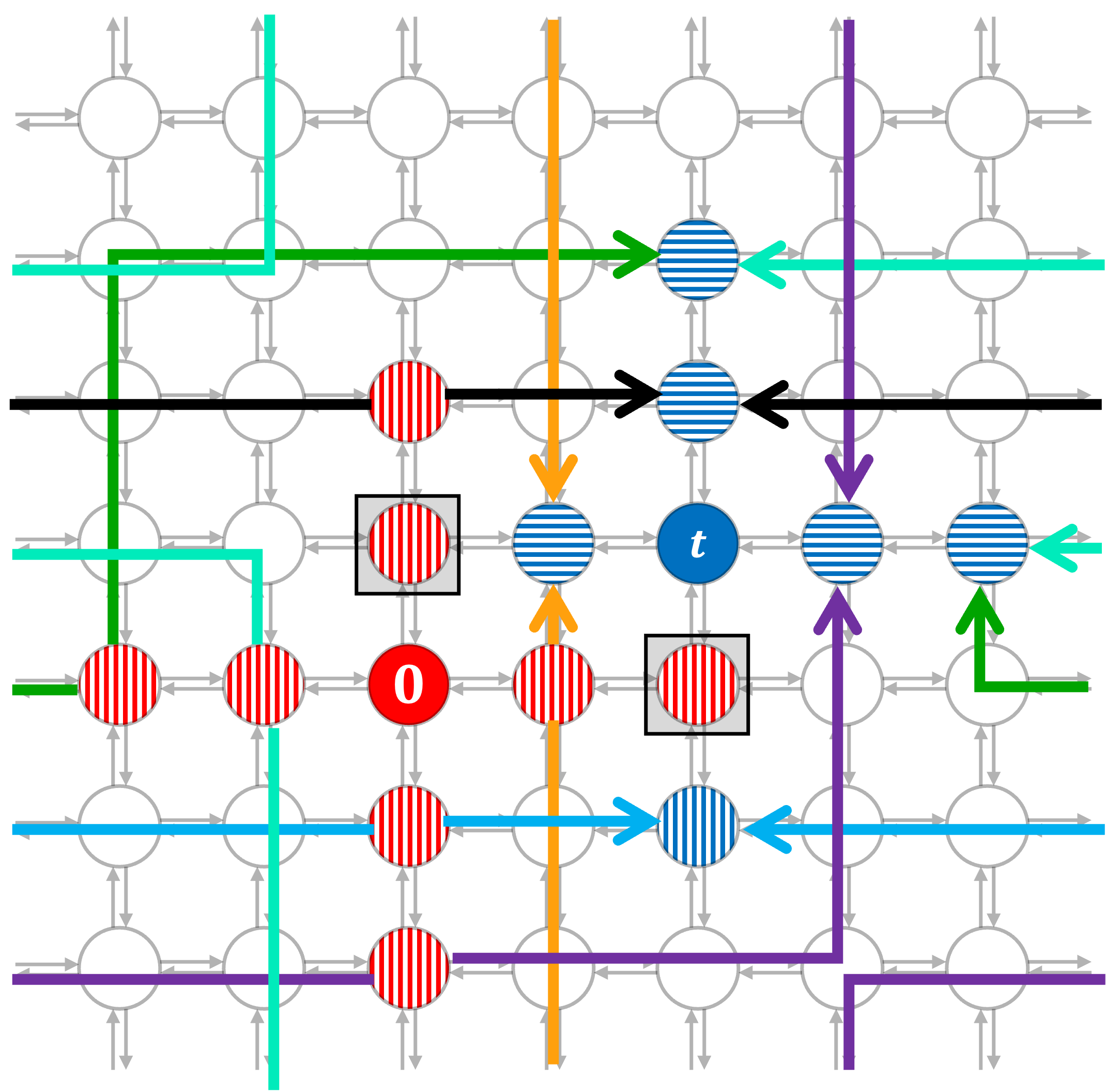}
         \caption{Second phase - Route traffic from $S_r(0)$ to $S_r(t)$ along edge-disjoint paths with two edge-disjoint paths emanating from each node in $S_r(0)$ and terminating at each node in $S_t(t)$. If a node in $S_r(0)$ also belongs to $S_r(t)$, then we do nothing.}
     \end{subfigure}
     \hfill
     \begin{subfigure}{0.3\textwidth}
         \centering
        \includegraphics[width=0.9\textwidth]{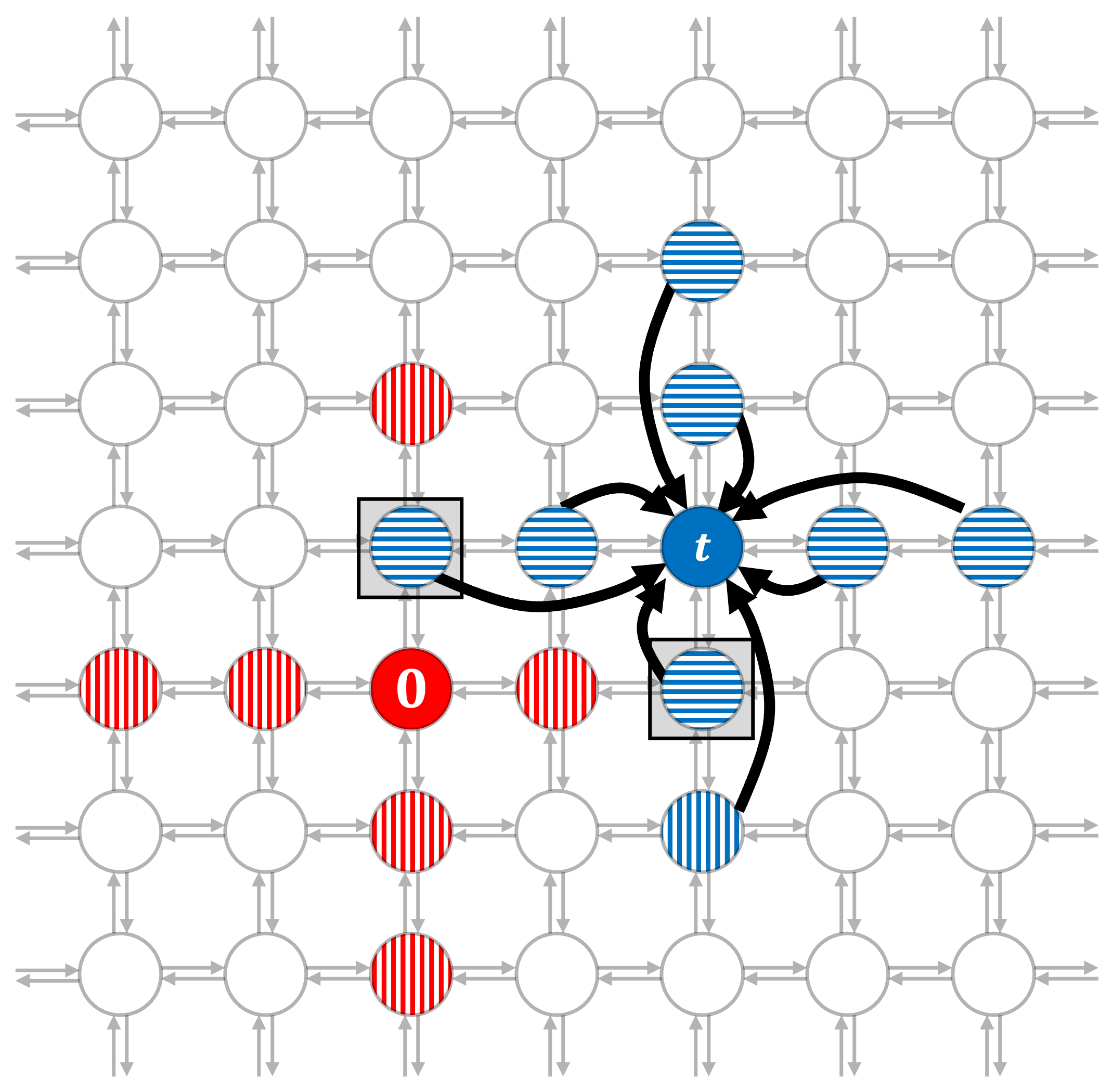}
         \caption{Third phase - Aggregate traffic from the stem nodes $S_r(t)$ to the destination node $t$ along the shortest path. The destination node is highlighted in dark blue and the stem nodes $S_r(t)$ are shaded with blue horizontal lines. The boxed nodes were common to both the initial stems $S_r(0)$ and $S_r(t)$.}
     \end{subfigure}
     \caption{LLB Routing Scheme with parameter $r=2$ with destination $t=(2,1)$ in a $7\times 7$ torus. Observe that the final flow from source to destination can have loops or redundant edges. These loops or redundant edges can be removed by revising the net flow in the forward direction.}
     \label{fig:cases_of_LLB_overlap_type_II}
\end{figure*}
As discussed in Section \ref{sec:routing_ub}, it suffices to describe the routing policy when $t_x \leq t_y \leq N/2$ as for other destinations, the routing can be obtained from automorphism transformations. In this section, we limit ourselves to $t_x \leq t_y \leq N/2$ for ease of presentation. 

When there is overlap between the stems $S_r(0)$ and $S_r(t)$, the above described routing policy cannot be used as is. If there are nodes $j$ such that $j \in S_r(0) \cap S_r(t)$, there needs to be minor modifications in the routing scheme.
Without loss in generality, there can only be the two types of overlap between stems, Type-I and type-II, as shown in Fig. \ref{fig:cases_of_LLB_overlap_type_I} and Fig. \ref{fig:cases_of_LLB_overlap_type_II}.  
\subsubsection{Type-I Overlap}
In Type-I overlap, the stems $S_r(0)$ and $S_r(t)$ overlap only along one of the four legs of the stem (Fig. \ref{fig:cases_of_LLB_overlap_type_I}). This only happens when $t = (t_x, 0)$ or $t=(0,t_y)$. Under Type-I overlap, we define modified stems $S^I_r(0)$ and $S^I_r(t)$,
\begin{align*}
    S^I_r(0) &= \{j\in V | j\in S_{r}(0) \text{ and }   \Delta(0,j) \leq \Delta(t,j)\}\\
    S^I_r(t) &= \{j \in V | j\in S_{r}(t) \text{ and }   \Delta(t,j) \leq \Delta(0,j)\}.
\end{align*}
Once we define the modified stem for Type-I overlap, the modifications in each routing phase is fairly straightforward. \\
$\bullet$ \underline{\textit{Modification in the first phase}}: Distribute $1/4r$ fraction of traffic to each of the nodes in the stems $S^I_r(0)$, except the last node in each of the legs. If the node is the last node in its leg and is $h$ hops away, it receives $(r-h+1)/4r$ fraction of the traffic.\\
$\bullet$ \underline{\textit{Modification in the second phase}}: Route traffic from each node in the stem $S^I_r(0)$ to each node in the stem $S^I_r(t)$ (or $S^I_r(t)$ along edge-disjoint paths, such that each node in $S^I_r(0)$ sends equal traffic along 2 unique edge-disjoint paths and each node in $S^I_r(t)$ receives traffic along 2 edge-disjoint paths. Each path carries $1/8r$ fraction of traffic. 
If there is residual traffic at a node in $S^I_r(0)$, then there must exist a node neighboring it that is in $S^I_r(t)$. The residual traffic must be directly sent to that neighbor. If a node belongs to both $S^I_r(0)$ and $S^I_r(t)$, then the node needs to do nothing in the second phase.\\
$\bullet$ \underline{\textit{Modification in the third phase}}: Aggregate traffic from the nodes in stem $S^I_r(t)$ to the final destination of the traffic, $t$, along the shortest path. As a consequence of the first two stages, $1/4r$ fraction of traffic needs to be aggregated from each node in $S^I_r(t)$ except the last nodes in each leg. A fraction of $(r-h+1)/4r$ of the traffic will be aggregated from the last node in each leg, where the last node is $h$ hops away.

\subsubsection{Type-II Overlap} In Type-II overlap, the stems $S_r(0)$ and $S_r(t)$ overlap exactly at two nodes at two of the four legs of the stem respectively (Fig. \ref{fig:cases_of_LLB_overlap_type_II}). There needs to be no modification to the stem or the routing policy. However, note that there is a subtlety in the second phase. If a node is in both $S_r(0)$ and $S_r(t)$, no traffic needs to be explicitly forwarded from that node in the second phase, as it is already in the destination stem.

\clearpage
\section{Proof Sketch of Lemma \ref{lemma:mengers_thm_for_routing}}\label{proof:mengers_thm}
According to Lemma \ref{lemma:mengers_thm_for_routing}, when the stems $S_r(0)$ and $S_r(t)$ are disjoint, then there exist $8r$ edge-disjoint paths from $S_r(0)$ to $S_r(t)$ such that there are two edge-disjoint paths emanating from each node in $S_r(0)$ and two edge-disjoint paths terminating at each node in $S_r(t)$. Define $S'_r(0) = S_{r}(0)\cup\{0\}$ and $S'_r(t) = S_{r}(t)\cup\{t\}$.

\begin{proof}[Proof Sketch of Lemma \ref{lemma:mengers_thm_for_routing}]
The crux of the proof lies in invoking Menger's Theorem \cite{lawler_combinatorial_1976}, which is a special case of the Max-Flow Min-Cut Theorem. Before we jump into the lemma, we define some preliminaries. For two mutually exclusive subsets of vertices, $S\subset V$ and $T \subset V$, we define the cut set (or simply, a cut) $C(S, T)$ as the set of edges that are directed from $S$ to $T$ such that, once removed, there exists no directed path from any node in $S$ to any node in $T$. We define the min-cut of $(S,T)$ as 
\newcommand{\MinCut}{\textsc{MinCut}}
$$
\MinCut(S,T) \triangleq \min |C(S,T)|.
$$
Menger's Theorem states that the maximum number of edge-disjoint paths from $S$ to $T$ is equal to $\MinCut(S,T)$. Therefore, to prove Lemma \ref{lemma:mengers_thm_for_routing}, we need to show that the $\MinCut(S'_r(0), S'_r(t))$ is at least $8r$. We will show that when $S'_r(0) \cap S'_r(t)$ is empty, the $\MinCut(S'_r(0),S'_r(t)) = 8r + 4$.

\begin{figure}[!hb]
    \centering
    \subfloat[Cutting out the shape]{
    \includegraphics[width=0.45\linewidth]{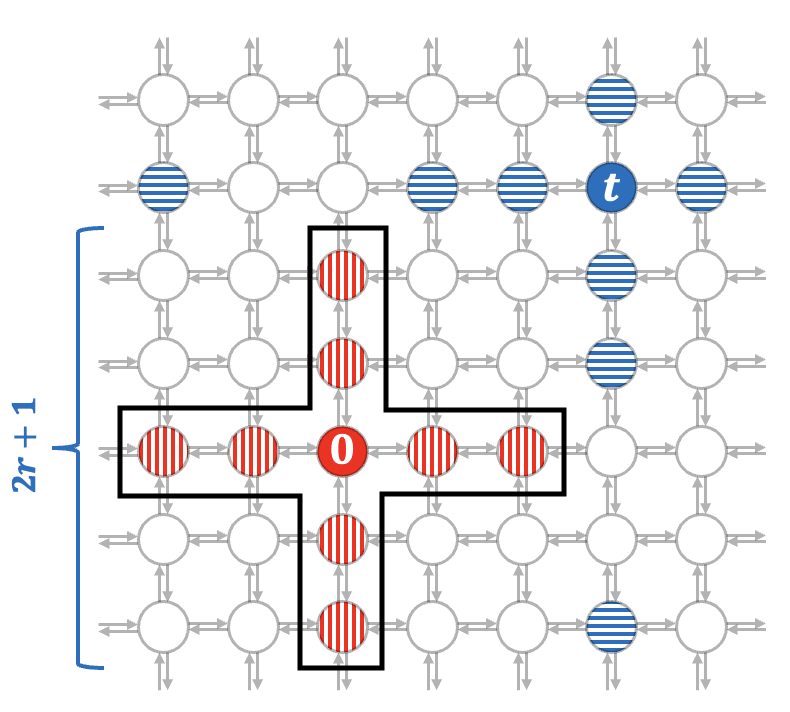}
    }
    \hfill
    \subfloat[Bisection cut and $2(\ell_1 + \ell_2)$]{
    \includegraphics[width=0.45\linewidth]{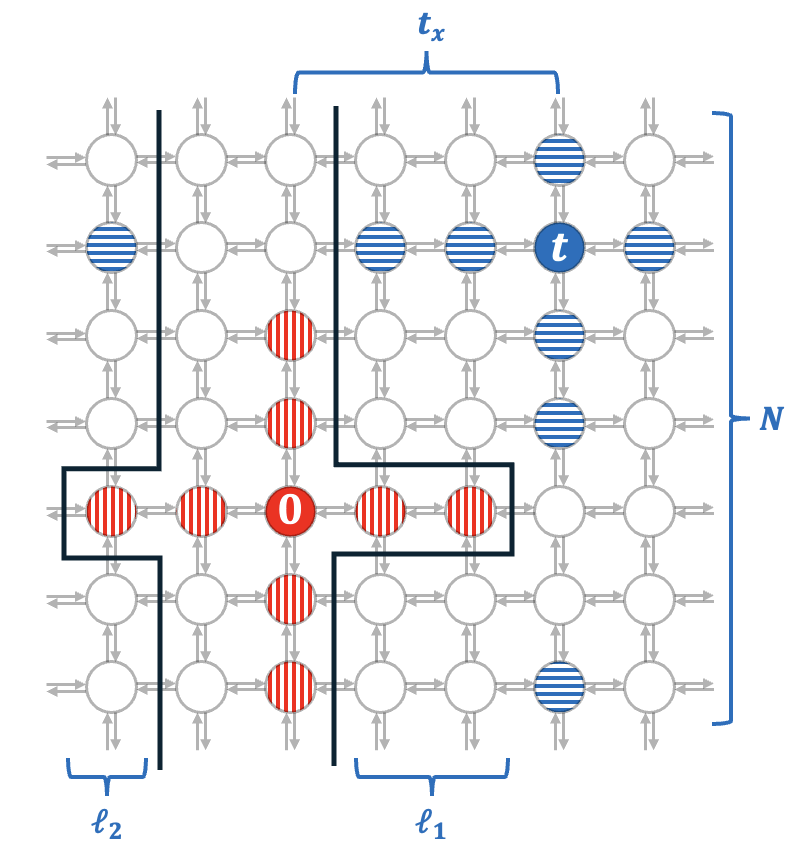}
    }
    \caption{Two possibilities for separating $S'_r(0)$ and $S'_r(t)$ in the $N\times N$ torus. In this example, $\ell_1$ and $\ell_2$ denote the lengths of the shadows in the horizontal axis}
    \label{fig:mengers_proof_figure_cut_possibilities}
\end{figure}
Geometrically, there are two minimal\footnote{Here, minimal means that adding the edges of neighboring nodes to the cutset leads to the cut size either increasing or not changing. Similarly, removing edges from the cutset would either create a path from $S'_r(0)$ to $S'_r(t)$ or would not to the cut size decrease.} ways to disconnect $S'_r(0)$ from $S'_r(t)$:\\
(a) cut along the perimeter of $S'_r(0)$,\\
(b) bisecting the torus and making a few additional cuts $2(\ell_1~+~\ell_2)$ to separate $S'_r(0)$ and $S'_r(t)$. Here, $\ell_1$ and $\ell_2$ denote the length of the regions where the $x-$coordinate of the nodes in $S_r(0)$ and $S_r(t)$ coincide. Specifically, 
\begin{align*}
    \ell_1 &= \max(2r + 1 - t_x, 0)\\
    \ell_2 &= \max(2r + 1 - N + t_x, 0).
\end{align*}
These cuts are shown in Figure~\ref{fig:mengers_proof_figure_cut_possibilities}. The cut size of method (a) is $8r + 4$ and the cut size of method (b) is $2N + 2\ell_1 + 2\ell_2$. Then,
$
\MinCut(S'_r(0), S'_r(t)) = \min(8r + 4, 2N + 2\ell_1 + 2\ell_2).
$
We break this into two cases.\\ \underline{\textit{Case 1}}: $r \leq N/4 - 1/2$\\
    In this case, we note that $8r + 4 \leq 2N \leq 2N + 2\ell_1 + 2\ell_2$. Hence, the $\MinCut$ is $8r + 4$.\\
\underline{\textit{Case 2}}: $r > N/4 - 1/2$\\
Since $r > N/4 - 1/2$, $2r + 1 > N/2$. Since $t_x \leq N/2 < 2r + 1$, it implies that $\ell_1 = 2r+1-t_x$. Moreover, $\ell_2 \geq 2r+1-N+t_x$. Therefore $\ell_1 + \ell_2 \geq 4r + 2 - N$. This implies that $2N + 2(\ell_1 + \ell_2) \geq 8r + 4$, and the $\MinCut$ is $8r + 4$. 

Since the size of the perimeter of $S'_r(0) = 8r + 4$, we can argue that the perimeter of $S'_r(0)$ will always form the $\MinCut$ of $S'_r(0)$ and $S'_r(t)$. In the perimeter of $S'_r(0)$, observe that each node in $S'_r(0)$ contributes to at least $2$ edges into the cut set. Consequently, there are at least two edge-disjoint paths emanating from each node in $S'_r(0)$. A similar argument follows for edge disjoint paths terminating at $S'_r(t)$ by considering the cut set to be the perimeter of $S'_r(t)$. 
\end{proof}



%

\section{Optimality of Local Load-Balancing Policy}\label{proof:optimal_oblivious_upper_bound}
\begin{proof}[Proof of Theorem \ref{thm:optimal_oblivious_upper_bound}] 
    For part (a), observe that in the LLB scheme with parameter $r$, the source node to first send $1/4r$ amount of the traffic to every node within $r$ hops in $S_r(0)$, and then, it has to aggregate  $1/4r$ amount of the traffic from every node within $r$ hops in $S_r(t)$, where $t$ is the destination. Moreover, since there are $8r$ paths from $S_r(0)$ to $S_r(t)$, any edge not in $S_r(0)$ or $S_r(t)$ would carry a maximum of $1/8r$ of the traffic. Consequently, for nodes along the vertical direction, 
    $$
    g^t_{i,i+e_1} \leq \begin{cases}
        \frac{1}{4r}\times(r - \Delta(0,i-e_1)) & i \in S_r(0)\\
        \frac{1}{4r}\times(r - \Delta(t,i)) & i \in S_r(t)\\
        \frac{1}{8r} & \text{else}.
    \end{cases}
    $$ 
    Now, observe that the amount of traffic carried by a link would be maximized when it carries traffic from the stem nodes as well as traffic from non-stem nodes for different source-destination pairs. A link can be in the stem of at most $2r$ source-destination pairs ($r$ unique sources and $r$ unique destinations). The remaining $k-2r$ traffic can contribute to the non-stem load. 
    Hence, the maximum load under LLB with parameter $r$ is 
    $$
        2\times\frac{1}{4r}\times\sum_{i=0}^{r-1}(r-i) + (k-2r)\times\frac{1}{8r} = \frac{r}{4} + \frac{k}{8r}.
    $$
    When $2k$ is a perfect square, the max load is minimized when $r = \sqrt{k/2}$. Consequently, the load is $\sqrt{2k}/4$.


    We can prove part (b) of Theorem \ref{thm:optimal_oblivious_upper_bound} by exploiting the result of part (a). If $m$ is the smallest integer such that $2(m-1)^2 < k \leq 2m^2$. Again, $g^{LLB(m)}$ is a feasible solution. Using arguments similar to that used in part (a), the maximum load would be bounded above by
    \begin{equation*}
        \frac{m}{4} + \frac{k}{8m} \leq \frac{m}{4} + \frac{m}{4} = \frac{m}{2}.\qedhere
    \end{equation*}
\end{proof}

\section{$\MLU$ Equivalence of $\D_k$ and $\D'_k$}\label{proof:vertices}
\begin{proof}[Proof of Lemma \ref{lemma:k_sparse_k_limited_equivalence}]
As discussed, it is easy to see that $\D'_k \subseteq \D_k$, which gives us $\max_{d \in \D'_k}\MLU(f,d) \leq \max_{d \in \D_k}\MLU(f, d)$ for any arbitrary routing $f$.
We are required to also prove that $\max_{d \in \D_k}\MLU(f,d) \leq \max_{d \in \D'_k}\MLU(f, d)$.
If we show that the vertices of the polytope of $\D_k$ are in $\D'_k$, i.e., $\text{vertices}(\D_k) \subset \D'_k$, then it is easy to see that 
\begin{align*}
    \max_{d \in \D_k}\MLU(f,d) &= \max_{d \in \text{vertices}(\D_k)} \MLU(f,d)\\
    &\leq \max_{d \in \D'_k}\MLU(f, d),
\end{align*}
as the optimal points of a linear objective over a convex polytope are at its vertices \cite{bertsimas_linear_optimization}.
To show that $\text{vertices}(\D_k) \subset \D'_k$, we invoke certain properties of vertices from  \cite{bertsimas_linear_optimization}. From \cite{bertsimas_linear_optimization}, any vertex (extreme point) $d^*$ of convex polytope $\D_k$ \textbf{cannot} be written as a convex combination of other points in $\D_k$. Now, consider a traffic matrix $d$ such that $d \in \D_k$ and $d \notin \D'_k$. We will show that it is possible to write $d$ as a convex combination of other points in $\D_k$. Since $d \notin \D'_k$ but $d \in \D_k$, either the number of sources with non-zero traffic in $d$ is strictly greater than $k$, or the number of sinks with non-zero traffic in $d$ is strictly greater than $k$. Without loss of generality, assume that the number of sources is at least $k+1$ (a similar line of reasoning will hold when we assume that the number of sinks is at least $k+1$ instead). As $d \in \D_k$, the total amount of traffic in the network must satisfy $\sum_{s,t} d_{s,s+t} \leq k$ and each source can produce at most 1 unit of traffic. We break it down into two cases.

\noindent
\underline{\textit{Case 1}}: $\sum_{s,t}d_{s,t} = k$\\
In this case, we observe that there must exist at least 2 sources that produce less than one unit of traffic, i.e., $\exists s_1, s_2$ such that $0<\sum_{t} d_{s_1, s_1 + t} < 1$ and $0<\sum_{t} d_{s_2, s_2 + t} < 1$. This can be proven by contradiction. Assume there is at most $1$ source that produces strictly less than 1 unit of traffic. Then, there are at least $k$ other sources that produce exactly 1 unit of traffic. This would imply that the total traffic in the network exceeds $k$ units. Consequently, there also exist $t_1$ and $t_2$ such that 
\begin{equation}
    \begin{aligned}
         0 < d_{s_1,s_1 + t_1} < 1\text{ and }
         0 < d_{s_2,s_2 + t_2} < 1. 
\end{aligned}\tag{*}\label{eqn:condition_for_inactive}
\end{equation}
Now, we construct two traffic matrices that also lie in $\D_k$. Define $d^{(1)}$ and $d^{(2)}$ as follows:
\begin{align*}
    d^{(1)}_{s,s+t} &= \begin{cases}
        d_{s,s+t} & (s, t) \notin \{(s_1, t_1), (s_2, t_2)\}\\
        d_{s,s+t} - \eps & (s,t) = (s_1, t_1)\\
        d_{s,s+t} + \eps & (s,t) = (s_2, t_2),
    \end{cases}\\
    d^{(2)}_{s,s+t} &= \begin{cases}
        d_{s,s+t} & (s, t) \notin \{(s_1, t_1), (s_2, t_2)\}\\
        d_{s,s+t} + \eps & (s,t) = (s_1, t_1)\\
        d_{s,s+t} - \eps & (s,t) = (s_2, t_2),
    \end{cases}
\end{align*}
where $\eps > 0$ is a small perturbation. As a consequence of condition \eqref{eqn:condition_for_inactive}, it is possible to find $\eps > 0$ such that 
\begin{align*}
    0 < d_{s,s+t} \pm \eps < 1 &&(s,t) \in \{(s_1,t_1), (s_2,t_2)\}. 
\end{align*}
One can easily verify that $d^{(1)}, d^{(2)} \in \D_k$, and $$d = \frac{d^{(1)} + d^{(2)}}{2},$$ 
which proves that $d$ is not a vertex of $\D_k$.

\noindent
\underline{\textit{Case 2}}: $\sum_{s,t}d_{s,s+t} < k$\\
In this case, we observe that there must be at least one source that produces less than one unit of traffic, i.e., $\exists s_0$ such that $0< \sum_t d_{s_0, s_0 + t} < 1$. Consequently, there also exists $t_0$ such that 

\begin{equation}
    0 < d_{s_0, s_0 + t_0} < 1 \tag{**}\label{eqn:cond_1_inactive}
\end{equation} 
Now, construct $d^{(1)}$ and $d^{(2)}$ as follows:
\begin{align*}
    d^{(1)}_{s,s+t} = \begin{cases}
        d_{s,s+t} & (s,t) \neq (s_0, t_0)\\
        d_{s,s+t} + \eps & (s,t) = (s_0, t_0),
    \end{cases}\\
    d^{(2)}_{s,s+t} = \begin{cases}
        d_{s,s+t} & (s,t) \neq (s_0, t_0)\\
        d_{s,s+t} - \eps & (s,t) = (s_0, t_0),
    \end{cases}
\end{align*}
where $\eps > 0$ is a small perturbation. As a consequence of \eqref{eqn:cond_1_inactive}, it is always possible to find $0 < d_{s_0,s_0+t_0} \pm \eps < 1$ and $\sum_{s,t} d_{s,s+t} < 1$. Finally, we observe that 
$$d = \frac{d^{(1)} + d^{(2)}}{2}.$$ 
Hence, $d$ is not a vertex of $\D_k$.
\end{proof}

\section{Reduced Version of the Oblivious Routing Program}\label{proof:reduced_problem}
\begin{proof}[Proof of Lemma \ref{lemma:reduced_problem}]
To prove the equivalence of both the problems, we first need to establish that it is sufficient to search within the class of automorphism-invariant policies to find the optimal solution to \eqref{eqn:main_problem1}. This comes from Lemma \ref{lemma:symmetry_sufficiency_condition}. Next, we need to show that the maximum load constraints \eqref{eqn:max_constraint1} and \eqref{eqn:max_constraint_sym} are equivalent, when restricted to automorphism-invariant policies.\\
(i) \underline{$\eqref{eqn:max_constraint1} \implies \eqref{eqn:max_constraint_sym}$}:
Proving this relationship is straightforward. Since $\max_{d \in \D_k}\sum_{s,t}d_{s,s+t}f^{s,s+t}_{i,i+e} \leq \theta$ $\forall i,e$, this condition has to be true at the origin node and at direction $e_1$. Therefore,
 $\sum_{s,t}d_{s,s+t}f^{s,s+t}_{0,e_1} \leq \theta$ for all $d \in \D_k$. Since $f$ is an automorphism-invariant routing policy, $f^{s,s+t}_{0,e_1} = g^t_{-s, -s+e_1}$. Hence, 
 \begin{equation}
     \sum_{s,t}d_{s,s+t}g^{t}_{-s,-s+e_1} \leq \theta \quad \forall d \in \D_k. \label{eqn:g_constraint}
 \end{equation}
 
\noindent
(ii) \underline{$\eqref{eqn:max_constraint_sym}\implies  \eqref{eqn:max_constraint1}$}:
Consider an arbitrary link in the network $(i,i+e)$. Since the routing policy is automorphism-invariant, $f^{s,s+t}_{i,i+e} = g^{t}_{i-s, i-s + e}$, where we have applied the translation automorphism. Now, for any arbitrary traffic $d^{(1)} \in \D_k$
\begin{align*}
    \sum_{s,t} d^{(1)}_{s,s+t}f^{s,s+t}_{i,i+e} &= \sum_{s,t} d^{(1)}_{s,s+t}g^{t}_{i-s,i-s+e}\\
    &= \sum_{s,t} d^{(1)}_{s+i,s+i+t}g^{t}_{-s,-s+e},
\end{align*}
where the last equality comes from reindexing the summation.
Let us define $d^{(2)}$ as $d^{(2)}_{s,s+t} = d^{(1)}_{s+i,s+i+t}$. Note that $d^{(2)} \in \D_k$, as the $d^{(2)}$ is just the translation of source-destination pairs of $d^{(1)}$.
Additionally, note that the set of directions $A = \{e_1, e_2, -e_1, -e_2\}$ can be completely produced using $e_1$ and the $I, R_{xy}, R_{0}$ and $R_{xy}R_0$ automorphisms respectively.
Therefore, from automorphism invariance, we get $$
g^{t}_{-s,-s+e} = g^{\phi(t)}_{\phi(-s),\phi(-s)+e_1},$$ where $\phi$ is the automorphism which maps $e$ to $e_1$, and,
\begin{align*}
    \sum_{s,t} d^{(1)}_{s,s+t}f^{s,s+t}_{i,i+e} 
    &= \sum_{s,t} d^{(1)}_{s+i, s+i+t}g^{t}_{-s,-s+e}\\
    &= \sum_{s,t} d^{(2)}_{s,s+t}g^{t}_{-s,-s+e}\\
    &= \sum_{s,t} d^{(2)}_{s,s+t}g^{\phi(t)}_{\phi(-s),\phi(-s)+e_1}\\
    &= \sum_{s,t} d^{(2)}_{\phi(s),\phi(s)+\phi(t)}g^{t}_{-s,-s+e_1}.
\end{align*}
Again, the last equality follows from a reindexing of the summation under the automorphism $\phi$. If we define $d^{(3)}$ to be a traffic matrix such that $d^{(3)}_{s,s+t} = d^{(2)}_{\phi(s),\phi(s)+\phi(t)},$ where $\phi \in \{I, R_{xy}, R_0, R_{xy}R_0\}$, then observe that $d^{(3)} \in \D_k$ also. Finally, 
\begin{align*}
    \sum_{s,t} d^{(1)}_{s,s+t}f^{s,s+t}_{i,i+e} &= \sum_{s,t} d^{(2)}_{\phi(s),\phi(s)+\phi(t)}g^{t}_{-s,-s+e_1}\\
    &= \sum_{s,t} d^{(3)}_{s,s+t}g^{t}_{-s,-s+e_1}\\
    &\leq \theta \qquad\text{(Condition \eqref{eqn:max_constraint_sym})}.
\end{align*}
Since this is true for any arbitrary traffic matrix $d^{(1)} \in \D_k$ and arbitrary $(i,i+e)$, we get $\max_{d \in \D_k}\sum_{s,t}d_{s,s+t}f^{s,s+t}_{i,i+e} \leq \theta$ for all links $(i,i+e)$.
\end{proof}

\section{Proof of Worst-Case Load Lower Bound }\label{proof:optimal_oblivious_lower_bound}
\begin{proof}[Proof Sketch of Theorem \ref{thm:optimal_oblivious_lower_bound}]
To prove part (a) of Theorem \ref{thm:optimal_oblivious_lower_bound}, we first consider the dual of the inner minimization problem $\min_{g \in \R_{sym}} \sum_{s,t} d_{s,s+t} g^{t}_{-s,-s+e_1}$ for any traffic matrix $d \in \D_{k}$. We focus on the case where $N$ is even for simplicity. The case when $N$ is odd is similar, with minor modifications. From strong duality, the problem $\min_{g \in \R_{sym}} \sum_{s,t}d_{s,s+t}g^{t}_{s,s+e_1}$ is equivalent to
{\thickmuskip=0mu
\thinmuskip=0mu
\medmuskip=0mu
\begin{equation}
    \begin{aligned}
        \min_{g \in \R_{sym}} \sum_{s,t}d_{s,s+t}g^{t}_{-s,-s+e_1}= \max_{\nu} &\quad \frac{1}{4}\sum_{t} \nu^t_t\\
        \text{s.t.} & \quad \nu^t_{i+e} - \nu^t_{i} \leq \lambda(d,t,i,e) &&\forall i, t\\
        &\quad \nu^t_0 = 0 && \forall t    \end{aligned}\label{eqn:min_dual}
\end{equation}
}
where $
\lambda(d,i,t,e) = d_{-\phi(i), -\phi(i)+\phi(t)}$ such that $e = \phi(e_1)$ and $\phi \in \{I, R_{xy}, R_0, R_0R_{xy}\}$.
Then, the problem $\max_{d \in \D_k} \min_{g \in \R_{sym}}\sum_{s,t}d_{s,s+t}g^{t}_{s,s+e_1}$ is equivalent to
\begin{equation}
    \begin{aligned}
        \max_{d, \nu} &\quad \frac{1}{4}\sum_{t} \nu^t_t\\
        \text{s.t.} & \quad \nu^t_{i+e} - \nu^t_{i} \leq \lambda(d,t,i,e) &&\forall i, t\\
        &\quad \nu^t_0 = 0 && \forall t\\
        &\quad d \in \D_k.
    \end{aligned}\label{eqn:min_dual2}
\end{equation}
This is very similar to the techniques used in \cite{applegate_making_2006}. Refer to \cite{applegate_making_2006} for more details on how to formulate the equivalent maximization problem.
In problem \eqref{eqn:min_dual2}, we interpret $\lambda(d,t,i,e)$ as the cost of using edge $(i,i+e)$ in a route  to $t$ under traffic matrix $d$. Due to the telescopic nature of constraint \eqref{eqn:min_dual2}, we can conclude that for any path from $0$ to $t$ (denoted as $\mathcal{P}_t$), 
$$\nu^t_t \leq \sum_{(i,i+e) \in \mathcal{P}_t} \lambda(d,t,i,e)$$ 
This has to be true for the minimum cost path as well, i.e., $\nu^t_t \leq \min_{\mathcal{P}_t}\sum_{(i,i+e) \in \mathcal{P}_t} \lambda(d,t,i,e)$. Since we need to maximize the sum of $\nu^t_t$, we can greedily set $\nu^t_t = \min_{\mathcal{P}_t}\sum_{(i,i+e) \in \mathcal{P}} \lambda(d,t,i,e)$, which consequently leads to
$$
\theta^* = \frac{1}{4}\max_{d \in \D_k}\min_{\bar{\mathcal{P}}} \sum_{t} \sum_{(i,i+e) \in \mathcal{P}_t}\lambda(d,t,i,e),
$$
where $\bar{\mathcal{P}}$ is the collection of paths from the node $0$ to every $t \in V \backslash\{0\}$. From the above formulation, observe that $\theta^*$ involves computing the maximum possible minimum sum cost of routing from $0$ to every node $t$.
Define $B_r(i) \triangleq \{j \in V | \Delta(i,j) \leq r-1\}$ to be the ball around node $i$ containing nodes which are at most $r-1$ hops away from $i$.
Observe that for a sufficiently ``distant" node $t$, any route from $0$ to $t$ must traverse through $B_r(0)$ and $B_r(t)$ for a sufficiently small $r$. In fact, it must traverse through at least $r$ edges emanating from nodes in $B_r(0)$ that are directed away from the origin, and traverse through at least $r$ edges emanating from nodes in $B_r(t)$ that are directed toward $t$. This idea is visualized in Figure \ref{fig:routing_ball_appendix}. 
\begin{figure}[!hbt]
    \centering
    \includegraphics[width=0.8\linewidth]{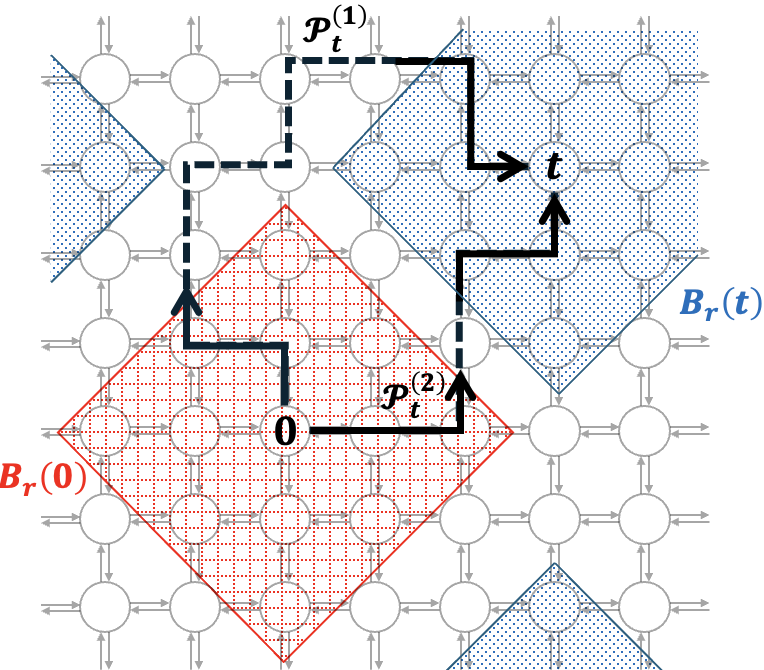}
    \caption{The ball around $0$ is highlighted in red. The ball around $t$ is highlighted in blue. Here, $r = 3$. Observe that any path from $0$ to $t$ must traverse through at least $r$ edges emanating from nodes in $B_r(0)$ that are directed away from the origin, and traverse through at least $r$ edges emanating from nodes in $B_r(t)$ that are directed toward $t$.}
    \label{fig:routing_ball_appendix}
\end{figure}

Based on this observation, if we select a traffic matrix $d \in \D_k$ such that the cost of using edge $(i,i+e)$ is constant, i.e.,
$\lambda(d,i,t,e) = \lambda$ 
for a particular choice of $t$ and $\forall i \in B_{r}(0), \forall e $ directed away from $0$ and $\forall i \in B_{r}(t), \forall e $ directed toward $t$, then we see that any path from $0$ to $t$ would incur a cost of at least $2r\lambda$. 

The split diamond traffic matrix of size $r$ does exactly this. It selects $t$ to be the farthest node from the origin, which is $( N/2, N/2)$, only when $k \leq N^2/2$. The sources and destinations of traffic are carefully selected such that every hop that is directed away from the origin in $B_r(0)$ incurs a cost of $1$ and every hop that is directed toward $t$ in $B_r(t)$ incurs a cost of $1$. Hence, $\lambda = 1$ and every path from $0$ to $t=(N/2, N/2)$ incurs a cost of $2r$. As mentioned earlier, for the split diamond traffic of size $r$, $\sum_{s,t}d^{SD(r)}_{s,s+t} = 2r^2$ must be lesser than or equal to $k$ for $d^{SD(r)}$ to belong to $\D_k$. The sum cost of the objective under Split-Diamond traffic of size $r$ is $\frac{1}{4} \times 2r = \frac{r}{2}$. We maximize this sum cost under the constraint $2r^2 \leq k, r \in \mathbb{N}$. When $2k$ is a perfect square, the best choice of $r$ would be $\sqrt{k/2}$ and the value of the objective is $\sqrt{2k}/4$. Since $d^{SD(r)}$ is a feasible solution, we arrive at $\theta^* \geq \sqrt{2k}/4$ when $2k$ is a perfect square.

The proof for part (b) of Theorem \ref{thm:optimal_oblivious_lower_bound} follows immediately from the result of part (a). When $2k$ is not a perfect square, we can find the greatest integer $m$ such that $2m^2 \leq k < 2(m+1)^2$. Then, we consider the Split-Diamond traffic matrix of size $m$, $d^{SD(m)}$ and the Split-Diamond traffic matrix of size $m+1$, $d^{SD(m+1)}$. Then, we define $\bar d$ to be a convex combination of $d^{SD(m)}$ and $d^{SD(m+1)}$ such that $\bar d \in \D_k$, i.e.,
$$\bar d = (1-\alpha) d^{SD(m)} + \alpha d^{SD(m+1)},$$
for $0 \leq \alpha < 1$ such that 
\begin{align*}
    \sum_{s,t}\bar d_{s,s+t} &= (1-\alpha)\sum_{s,t}d^{SD(m)} + \alpha \sum_{s,t}d^{SD(m+1)}\\
    &= 2(1-\alpha)m^2 + 2\alpha(m+1)^2\\
    &= 2m^2 + 2\alpha(2m + 1) \leq k.
\end{align*}
This is clearly satisfied when $\alpha = \frac{k-2m^2}{4m + 2}.$
Finally,
\begin{align*}
    \theta^* &= \max_{d \in \D_{k}}\min_{g \in \R_{sym}} \sum_{s,t}d_{s,s+t}g^{t}_{s,s+e_1}\\
    &\geq \min_{g \in \R_{sym}} \sum_{s,t}\bar d_{s,s+t}g^{t}_{s,s+e_1}\\
    &\geq (1-\alpha)\min_{g \in \R_{sym}} \sum_{s,t} d^{SD(m)}_{s,s+t}g^{t}_{s,s+e_1}\\
    &\quad+ \alpha \min_{g \in \R_{sym}} \sum_{s,t} d^{SD(m+1)}_{s,s+t}g^{t}_{s,s+e_1}\\
    &\geq (1-\alpha)\frac{m}{2} + \alpha \frac{m+1}{2}= \frac{m+\alpha}{2}.
\end{align*}
By noticing that $2m^2 \leq k \leq 2(m+1)^2$, we can conclude that $\theta^* \gtrsim \frac{\sqrt{2k}}{4}$.
\end{proof}


\section{Proof of Worst-Case Lower Bound in General $N\times M$ Tori}\label{proof:asymmetric_tori_lb}
\noindent
Before we begin the proof, we need certain definitions that we will use in the proof. We define the following distance metric $\Delta_{(\lambda_1, \lambda_2)}(.,.)$ on $V$.
{\thickmuskip=0mu
\thinmuskip=0mu
\medmuskip=0mu
\begin{align}
\Delta_{(\lambda_1, \lambda_2)}(u,v) = &\lambda_1\min\{(u_y - v_y) \bmod N, (v_y - u_y) \bmod N \}\nonumber\\
+ &\lambda_2\min\{(u_x - v_x) \bmod M, (v_x - u_x) \bmod M \}.
\end{align}
}
The distance metric $\Delta_{(\lambda_1, \lambda_2)}(\cdot,\cdot)$ counts the minimum number of weighted hops required to go from node $u$ to node $v$, where the vertical hops are weighted by $\lambda_1$ and horizontal hops are weighted $\lambda_2$. \\
Next, we define $B^{(\lambda_1,\lambda_2)}_r(i)$ to be the ball around node $i$ containing nodes which are under $r$ ``generalized" hops away from $i$. In other words,
$$B^{(\lambda_1,\lambda_2)}_r(i) \triangleq \{j \in V | \Delta_{(\lambda_1,\lambda_2)}(i,j) < r\}.$$
We are now ready to begin the proof.
\begin{proof}[Proof of Theorem \ref{thm:asymmetric_tori_lb}a)]
 We consider the dual of the problem \eqref{eqn:min_max_concise_NxM}. We focus on the case where $N$ is even for simplicity. The case when $N$ is odd is similar, with minor modifications. From weak duality, problem \eqref{eqn:min_max_concise_NxM} can be lower bounded by
{
\begin{equation}
    \begin{aligned}
        \theta^* \geq \max_{\nu, \lambda} &\quad \frac{1}{2}\sum_{t} \nu^t_t\\
        \text{s.t.} & \quad \nu^t_{i+e} - \nu^t_{i} \leq \begin{cases}
            \lambda_1 d^{(1)}_{i,i+t} & e = e_1\\
            \lambda_1 d^{(1)}_{-i,-i-t} & e = -e_1\\
            \lambda_2 d^{(2)}_{i,i+t} & e = e_2\\
            \lambda_2 d^{(2)}_{-i,-i-t} & e = -e_2
        \end{cases} &&\forall i, t \\
        &\quad c_1\lambda_1 + c_2\lambda_2 \leq 1, \quad \lambda_1,\lambda_2 \geq 0\\
        &\quad \nu^t_0 = 0 && \forall t\\
        &\quad d^{(1)}, d^{(2)} \in \D_k\end{aligned}\label{eqn:min_dual_NxM}
\end{equation}
}
Due to the telescopic nature of constraint in \eqref{eqn:min_dual_NxM}, we can conclude that for any path from $0$ to $t$ (denoted as $\mathcal{P}_t$), 
$$\nu^t_t \leq \sum_{(i,i+e) \in \mathcal{P}_t} \lambda(t,i,e),$$ 
where 
$$\lambda(t,i,e) = \begin{cases}
            \lambda_1 d^{(1)}_{i,i+t} & e = e_1\\
            \lambda_1 d^{(1)}_{-i,-i-t} & e = -e_1\\
            \lambda_2 d^{(2)}_{i,i+t} & e = e_2\\
            \lambda_2 d^{(2)}_{-i,-i-t} & e = -e_2
        \end{cases},$$
and variable $\lambda(t,i,e)$ is to be interpreted as the cost of using link $(i,i+e)$.
This has to be true for the minimum cost path as well, i.e., $\nu^t_t \leq \min_{\mathcal{P}_t}\sum_{(i,i+e) \in \mathcal{P}_t} \lambda(t,i,e)$. Since we need to maximize the sum of $\nu^t_t$, we can greedily set $\nu^t_t = \min_{\mathcal{P}_t}\sum_{(i,i+e) \in \mathcal{P}} \lambda(t,i,e)$, which consequently leads to the objective function
$$
\theta^* = \frac{1}{2}\max_{\lambda_1,\lambda_2,d^{(1)}, d^{(2)}}\min_{\bar{\mathcal{P}}} \sum_{t} \sum_{(i,i+e) \in \mathcal{P}_t}\lambda(t,i,e),
$$
where $\bar{\mathcal{P}}$ is the collection of paths from the node $0$ to every $t \in V \backslash\{0\}$. From the above formulation, observe that $\theta^*$ involves computing the minimum sum cost of routing from $0$ to every node $t$, with costs that are determined by traffic matrices $d^{(1)}$ and $d^{(2)}$.

However, there is a property that any path from $0$ to any destination $t$ must satisfy, that we can use for the lower bound. 
Observe that for a sufficiently ``distant" node $t$, any route from $0$ to $t$ must traverse through $B^{(\lambda_1,\lambda_2)}_r(0)$ and $B^{(\lambda_1,\lambda_2)}_r(t)$ for a sufficiently small $r$. In fact, it must must traverse through edges emanating from nodes in $B^{(\lambda_1,\lambda_2)}_r(0)$ that are directed away from the origin, and traverse through edges emanating from nodes in $B^{(\lambda_1,\lambda_2)}_r(t)$ that are directed toward $t$. This idea is visualized in Figure \ref{fig:routing_ball_appendix_NxM}. 
\begin{figure}[!hbt]
    \centering
    \includegraphics[width=\linewidth]{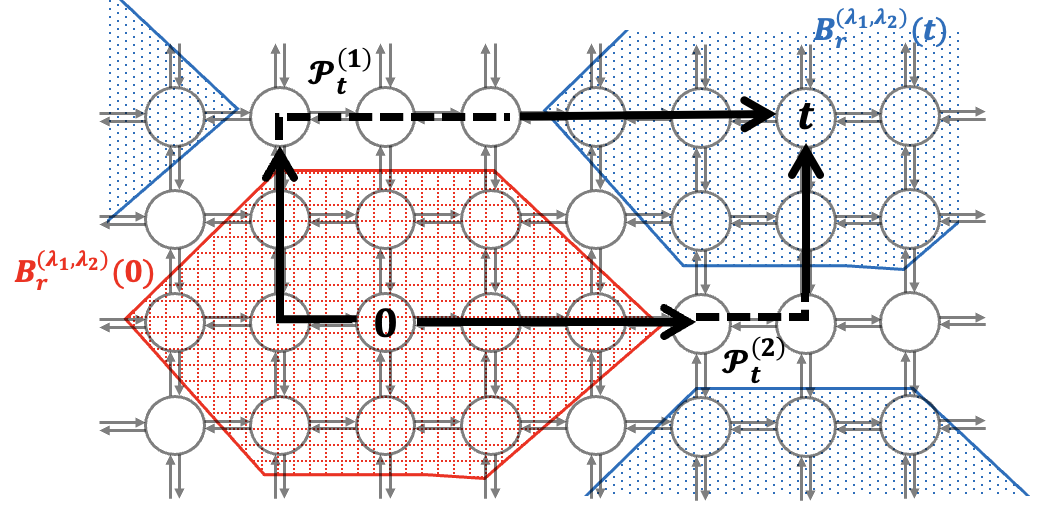}
    \caption{The ball around $0$ is highlighted in red. The ball around $t$ is highlighted in blue. Here, $N=4, M=10, r = 2$, $\lambda_1 = 1, $ and $\lambda_2 = \frac{1}{2}$. Observe that any path from $0$ to $t$ must traverse through at least $r$ edges emanating from nodes in $B^{(\lambda_1,\lambda_2)}_r(0)$ that are directed away from the origin, and traverse through at least $r$ edges emanating from nodes in $B^{(\lambda_1,\lambda_2)}_r(t)$ that are directed toward $t$.}
    \label{fig:routing_ball_appendix_NxM}
\end{figure}

Based on this observation, if we select traffic matrices $d^{(1)},d^{(2)}  \in \D_k$ such that the following is true-
\begin{enumerate}[(i)]
    \item The cost of using edge $(i,i\pm e_1)$ is $\lambda_1$
 for all $ i \in B^{(\lambda_1,\lambda_2)}_{r}(0), \forall e $ directed away from $0$ and for all $i \in B^{(\lambda_1,\lambda_2)}_{r}(t), \forall e $ directed toward $t$
 \item The cost of using edge $(i,i\pm e_2)$ is $\lambda_2$
 for all $i \in B^{(\lambda_1,\lambda_2)}_{r}(0), \forall e $ directed away from $0$ and for all $i \in B^{(\lambda_1,\lambda_2)}_{r}(t), \forall e $ directed toward $t$,
\end{enumerate}
 then we see that any path from $0$ to $t$ would incur a cost of at least $2r$. We aim to do exactly this using two traffic matrices defined below. 

We define the traffic matrices $d^{(1),(\lambda_1,\lambda_2, r)}$ and $d^{(2),(\lambda_1,\lambda_2, r)}$  (which are parametrized by $\lambda_1, \lambda_2$ and $r$) so as to set the cost of each edge. Let $t^* = \left(M/2, N/2\right)$. Then,
\begin{align*}
    d^{(1),(\lambda_1,\lambda_2, r)}_{s,s+t} &= \begin{cases}
        1 & \Delta_{(\lambda_1, \lambda_2)}(0,s) < r \text{ and } \\
        & s_y \leq N/2-1 \text{ and } t = t^*\\
        1 & \Delta_{(\lambda_1, \lambda_2)}(t^*,s) < r \text{ and }\\
        & s_y \leq N/2-1
        \text{ and } t = t^*\\
        0 & \text{Otherwise}
    \end{cases},\\
    d^{(2),(\lambda_1,\lambda_2, r)}_{s,s+t} &= \begin{cases}
        1 & \Delta_{(\lambda_1, \lambda_2)}(0,s) < r \text{ and }\\
        &s_x \leq M/2-1 \text{ and } t = t^*\\
        1 & \Delta_{(\lambda_1, \lambda_2)}(t^*,s) < r \text{ and }\\
        &s_x \leq M/2-1 \text{ and } t = t^*\\
        0 & \text{Otherwise}
    \end{cases}.
\end{align*}
An example of this traffic is plotted in Figure \ref{fig:traffic_NxM}. Observe that traffic is destined only to the farthest node from the every source, i.e., $t = t^* = ( M/2, N/2)$. The sources and destinations of traffic are carefully selected such that every hop along $\pm e_1$ and $\pm e_2$ that is directed away from the origin in $B^{(\lambda_1,\lambda_2)}_r(0)$ incurs a cost of $\lambda_1$ and $\lambda_2$ respectively. It also ensures the same for $B^{(\lambda_1,\lambda_2)}_r(t^*)$. Additionally, this is true only when $k \leq L^2/2$.

\begin{figure}[!hbt]
    \centering
     \begin{subfigure}{0.45\textwidth}
        \centering
        \includegraphics[width=\textwidth]{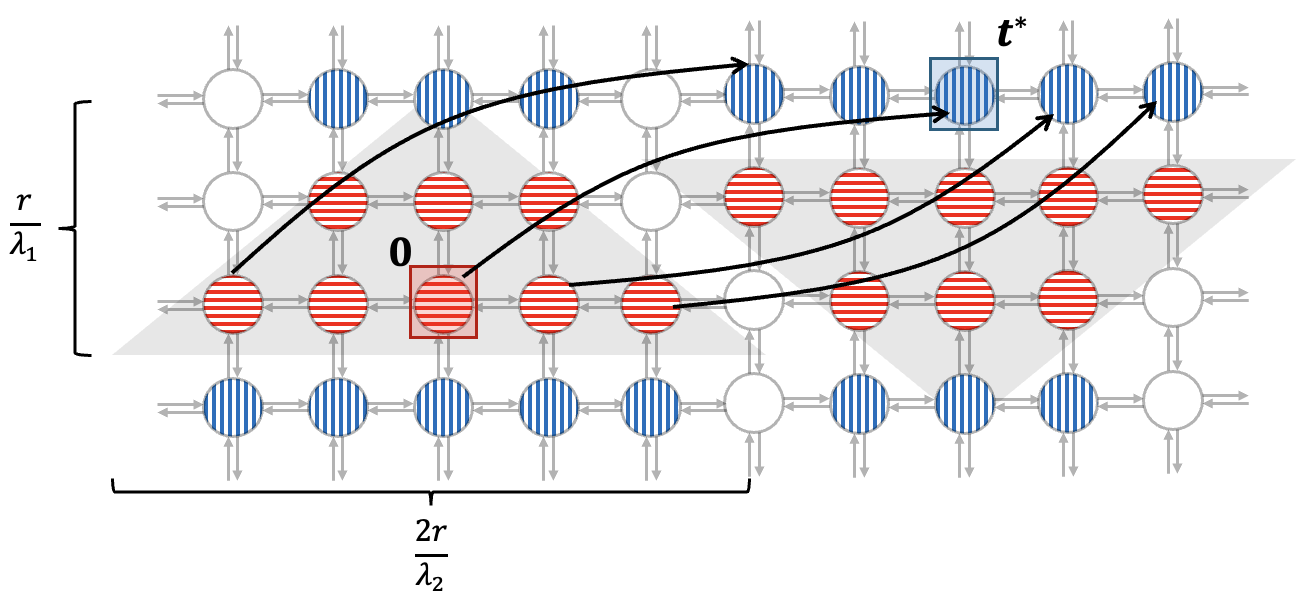}
        \caption{Traffic Matrix $d^{(1),(\lambda_1,\lambda_2,r)}$ with $r=2, \lambda_1=1, \lambda_2 = \frac{1}{2}$ for an $4\times10$ torus network. The origin $0=(0,0)$ and node $t^*$ are marked by boxes. Here, $t^* = (5,2)$. The source nodes of traffic are shaded with red horizontal lines and the destinations are shaded with blue vertical lines. Few of the source-destination pairs are marked with dark arrow lines. Every source node $s$ has traffic to destination $s+t^*$.  There is a total of $16$ units of traffic in this traffic matrix, which is closely upper bounded by $\frac{2r^2}{\lambda_1\lambda_2} = 16$. The shaded triangles represent the approximation.}
        \label{fig:traffic_d1_NxM}
    \end{subfigure}
    \hfill 
    \begin{subfigure}{0.45\textwidth}
        \centering
        \includegraphics[width=\textwidth]{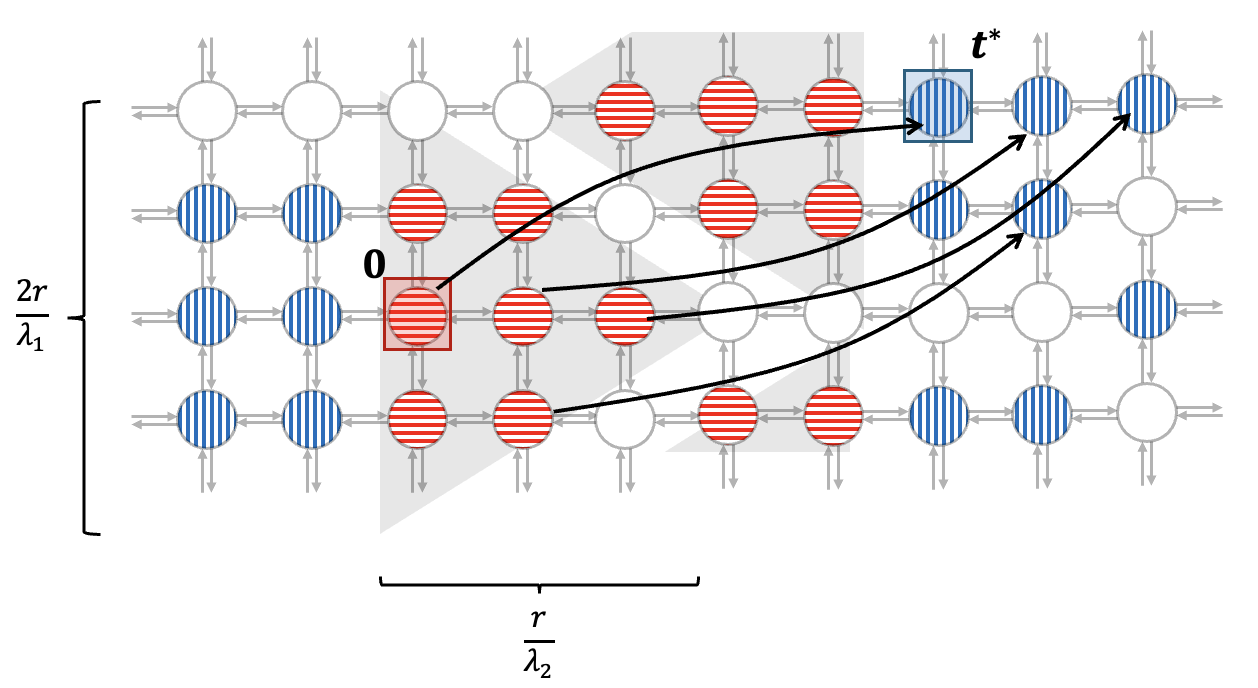}
        \caption{Traffic Matrix $d^{(2),(\lambda_1,\lambda_2,r)}$ with $r=2, \lambda_1=1, \lambda_2 = \frac{1}{2}$ for an $4\times10$ torus network. The origin $0=(0,0)$ and node $t^*$ are marked by boxes. Here, $t^* = (5,2)$. The source nodes of traffic are shaded with red horizontal lines and the destinations are shaded with blue vertical lines. Few of the source-destination pairs are marked with dark arrow lines. Every source node $s$ has traffic to destination $s+t^*$.  There is a total of $14$ units of traffic in this traffic matrix, which is closely upper bounded by $\frac{2r^2}{\lambda_1\lambda_2} = 16$. The shaded triangle represent the approximation.}
        \label{fig:traffic_d2_NxM}
    \end{subfigure}
    \caption{}
    \label{fig:traffic_NxM}
\end{figure}

There are approximately $2\times\frac{r^2}{\lambda_1\lambda_2}$ nodes that produce traffic in $d^{(1),(\lambda_1,\lambda_2, r)}$ and $d^{(2),(\lambda_1,\lambda_2, r)}$ each. This can be explained by observing that the set $\{ s \in V | \Delta_{(\lambda_1, \lambda_2)}(0,s) < r$ and $s_y \geq 0\}$ can be approximated by the triangular region formed by  $\{(x,y) \in \mathbb{R}^2 | \lambda_1y + \lambda_2x < r$, $\lambda_1y - \lambda_2x < r$ and $y \geq 0\}$. Similar arguments hold for the other sets. These approximations for counting are displayed in Fig. \ref{fig:traffic_NxM}. 
Hence, for $ d^{(1),(\lambda_1,\lambda_2, r)}, d^{(2),(\lambda_1,\lambda_2, r)} \in \D_k$, we require $\frac{2r^2}{\lambda_1\lambda_2} \leq k$, because the total traffic is limited by $k$. Hence, we can set $r = \sqrt{k\lambda_1\lambda_2/2}$.

As mentioned, the cost of any path from $0$ to $t$ is at least $\sum_t\nu^t_t = 2r = \sqrt{2k\lambda_1\lambda_2}$. We need to pick $\lambda_1$ and $\lambda_2$ in order to maximize the cost. To that end, note that {\thickmuskip=0mu
\thinmuskip=0mu
$$\sqrt{\lambda_1\lambda_2} = \frac{1}{\sqrt{c_1c_2}}\sqrt{c_1\lambda_1c_2\lambda_2} \stackrel{(a)}{\leq} \frac{1}{\sqrt{c_1c_2}} \frac{c_1\lambda_1 + c_2\lambda_2}{2} \stackrel{(b)}{\leq} \frac{1}{2\sqrt{c_1c_2}},$$}
where inequality $(a)$ comes from the AM-GM inequality, and $(b)$ comes from \eqref{eqn:min_dual_NxM}. Hence, we can set $\lambda_1, \lambda_2$ accordingly in order to maximize the cost of any path to $\frac{\sqrt{2k}}{2\sqrt{c_1c_2}}$. Finally, we observe that $\theta^*$ is only half of the cost $\sum_t\nu^t_t$. Therefore, $\theta^* \geq \frac{\sqrt{2k}}{4\sqrt{c_1c_2}} - \eps$, where $\eps$ is the ``small" correction term that accounts for the deviation from the  approximation. The error in approximation can be bounded using Pick's Theorem.
\end{proof}
Parts (b) and (c) of Theorem \ref{thm:asymmetric_tori_lb} are much simpler to prove. 
\begin{proof}[Proof of Theorem \ref{thm:asymmetric_tori_lb}b)]
Let us split the nodes in the network $V$ into two sets $S$ and $T = V\backslash S$. Clearly, all the traffic from nodes in $S$ destined to nodes in $T$ must flow through the cut-set of $S$ and $T$, denoted as $C(S,T)$. Formally, a cut set $C(S,T)$ is a set of edges that are directed from $S$ to $T$ such that, once removed, there exists no directed path from any node in $S$ to any node in $T$. Here, we define the minimum capacity cut from $S$ to $T$ denoted as 
\newcommand{\MinCapCut}{\textsc{MinCapCut}}
$\MinCapCut(S,T)$ as
$$
\MinCapCut(S,T) = \min_{C(S,T)}\sum_{(i,i+e) \in C(S,T)} c_{i,i+e},
$$
where $c_{i,i+e}$ is the capacity of the link $(i,i+e)$ and $$
c_{i,i+e} = \begin{cases}
    c_1 & e = \pm e_1\\
    c_2 & e = \pm e_2
\end{cases}.
$$
We define the bisection bandwidth as 
\begin{align*}
   &\textsc{BisectionBW} = \min_{S, |S| = NM/2, T=V\backslash S}\MinCapCut(S,T). 
\end{align*}
Then, observe that the bisection bandwidth of the general $N\times M$ torus is $\min(2c_1M, 2c_2N)$. Note that $$
L = \min\left(\sqrt{\frac{c_2}{c_1}}N, \sqrt{\frac{c_1}{c_2}}M\right) =  \frac{\textsc{BisectionBW}}{2\sqrt{c_1c_2}}.$$
Let us pick a traffic matrix in which all traffic originates from one half of the network to the other half of the network. This is possible when $k \leq NM/2$. All the traffic must flow through the bisection cut which separates one half from the other. Hence, for any routing policy, the load must no lesser than $\frac{k}{\textsc{BisectionBW}} = \frac{k}{2\sqrt{c_1c_2}L}$.
\end{proof}
\begin{proof}[Proof of Theorem \ref{thm:asymmetric_tori_lb}c)]
Note that when $k = NM/2$, the load must be greater than or equal to $\frac{NM}{4\sqrt{c_1c_2}L}$. This implies that the load is greater than or equal to $\frac{NM}{4\sqrt{c_1c_2}L}$ for $k \geq NM/2$ as the load cannot reduce as the total traffic increases.  
\end{proof}

\section{Generalized Local Load-Balancing Scheme}\label{appendix:GLLB}
Before the description of GLLB, we define some terminology. We define a \textit{general stem} set, similar to that in Section \ref{sec:routing_ub}. A \textit{stem} $S_{r_1,r_2}(i)$ of size $r_1, r_2$ at node $i$ is the set of nodes that differ from node $i$ only in one coordinate and is at most $r_1$ and $r_2$ hops away from node $i$ in the vertical and horizontal directions respectively. Mathematically, the stem is the set of nodes
$$S_{r_1,r_2}(i) \triangleq \left\{j \in V \;\bigg|\; \parbox{5cm}{$0 < \Delta(i,j) \leq r_1 \text{ and } j_x = i_x$, or \\
 $\; 0 < \Delta(i,j) \leq r_2 \text{ and } j_y = i_y$}\right\}.$$
The stem of a node is displayed in the example in Figure \ref{fig:GLLB_phase1} and \ref{fig:GLLB_phase3}. For the $N \times M$ torus, the stem of size $r_1,r_2$ has $2r_1 + 2r_2$ nodes. \\
\textbf{Remark.} \textit{One can also view the \textit{stem} $S_{r_1,r_2}(i)$ as the collection of $r_1$ and $r_2$ node neighbors of node $i$ along the positive and negative vertical and horizontal directions respectively. We will henceforth call the nodes in the stem along each of these directions as ``legs". There are 4 legs in the stem.}

The routing policy can be viewed as being split into three phases in general. We will only describe the set of routes from the origin to a destination $t$ such that $0 \leq t_x \leq M/2$ and $0 \leq t_y \leq N/2$. Recall that it is sufficient for us to do so because, from the consequences of automorphism invariance, the route from any source $s$ to any destination $s+t$ can be obtained from the transformation of the aforementioned set of routes. We describe the Generalized Local Load Balancing (GLLB) scheme with parameter $r_1, r_2$ for different cases. Without loss of generality, assume $M > N$ and both are even integers. The cases when $M\leq N$ and when either integer is odd can be similarly described.

\newcommand{\MinCut}{\textsc{MinCut}}
\subsection*{\underline{\textbf{Case 1}}: $2(r_1 + r_2) \leq N + \max(2r_2+1-M/2,0)$}
In this case, the $\MinCut(S_{r_1,r_2}(0), S_{r_1,r_2}(t))$ is at least $4(r_1 + r_2)$ when $S_{r_1,r_2}(0)$ and $S_{r_1,r_2}(t)$ do not intersect with each other. This happens in the case when $c_2N$ and $c_1M$ are of similar sizes (one dimension is not much smaller than the other). 
\subsubsection*{No overlap between $S_{r_1,r_2}(0)$ and $S_{r_1,r_2}(t)$}
\begin{figure}[!ht]
\centering
     \begin{subfigure}{0.45\textwidth}
        \centering
     \includegraphics[width=0.8\textwidth]{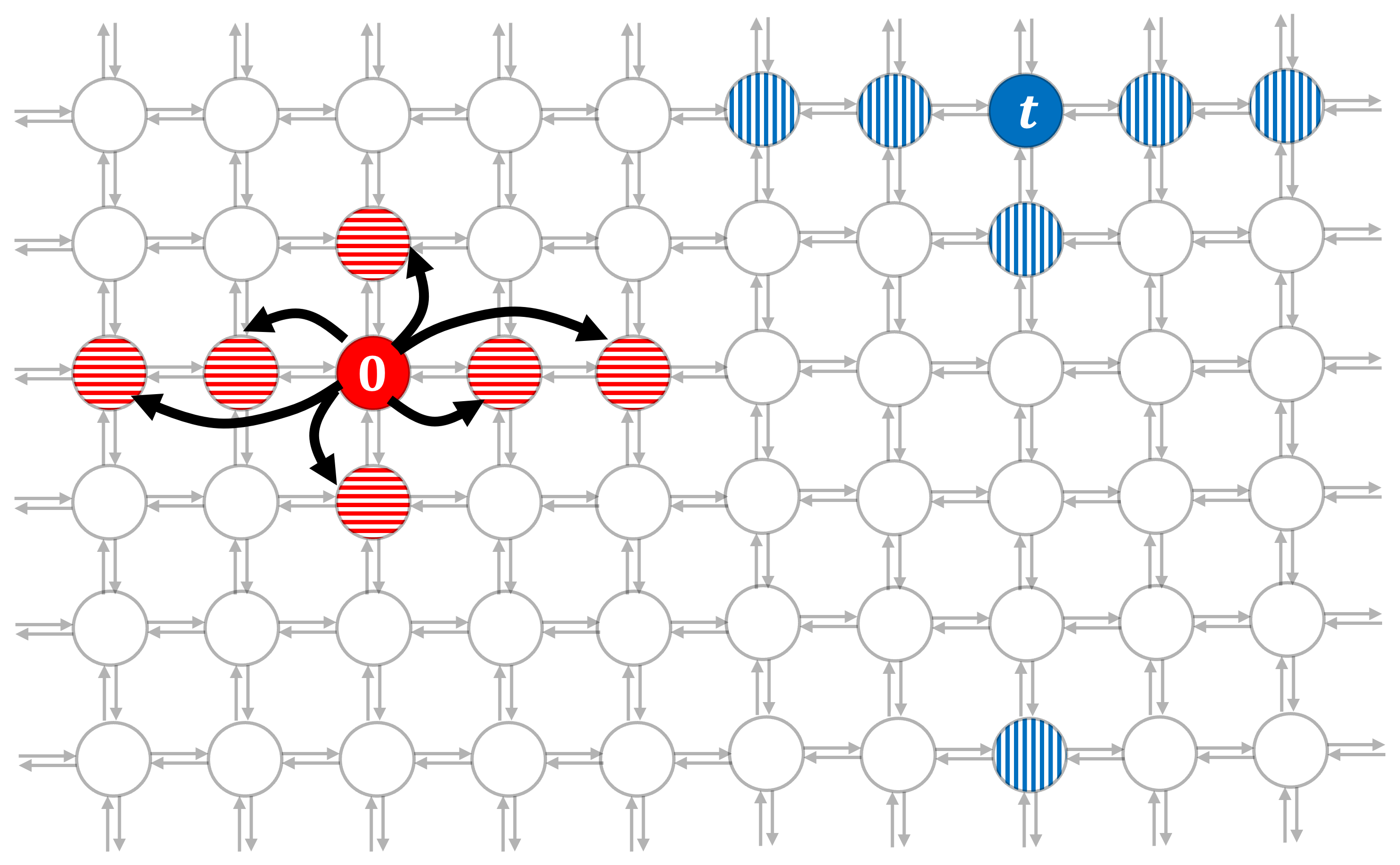}
         \caption{First phase - Distribute source traffic among stem nodes along the shortest path. The source (origin) is highlighted in dark red and the stem nodes $S_{r_1,r_2}(0)$ are shaded with red vertical lines.}
         \label{fig:GLLB_phase1}
     \end{subfigure}
     \hfill
     \begin{subfigure}{0.45\textwidth}
         \centering
        \includegraphics[width=0.8\textwidth]{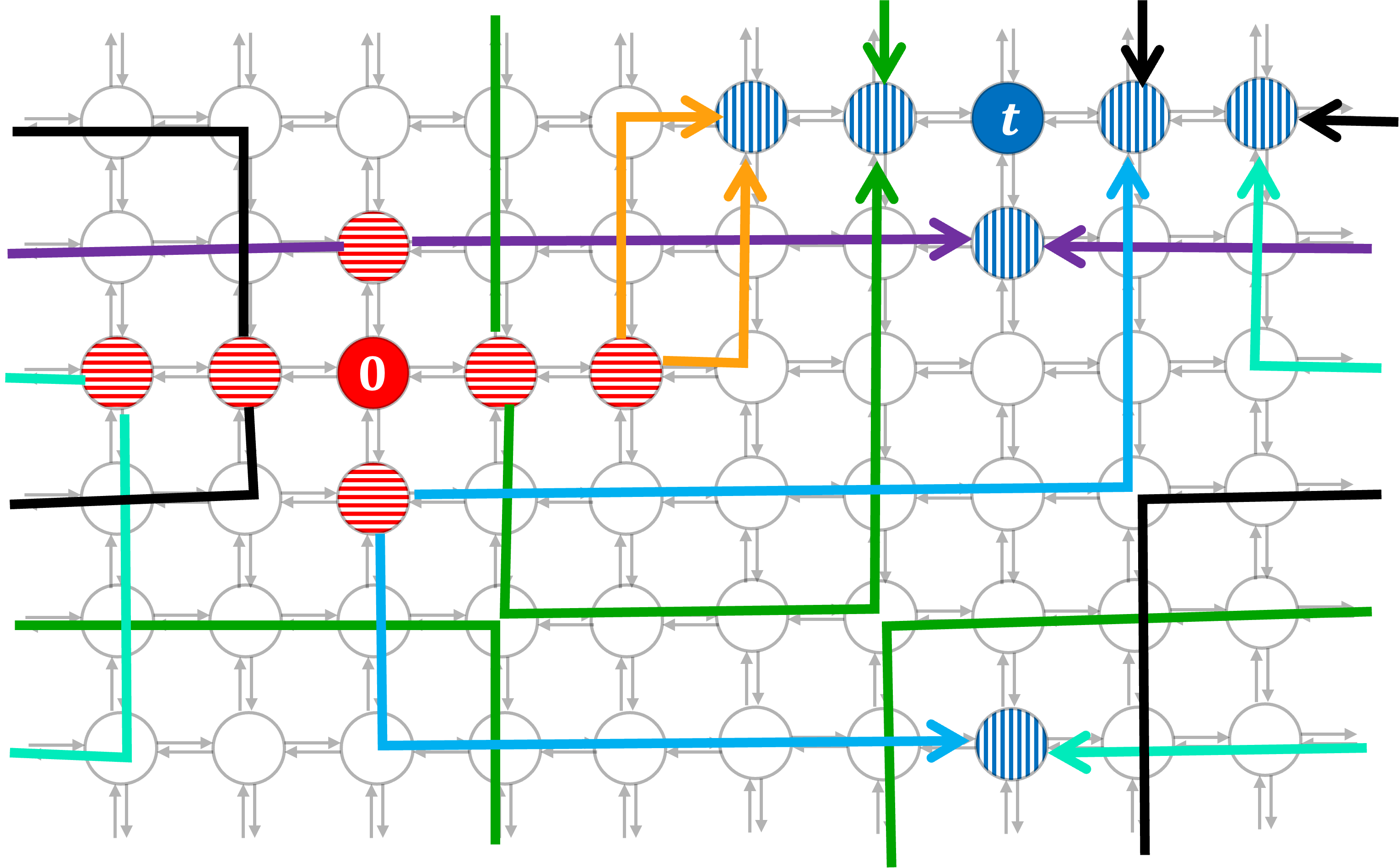}
         \caption{Second phase - Route traffic from $S_{r_1,r_2}(0)$ to $S_{r_1,r_2}(t)$ along edge-disjoint paths with two edge-disjoint paths emanating from each node in $S_{r_1,r_2}(0)$ and terminating at each node in $S_{r_1,r_2}(t)$.}
         \label{fig:GLLB_phase2}
     \end{subfigure}
     \begin{subfigure}{0.45\textwidth}
         \centering
        \includegraphics[width=0.8\textwidth]{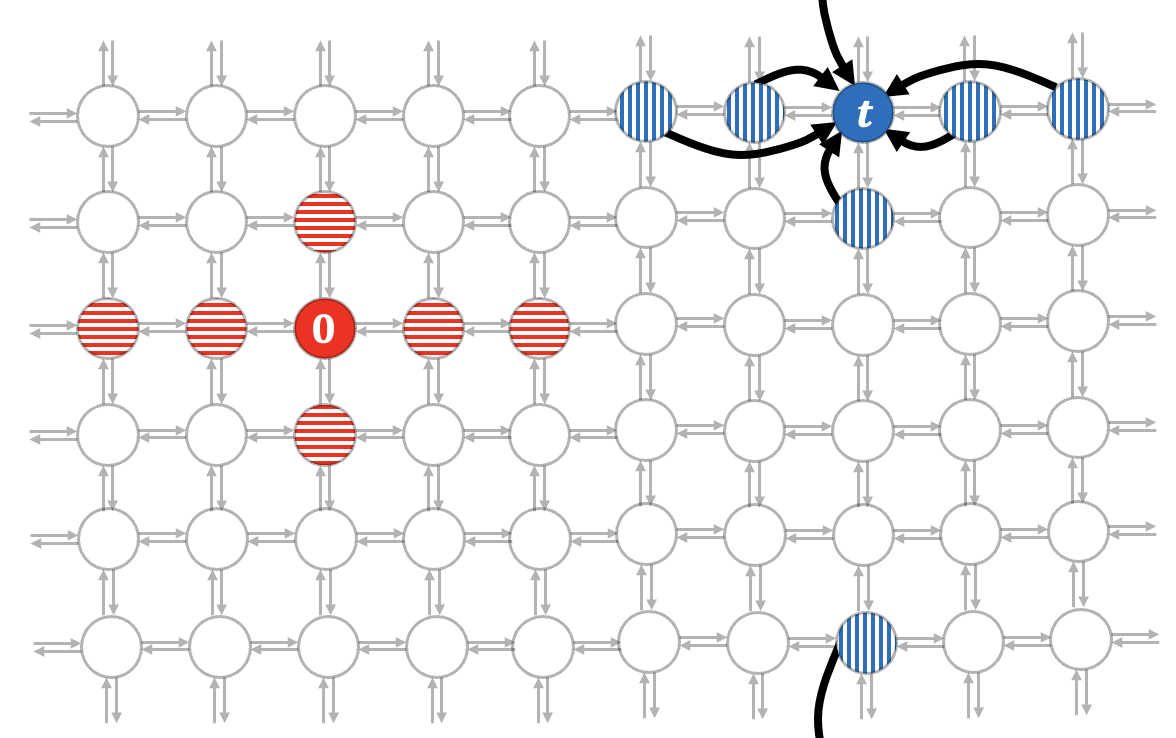}
         \caption{Third phase - Aggregate traffic from the stem nodes $S_{r_1,r_2}(t)$ to the destination node $t$ along the shortest path. The destination node is highlighted in dark blue and the stem nodes $S_{r_1,r_2}(t)$ are shaded with blue horizontal lines.}
         \label{fig:GLLB_phase3}
     \end{subfigure}
     \caption{Example of Generalized Local Load Balancing Routing Scheme with parameter $r_1=1, r_2=2$.}
     \label{fig:GLLB_example}
\end{figure}

 The $\MinCut$ condition tells us that there are at least $4r_1 + 4r_2$ edge-disjoint paths from the stem $S_{r_1,r_2}(0)$ to the stem $S_{r_1,r_2}(t)$ with each node in $S_{r_1,r_2}(0)$ having two paths emanating from it and  each node in $S_{r_1,r_2}(t)$ having two paths terminating at it (Menger's Theorem).
 In this case, the phases of the GLLB are as follows.\\
$\bullet$ \underline{\textit{First phase}}: Distribute traffic equally among the nodes in $S_{r_1,r_2}(0)$ along the shortest path. In other words, $1/(2r_1 + 2r_2)$ fraction of traffic is first sent from the origin to every node in $S_{r_1,r_2}(0)$.\\
$\bullet$ \underline{\textit{Second phase}}: Route traffic from each node in the stem $S_{r_1,r_2}(0)$ to each node in the stem $S_{r_1,r_2}(t)$ along edge-disjoint paths, such that each node in $S_{r_1,r_2}(0)$ sends equal traffic along 2 unique edge-disjoint paths and each node in $S_{r_1,r_2}(t)$ receives traffic along 2 edge-disjoint paths. Each path carries $1/(4r_1+4r_2)$ fraction of traffic. These paths can be found using Breadth-First-Search.\\
$\bullet$ \underline{\textit{Third phase}}: Aggregate traffic from the nodes in stem $S_{r_1,r_2}(t)$ to the final destination of the traffic, $t$, along the shortest path. As a consequence of the first two stages, $1/(2r_1+2r_2)$ fraction of traffic needs to be aggregated from each node in $S_{r_1,r_2}(t)$.
The GLLB scheme with $r_1=1, r_2=2$ is visualized in Figure \ref{fig:GLLB_example} for a $N\times M$ torus with $N=6, M=10$. In this example, there is no overlap between the stems and the $\MinCut$ condition is satisfied.

\subsubsection*{Overlap between $S_{r_1,r_2}(0)$ and $S_{r_1,r_2}(t)$}
When $S_{r_1,r_2}(0)$ and $S_{r_1,r_2}(t)$ overlap, we apply modifications to the above routing policy, just as in the case in Appendix \ref{appendix:llb_modification}.
Again, there are two kinds of overlap--

\textit{(i) Type-I Overlap}:
In Type-I overlap, the stems $S_{r_1,r2}(0)$ and $S_{r_1,r_2}(t)$ overlap only along one of the four legs of the stem. This only happens when $t = (t_x, 0)$ or $t=(0,t_y)$. Under Type-I overlap, we define modified stems $S^I_{r_1,r_2}(0)$ and $S^I_{r_1,r_2}(t)$,
\begin{align*}
    S^I_{r_1,r_2}(0) &= \{j\in V | j\in S_{r_1,r_2}(0) \text{ and }   \Delta(0,j) \leq \Delta(t,j)\}\\
    S^I_{r_1,r_2}(t) &= \{j \in V | j\in S_{r_1,r_2}(t) \text{ and }   \Delta(t,j) \leq \Delta(0,j)\}.
\end{align*}
Once we define the modified stem for Type-I overlap, the modifications in each routing phase is fairly straightforward. \\
$\bullet$ \underline{\textit{Modification in the first phase}}: Distribute $1/2(r_1+r_2)$ fraction of traffic to each of the nodes in the stems $S^I_{r_1,r_2}(0)$, except the last node in each of the legs. If the node is the last node in its leg and is $h$ hops away, it receives $(r_i-h+1)/2(r_1+r_2)$ fraction of the traffic, $i = 1,2$.\\
$\bullet$ \underline{\textit{Modification in the second phase}}: Route traffic from each node in the stem $S^I_{r_1,r_2}(0)$ to each node in the stem $S^I_{r_1,r_2}(t)$ (or $S^I_{r_1,r_2}(t)$ along edge-disjoint paths, such that each node in $S^I_{r_1,r_2}(0)$ sends equal traffic along 2 unique edge-disjoint paths and each node in $S^I_{r_1,r_2}(t)$ receives traffic along 2 edge-disjoint paths. Each path carries $1/4(r_1 + r_2)$ fraction of traffic. 
If there is residual traffic at a node in $S^I_{r_1,r_2}(0)$, then there must exist a node neighboring it that is in $S^I_{r_1,r_2}(t)$. The residual traffic must be directly sent to that neighbor. If a node belongs to both $S^I_{r_1,r_2}(0)$ and $S^I_{r_1,r_2}(t)$, then the node needs to do nothing in the second phase.\\
$\bullet$ \underline{\textit{Modification in the third phase}}: Aggregate traffic from the nodes in stem $S^I_{r_1,r_2}(t)$ to the final destination of the traffic, $t$, along the shortest path. As a consequence of the first two stages, $1/2(r_1 + r_2)$ fraction of traffic needs to be aggregated from each node in $S^I_{r_1,r_2}(t)$ except the last nodes in each leg. A fraction of $(r_i-h+1)/2(r_1+r_2)$ of the traffic will be aggregated from the last node in each leg, where the last node is $h$ hops away, $i = 1,2$.

\textit{(ii) Type-II Overlap}: In Type-II overlap, the stems $S_{r_1,r_2}(0)$ and $S_{r_1,r_2}(t)$ overlap exactly at two nodes at two of the four legs of the stem respectively. There needs to be no modification to the stem or the routing policy. However, note that there is a subtlety in the second phase. If a node is in both $S_{r_1,r_2}(0)$ and $S_{r_1,r_2}(t)$, no traffic needs to be explicitly forwarded from that node in the second phase, as it is already in the destination stem.

\subsection*{\underline{\textbf{Case 2}}: $2(r_1 + r_2) > N + \max(2r_2+1-M/2,0)$}
In this case, the $\MinCut(S_{r_1,r_2}(0), S_{r_1,r_2}(t))$ is smaller than $4(r_1 + r_2)$ when $S_{r_1,r_2}(0)$ and $S_{r_1,r_2}(t)$ do not intersect with each other. This happens in the case when $c_2N$ and $c_1M$ are of different sizes (one dimension is much smaller than the other). In this case, the $\MinCut(S_{r_1,r_2}(0), S_{r_1,r_2}(t))$ would be close to the Bisection cut of the torus. We exploit this fact in constructing the paths.

\subsubsection*{$S_{r_1,r_2}(0)$ and $S_{r_1,r_2}(t)$ do not intersect}
\begin{figure}[!ht]
\centering
     \begin{subfigure}{0.45\textwidth}
        \centering
     \includegraphics[width=0.8\textwidth]{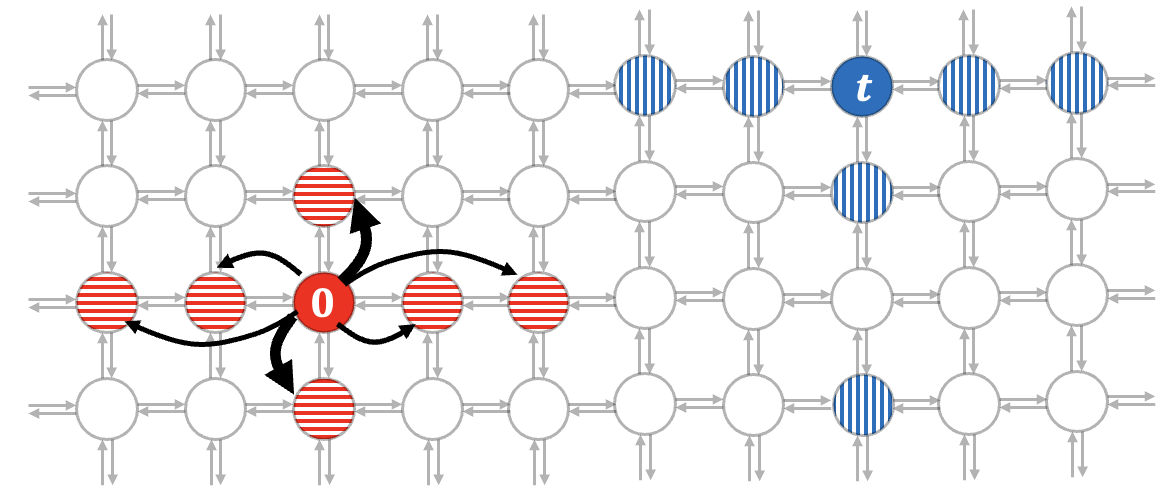}
         \caption{First phase - Distribute $2\lambda_2$ of traffic equally among the nodes in the vertical \textit{legs} of the $S_{r_1,r_2}(0)$ along the shortest paths, and,  distribute $2\lambda_1$ of traffic equally among the nodes in the horizontal \textit{legs} of the $S_{r_1,r_2}(0)$ along the shortest paths.}
         \label{fig:GLLB2_phase1}
     \end{subfigure}
     \hfill
     \begin{subfigure}{0.45\textwidth}
         \centering
        \includegraphics[width=0.8\textwidth]{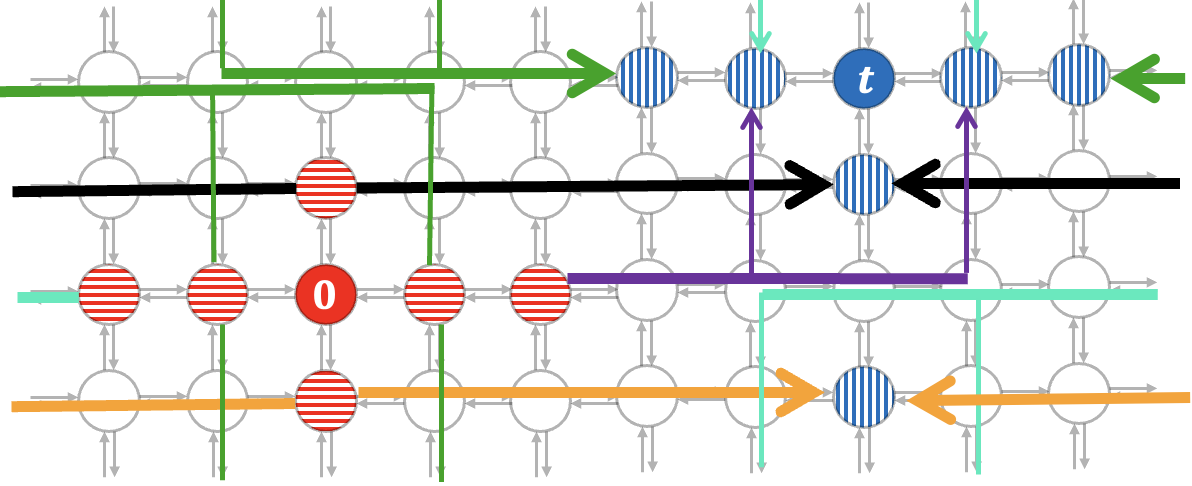}
         \caption{Second phase - Route traffic from $S_{r_1,r_2}(0)$ to $S_{r_1,r_2}(t)$ along carefully chosen paths. The thinner lines denote links that carry a maximum of $\lambda_1$ of the traffic and the thicker lines denote links that carry a maximum of $\lambda_2$ of the traffic. In our example, $\lambda_2/\lambda_1 = 2r_2/(N-2r_1) = 2$. Therefore, we see that at most 2 vertical paths aggregate into any one horizontal path.}
         \label{fig:GLLB2_phase2}
     \end{subfigure}
     \begin{subfigure}{0.45\textwidth}
         \centering
        \includegraphics[width=0.8\textwidth]{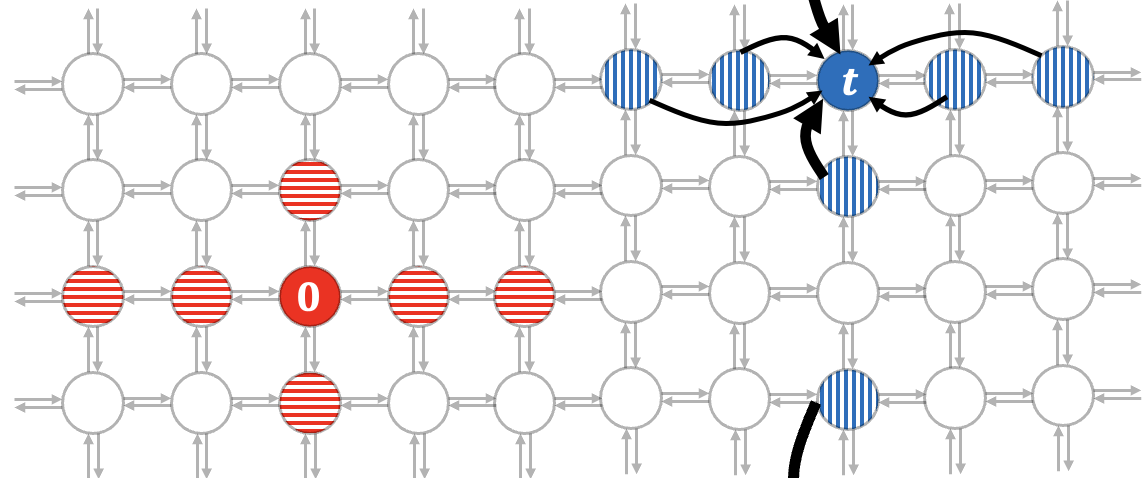}
         \caption{Third phase - Aggregate $2\lambda_2$ of traffic from the nodes in the vertical \textit{legs} of the $S_{r_1,r_2}(t)$ along the shortest paths, and,  aggregate $2\lambda_1$ of traffic from the nodes in the horizontal \textit{legs} of the $S_{r_1,r_2}(t)$ along the shortest paths.}
         \label{fig:GLLB2_phase3}
     \end{subfigure}
     \caption{Example of Generalized Local Load Balancing Routing Scheme with parameter $r_1=1, r_2=2$, and $N = 4, M = 10$. Since $\lambda_2 > \lambda_1$, the thinner lines show links carrying a maximum of $\lambda_1$ of the traffic and the thicker lines show links carrying a maximum of $\lambda_2$ of the traffic.}
     \label{fig:GLLB_example-2}
\end{figure}

We describe the GLLB routing policy in this case below:\\
$\bullet$ \underline{\textit{First phase}}: Distribute $1/N$ of traffic equally among the nodes in the vertical \textit{legs} of the $S_{r_1,r_2}(0)$ along the shortest paths, and,  distribute the remaining $1-2r_1/N$ traffic equally among the horizontal legs. In other words, each node in  the horizontal \textit{legs} of the $S_{r_1,r_2}(0)$ gets $(1/(2r_2) - r_1/(r_2N))$ of traffic equally among the nodes along the shortest path to them respectively.\\
$\bullet$ \underline{\textit{Second phase}}: In this phase, we route traffic from the source stem $S_{r_1,r_2}(0)$ to the destination stem $S_{r_1,r_2}(t)$ such that the vertical and horizontal links outside the stem carry a maximum fraction of $\lambda_1$ and $\lambda_2$ of the traffic respectively. We choose $\lambda_2 = 1/2N$ and $\lambda_1 = 1/(4r_2) - r_1/(2r_2N)$. 
Intuitively, since the $\MinCut(S_{r_1,r_2}(0), S_{r_1,r_2}(t)) < 4(r_1 + r_2)$, the $\MinCut$ is ``close" to the bisection cut of the network $2N$. We want to uniformly distribute of traffic along the bisection, which would be close to the $\MinCut$. Additionally, we \textit{claim} that there are $2N$ edge-disjoint paths from nodes in $S_{r_1,r_2}(0)$ to nodes in $S_{r_1,r_2}(t)$. The traffic from $(\lambda_2/\lambda_1)$ vertical edges will combine at some intermediate nodes in order to maintain the maximum of $\lambda_2$ fraction of traffic along horizontal edges. It is possible to find such a set of paths based on the choice of our $\lambda_1$ and $\lambda_2$ due to the grid-like geometry of the network, and fractional splitting of traffic.\\ 
$\bullet$ \underline{\textit{Third phase}}:
Aggregate traffic from the nodes in stem $S_{r_1,r_2}(t)$ to the final destination of the traffic, $t$, along the shortest path. As a consequence of the first two stages, $1/N$ fraction of traffic needs to be aggregated from each node in the vertical legs of $S_{r_1,r_2}(t)$, and $(1/(2r_2) - r_1/(r_2N))$ fraction of traffic needs to be aggregated from each node in the horizontal legs of $S_{r_1,r_2}(t)$.
The GLLB scheme in the second case is visualized in Figure \ref{fig:GLLB_example-2} for a $N\times M$ torus with $N=4, M=10$. In this example, there is no overlap between the stems and $\MinCut(S_{r_1,r_2}(0), S_{r_1,r_2}(t)) = 8 < 4r_1 + 4r_2 = 12$.
\subsubsection*{$S_{r_1,r_2}(0)$ and $S_{r_1,r_2}(t)$ intersect with each other}
When $S_{r_1,r_2}(0)$ and $S_{r_1,r_2}(t)$ intersect with each other, similar modifications (for Type-I and Type-II overlap) as in the previous case follow, while ensuring that the fraction of traffic flowing through the non-stem horizontal links is upper capped to $1/2N$, and the fraction of traffic flowing through the non-stem vertical links is $1/(4r_2) - r_1/(2r_2N)$. We omit the details in the interest of space.

\section{Optimality of the Generalized LLB Scheme}\label{proof:asymmetric_tori_ub}
\begin{proof}[Proof of Theorem \ref{thm:optimal_oblivious_upper_bound}a)] 
    Without loss in generality, assume $M > N$. We choose $r_1 = \left\lceil\sqrt{\frac{c_1k}{2c_2}}\right\rceil$ and $r_2 = \left\lceil\sqrt{\frac{c_2k}{2c_1}}\right\rceil$. Since $k \leq \min\left(c_2N^2/(2c_1), c_1M^2/(2c_2)\right)$, there will be totally $2r_1 + 2r_2$ nodes in the stems.
    Let us define \begin{align*}
        (\Lambda_1, \Lambda_2) &\triangleq \left(\frac{1}{4(r_1+r_2)}, \frac{1}{4(r_1+r_2)}\right) \text{ or }\\
        (\Lambda_1, \Lambda_2) &\triangleq \left( \frac{1}{4r_2}-\frac{r_1}{2r_2N}, \frac{1}{2N} \right).
    \end{align*} 
    These are the maximum fractions of traffic a non-stem link will carry along the vertical and horizontal direction respectively.
   Consequently, in the GLLB scheme with the chosen parameters $r_1,r_2$, the source and destination nodes must send and aggregate at most
    \begin{enumerate}[(i)]
        \item $2\Lambda_2$ fraction of traffic to every node in the vertical legs of $S_{r_1,r_2}(0)$ and $S_{r_1,r_2}(t)$ respectively,
       \item $2 \Lambda_1$ fraction of traffic to every node in the horizontal legs of $S_{r_1,r_2}(0)$ and $S_{r_1,r_2}(t)$ respectively,
    \end{enumerate}
    where $t$ is the destination.
 Additionally, by design of the GLLB scheme, any vertical or horizontal link neither in $S_{r_1,r_2}(0)$ nor $S_{r_1,r_2}(t)$ would carry a maximum of $\Lambda_1$ or $\Lambda_2$ of the traffic respectively. Therefore, 
    \begin{align*}
        g^t_{i,i+e_1} &\leq \begin{cases}
        2\Lambda_2\times(r_1 - \Delta(0,i-e_1)) & i \in S_{r_1,r_2}(0)\\
        2\Lambda_2\times(r_1 - \Delta(t,i)) & i \in S_{r_1, r_2}(t)\\
        \Lambda_1 & \text{else}
    \end{cases}, \text{ and}\\
    g^t_{i,i+e_2} &\leq \begin{cases}
        2\Lambda_1\times(r_2 - \Delta(0,i-e_2)) & i \in S_{r_1,r_2}(0)\\
        2\Lambda_1\times(r_2 - \Delta(t,i)) & i \in S_{r_1, r_2}(t)\\
        \Lambda_2 & \text{else}
    \end{cases}.
    \end{align*}
    Now, observe that the amount of traffic carried by any link would be maximized when it carries traffic from the stem nodes as well as traffic from non-stem nodes for different source-destination pairs. 
    Let us first consider the load of a vertical link. A vertical link can be in the stem of at most $2r_1$ source-destination pairs ($r_1$ unique sources and $r_1$ unique destinations). The remaining $k-2r_1$ traffic can contribute to the non-stem load. 
    Hence, the maximum load for a vertical link under GLLB is 
    \begin{align*}
        \textsc{VertLoad} &= \frac{1}{c_1}\left(2\times 2\Lambda_2 \sum_{i=0}^{r_1 - 1} (r - i) + (k-2r_1)\Lambda_1\right)\\
        &= \frac{2\Lambda_2r_1^2}{c_1} + 2\frac{r_1}{c_1}(\Lambda_2 - \Lambda_1) + \frac{k\Lambda_1}{c_1}\\
        &\leq \frac{\sqrt{2k}}{4\sqrt{c_1c_2}} + \frac{1}{\min(c_1, c_2)},
    \end{align*}
    where the last inequality comes from substituting our choices for $r_1, r_2, \Lambda_1$ and $\Lambda_2$ respectively.
    
    Similarly, a horizontal link can be in the stem of at most $2r_2$ source-destination pairs ($r_2$ unique sources and $r_2$ unique destinations). The remaining $k-2r_2$ traffic can contribute to the non-stem load. 
    Hence, the maximum load for a horizontal link under GLLB is 
    \begin{align*}
        \textsc{HorLoad} &= \frac{1}{c_2}\left(2\times 2\Lambda_1 \sum_{i=0}^{r_2 - 1} (r - i) + (k-2r_2)\Lambda_2\right)\\
        &= \frac{2\Lambda_1 r_2^2}{c_2} + 2\frac{r_2}{c_2}(\Lambda_1 - \Lambda_2) + \frac{k\Lambda_2}{c_2}\\
        &\leq \frac{\sqrt{2k}}{4\sqrt{c_1c_2}} + \frac{1}{\min(c_1, c_2)},
    \end{align*}
    where the last inequality comes from substituting our choices for $r_1$ and $r_2$, and either choice of $(\Lambda_1, \Lambda_2)$ respectively.
    
\end{proof}
\begin{proof}[Proof of Theorem \ref{thm:optimal_oblivious_upper_bound}b)]
Without loss of generality, let us assume $c_2N < c_1M$. In other words, we assume that the horizontal bisection bandwidth $2c_2N$ is lesser than the vertical bisection bandwidth $2c_1M$, and $L = \sqrt{c_2/c_1} N$. In the other case when $c_2N > c_1M$, we shall interchange the role of vertical and horizontal edges. 
Without loss of generality, let us assume the source node to be the origin $(0,0)$, and let the destination be $t=(t_x,t_y)$.

We distribute the traffic uniformly among the vertical nodes in the vertical ring of the source node (each node with coordinates $(0,y)$, i.e., node with the same $x$ coordinate as the source node gets $1/N$ of the traffic). Next, we forward the traffic along the horizontal edges from the vertical ring of the source node to the vertical ring of the destination using the $2N$ paths. In other words, we forward traffic from node $(0,y)$ to node $(t_x, y)$ along the two horizontal paths in the horizontal ring containing $(0,y)$ and $(t_x, y)$. Finally, we aggregate the traffic from each node in the vertical ring of the destination to the destination itself. See Figure \ref{fig:ring_load_balancing} for an example.

\begin{figure}[h]
    \centering
    \begin{subfigure}{0.45\textwidth}
        \centering
        \includegraphics[width=0.6\textwidth]{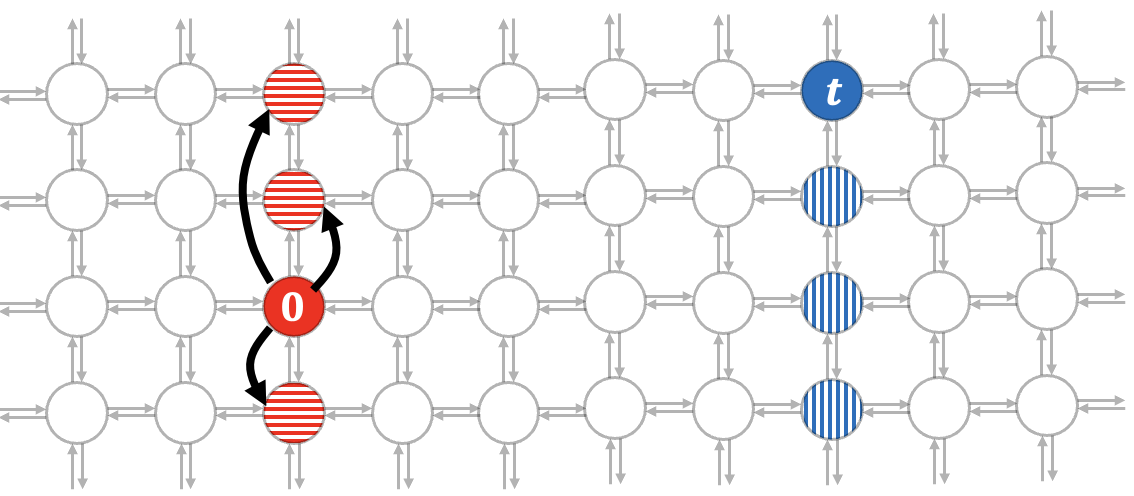}
        \caption{Distribute traffic uniformly across vertical ring of the source node.}
        \label{fig:ring_lb_phase1}
    \end{subfigure}
    \hfill 
    \begin{subfigure}{0.45\textwidth}
        \centering
        \includegraphics[width=0.6\textwidth]{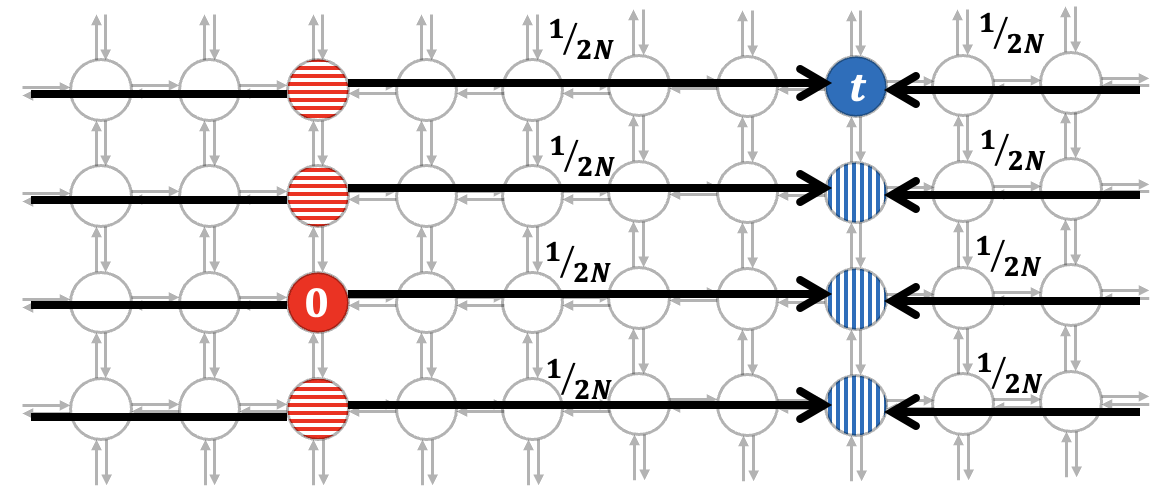}
        \caption{Forward traffic from the vertical ring of the source to the vertical ring of the destination node.}
        \label{fig:ring_lb_phase2}
    \end{subfigure}
    \hfill
    \begin{subfigure}{0.45\textwidth}
        \centering
        \includegraphics[width=0.6\textwidth]{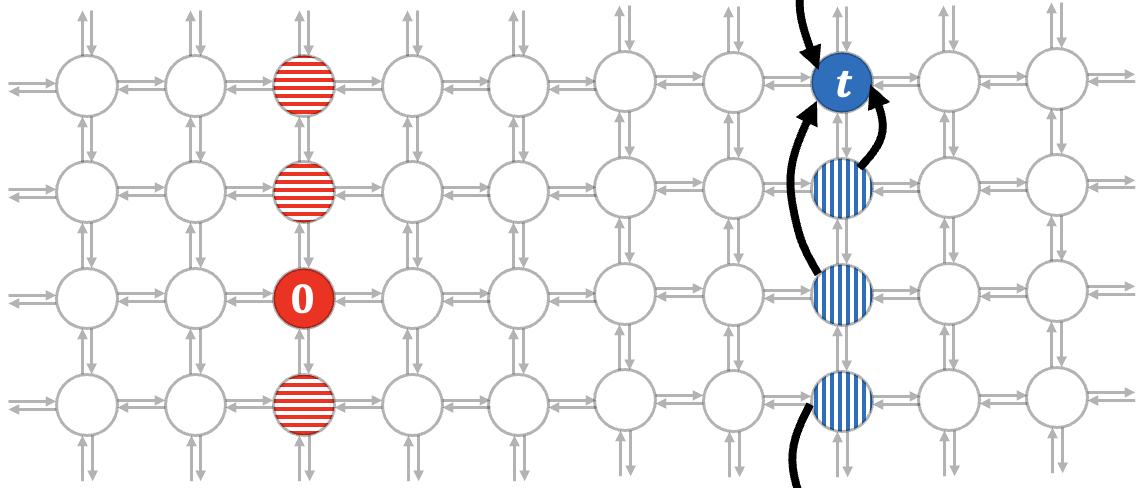}
        \caption{Aggregate traffic uniformly from vertical ring of the destination node.}
        \label{fig:ring_lb_phase3}
    \end{subfigure}
    \caption{Ring Load Balancing for $k \geq L^2/2$.}
    \label{fig:ring_load_balancing}
\end{figure}

Based on this routing policy, the load on any vertical link is at most $N/4c_1$. This is because the load from uniformly distributing unit traffic from each node in a ring of size $N$ with link capacity $c_1$ is $N/8c_1$ \cite{applegate_making_2006, dally_principles_2004}. This load appears twice, once due to distribution phase and once due to the aggregation phase.

Since traffic is uniformly distributed among the nodes in the vertical rings, the maximum load due to any one source destination pair on a horizontal link is at most $1/2Nc_2$. The worst case load on any one horizontal link is when sources situated in one half of the network need to forward traffic to destinations in the other half of the network. When $L^2/2 \leq k \leq NM/2$, this load is $k/2Nc_2 = k/2L\sqrt{c_1c_2}$. When $k \geq NM/2$, the maximum number of nodes in a particular bisection of the network would be $NM/2$. Consequently, when $k \geq NM/2$, the maximum load on a horizontal link is $NM/4Nc_2 = NM/4L\sqrt{c_1c_2}$. 
\end{proof}

\end{document}